\documentclass[twoside,onecolumn,11pt,a4paper]{book}
\usepackage{anuthesis}
\usepackage{fancyhdr}
\usepackage{graphicx}
\usepackage{mathrsfs}
\usepackage{makeidx}         
\usepackage{graphicx}        
\usepackage{multicol}        
\usepackage{cite}            
\usepackage[bottom]{footmisc}
\usepackage{bm}
\usepackage{amsmath}
\usepackage{amssymb}
\usepackage{color}
\usepackage[final]{pdfpages}
\usepackage[hidelinks]{hyperref}
\usepackage[framemethod=TikZ]{mdframed}
\usepackage{enumitem}
\usepackage{float}
\usepackage{titling}
\usepackage{xfrac}
\usepackage{newfloat}
\usepackage{caption}

\mdfdefinestyle{BenFrame}{%
    outerlinewidth=0pt,
    roundcorner=10pt,
    innertopmargin=\baselineskip,
    innerbottommargin=\baselineskip,
    innerrightmargin=20pt,
    innerleftmargin=20pt,
    backgroundcolor=gray!30!white}

\newcommand*{\scriptbelow}[3][1pt]{%
  \begingroup
    \renewcommand*{\arraystretch}{0}%
    \begin{array}[t]{@{}c@{}}%
      #2\\[{#1}]%
      \scriptstyle #3%
    \end{array}%
  \endgroup
}

\newcommand*{\bbar}[1]{\bar{\bar{#1}}}

\makeatletter
\newcommand{\raisemath}[1]{\mathpalette{\raisem@th{#1}}}
\newcommand{\raisem@th}[3]{\raisebox{#1}{$#2#3$}}
\makeatother

\DeclareFloatingEnvironment[fileext=frm,placement={!ht},name=Box]{myfloat}
\begin{document}

\pagenumbering{roman}  

\begin{titlepage}
\title{\LARGE \textbf{Collective Resonances in Nanoparticle Oligomers}\\[2cm]}
 \author{\Large \textmd{A thesis submitted for the degree of}\\[1ex]
 \Large\textmd{Doctor of Philosophy} \\[1ex]
\Large \textmd{The Australian National University}\\[3cm]
 \Large\textmd{Ben Hopkins}\\[3cm]
 }
 \date{\Large\textmd{August 2017}}
\predate{\begin{center}}
\postdate{
\\\vspace{20mm} \includegraphics[width=0.5\textwidth]{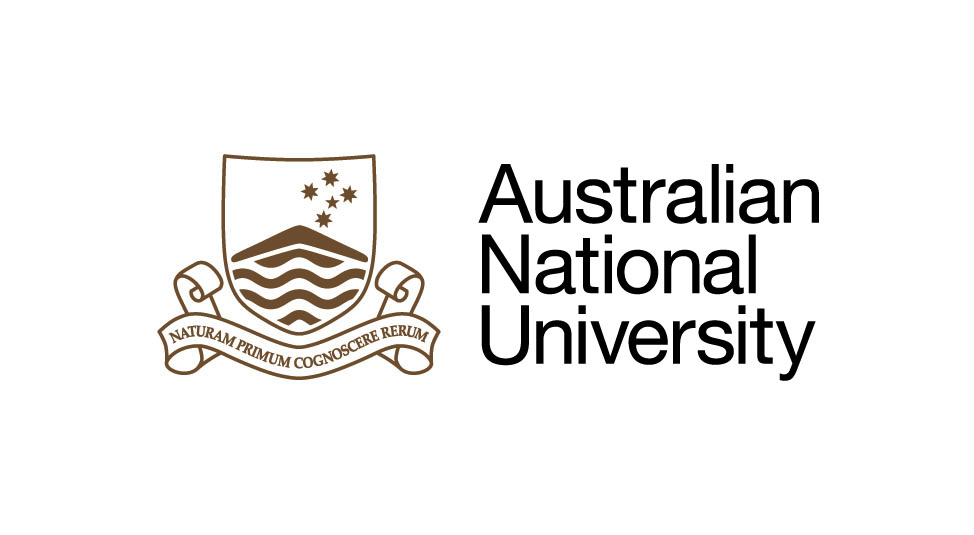}\end{center}
}
\maketitle
 \end{titlepage}
 
 \sloppy

\newpage
\thispagestyle{empty}

\chapter*{Declaration}
\addcontentsline{toc}{section}{Declaration}
\vspace{20mm}  
This thesis is an account of research undertaken at the Nonlinear Physics Centre between January 2013 and November 2016, while I was enrolled in the \mbox{degree} of \mbox{Doctor of Philosophy} at the Australian National University.
The research was conducted under the supervision of \mbox{Andrey Miroshnichenko} within the Research School of Physics and Engineering.  
I declare that the material presented within this thesis is my own work except where stated otherwise, and has never been submitted for any degree at this or any other institution of learning.
\vspace{5mm}  

\hspace{80mm}\rule{40mm}{.15mm}\par   
\vspace{18mm}  
\hspace{80mm} Ben Hopkins\par
\hspace{80mm} August, 2017

\setcounter{page}{1}

\chapter*{Acknowledgements}
\addcontentsline{toc}{section}{Acknowledgements}
\vspace{20mm}  

My most clear acknowledgement needs to go to my supervisor Andrey Miroshnichenko, who has driven my interest in broad aspects of physical theory, and has freely offered more of his time than should ever be expected of any supervisor. 
The many discussions, the subtle guidance and steering, but also the free pursuit of my own research directions despite regularly unproductive ends; these will be the most enduring memories of my time at NLPC.
Substantial recognition should also go to the advice and guidance offered by \mbox{Yuri Kivshar}, and for the multiple opportunities to travel and learn from other groups.
Indeed I am very grateful to \mbox{Brian Stout}, \mbox{Andrea Al\`u}, and \mbox{Costantino de Angelis}, for each hosting me in their own research groups.
Otherwise, there is an extensive list of colleagues I should thank, both from NLPC and visiting from elsewhere, each of whom have influenced different parts of my studies: from the experimental projects I had the opportunity to be involved in, to the like-minded pub companions. 
Though I particularly should acknowledge \mbox{David Powell} and \mbox{Mingkai Liu} for the sheer number of red couch discussions over lunch, but also \mbox{Wei Liu} and \mbox{Alexander Poddubny} for their respective roles in guiding me onto the path of the research presented herein.
Now, in concluding, I could not overlook my officemate \mbox{Guangyao Li}, who has both been my company and reliable counsel on many, often poorly defined, physics problems.
I am grateful to all for the time I spent pursuing my studies presented herein.

\chapter*{Abstract}
\addcontentsline{toc}{section}{Abstract}
\vspace{20mm}  


The study of nanostructured artificial media for optics has expanded rapidly over the last few decades, coupled with improvements of fabrication technology that have enabled investigation of previously unrealisable optical scattering systems.
Such development is complemented by renewed impetus to understand the physics of optical scattering from complex subwavelength geometry and nanoparticle systems.  
Here I investigate specifically the optical properties of closely packed arrangements of  nanoparticles, known as {\it nanoparticle oligomers}, which provide an intuitive platform for analytical and numerical study on the formation and interplay of collective resonances.
I consider both plasmonic nanoparticles, and also high-refractive-index dielectric nanoparticles that support Mie-type electric \textsl{and magnetic} dipole resonances.  
Specific outcomes of this study are listed as follows.
(i)~A new model is presented for optical Fano resonances, which is based on interference between nonorthogonal eigenmodes of the associated scattering object.
This is demonstrated to correctly describe Fano resonances in both plasmonic and high-refractive-index dielectric nanoparticle oligomers; it also revealed capacity for two-channel Fano interference in the magnetic dipolar response from the dielectric oligomers. 
(ii)~Polarisation-independent scattering and absorption losses are shown to be enforced by $n$-fold discrete rotational symmetry, $C_n$ ($n\geq3$), and reciprocal degeneracy of eigenmodes.
(iii)~A new form of circular dichroism is presented, which occurs due to the interaction of nonorthogonal resonances, and impacts the ratio of radiative scattering loss to dissipative absorption loss experienced by reciprocal plane waves.  
Geometric asymmetry and optical chirality are also reviewed to quantify the minimum symmetries that must be broken to allow other circular dichroism effects in chiral and achiral scattering objects.
The sequence of general theoretical conclusions (i)-(iii) serve to build the understanding of optical scattering from nanoparticle systems while removing existing ambiguities.
\newpage
\chapter*{Notation}
\addcontentsline{toc}{section}{Notation}
\subsection*{Use of fonts}
\begin{tabular}{ll}
$u$ & scalar\\
$\mathbf u$ & vector (function of frequency)\\
$\boldsymbol u$ & vector (function of time)\\
$\mathbf{\hat u}$ & unit vector\\
$\mathbf{\bbar u}$ & matrix\\
$\hat u$ & operator
\end{tabular}

\subsection*{Generic terms}
\begin{tabular}{ll}
$t$ & time\\
$\omega$ & angular frequency\\
$\mathbf r$ & position vector \\
$V$, $\Omega_V$ & a volume and its surface 
\end{tabular}

\subsection*{Electromagnetism}
\begin{tabular}{ll}
$\epsilon_0, \mu_0$ & permittivity and permeability of background medium\\
$\mathbf E(\mathbf r, \omega), \boldsymbol E(\mathbf r, t)$ & electric field\\
$\mathbf J(\mathbf r, \omega), \boldsymbol J(\mathbf r, t)$ & electric current\\
$\mathbf P(\mathbf r, \omega), \boldsymbol P (\mathbf r, t)$ & electric polarization\\
$\mathbf p(\omega), \boldsymbol p(t)$ & electric dipole moment\\
$\mathbf B(\mathbf r, \omega), \boldsymbol B(\mathbf r, t)$ & magnetic B-field\\
$\mathbf H(\mathbf r, \omega), \boldsymbol H(\mathbf r, t)$ & magnetic H-field\\
$\mathbf m(\omega), \boldsymbol m(t)$ & magnetic dipole moment\\
$P(\omega), \sigma(\omega)$ & power and cross-section\\
$\mathbf j_v(\mathbf r, \omega), \mathbf p_v(\omega), \mathbf m_v(\omega)$ & currents and dipole moments of an eigenmode $|v\rangle$\\
$\lambda_v(\omega)$ & eigenvalue of an eigenmode $|v\rangle$
\end{tabular}

\subsection*{Symmetry operations}
\begin{tabular}{ll}
${\hat C_n}$ & rotation about a principle axis by ${\frac{2\pi}{n}}$ radians \\
${\hat \sigma_v, \hat \sigma_d}$ & reflection plane parallel to the principle axis \\
${\hat \sigma_h}$ & reflection plane perpendicular to the principle axis  \\
${\hat S_n}$ &  improper rotation: both ${\hat C_n}$ and  ${\hat \sigma_h}$  \\
${\hat i}$ & point inversion through the origin 
\end{tabular}
\newpage
\clearpage\setcounter{page}{0}

\tableofcontents

\pagenumbering{arabic} 
\setcounter{page}{1}  

\chapter{Introduction}

The study of artificial optical media grew from a desire to capitalise on the broad advancements in nanoscale fabrication capacity, and to enable new optical functionality that cannot be realised with conventional materials.
Here I will introduce, first: the reason and motivation for research into artificial media and nanoparticle optical systems generally, and second: the physics involved in the optical scattering from nanoparticles.  
Discussion on these topics is not intended to be comprehensive, but to provide sufficient background and context for the technical Chapters~\ref{chapterModels}-\ref{chapterChiral}. 
This introductory Chapter will then end with a brief overview of the broader goals of the thesis and each of the coming chapters.

\section{Artificial nanostructured materials for optics}

\begin{figure}[!ht]
\centerline{\includegraphics[width=0.88\textwidth]{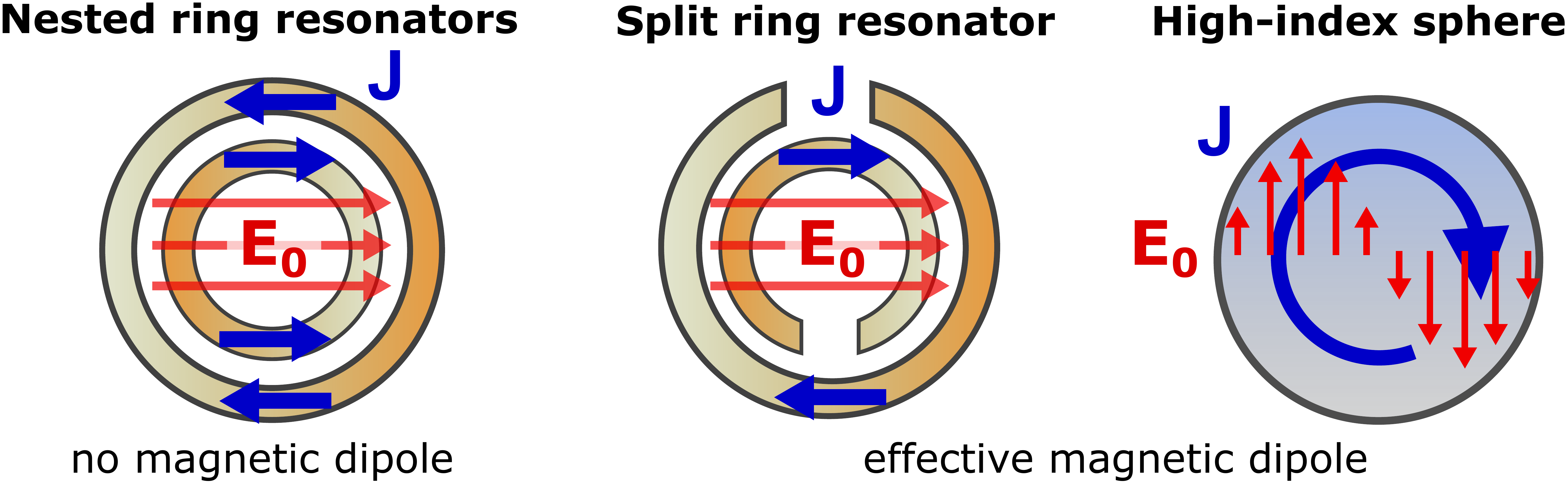}}
\caption{Qualitative illustration of magnetic response from nanoparticles.  A pair of nested ring resonators with different resonant frequencies, which produce induced currents $\mathbf J$ that are $\pi$ out of phase, will not allow a net circulation of current (left).  However, introducing two splits to suppress part of the current in each ring leads to net circulating current (centre).  Alternatively, a high-refractive-index dielectric sphere can instead use the retardation of an applied plane wave's electric field to produce a circulating polarization current (right).    }
\label{fig:SRR}
\end{figure}

Optical technology is a well-established tool throughout research, manufacturing, and technology; its mark found on some of the foremost achievements of society, from optical microscopy in the discovery of bacteria and microbiology, to the fabrication of semiconductor microprocessors that now saturate our digital age. 
Yet contemporary demands increasingly call for optical functionality that would lie beyond the limits of conventional, diffractive optics. 
Indeed, many elusive facets of nature would become readily accessible once we can observe the structure and dynamics of subcellular networks, or protein structure and its folding~\cite{Huang2010, Sydor2015}; or the greater part of energy consumed by computation could be suppressed once we can miniaturise optical communication networks to replace their electrical counterparts within computing infrastructure, and microchips themselves~\cite{Miller2009, Agrell2016}. 
As part of this push away from conventional optics, the development of artificial media for optics has been receiving a surge of renewed interest over the last two decades. 
The growing capacity to accurately deposit~\cite{Xia2011}, etch~\cite{Huang2011} or assemble~\cite{Liu2013_ChemRev,Vogel2015} nanostructured materials has enabled investigation of previously unimagined optical systems, particularly those grouped largely as {\it metamaterials}: artificial media made by massed concatenation of subwavelength-sized resonant nanoparticles or general nanostructured inclusions.    
Conceptually, metamaterials aim to mimic the construction of natural materials from atoms, or molecules, by substituting nanoparticles or analogous inclusions as artificial {\it meta-atoms} that remain small relative to the operating wavelength of light. 
With control over the geometry of the constituent meta-atoms, and their arrangement, it becomes possible to design the optical properties of the collective media. 
This freedom has allowed artificial media, or nanostructures generally, to be employed for rapidly diversifying research pursuits.
Some particular examples of note are: the enhancement of spontaneous emission from nearby molecules~\cite{Purcell1946, Tam2007, Yoo2015, Fernandez-Corbaton2016}, combined with sharp frequency selective sensing~\cite{Stewart2008, Li2015} of even \textsl{single} molecules~\cite{Zijlstra2012}; the imaging of light~\cite{Bauer2014, Rotenberg2014} or objects~\cite{Park2014}; control over propagation and state of light~\cite{Bomzon2002, Yu2011, Arbabi2015, Kruk2016}; and the enhancement of nonlinear harmonic generation, wavemixing and switching effects~\cite{Shcherbakov2014, Celebrano2015, Shcherbakov2015, Yang2015}.
However, until very recently, the original concept of bulk metamaterials was impeded at optical frequencies by the limited capacity of three-dimensional material fabrication at application-relevant volumes. 
Seemingly in a bid to bypass this challenge, optical metamaterials research made a shift toward single- or few-layer realisations of metamaterials, so-called {\it metasurfaces}, as two-dimensional analogues of the existing metamaterials.  
The promise underlying optical metasurfaces was perhaps conveyed most concisely by Pfeiffer and Grbic~\cite{Pfeiffer2013}, recognising that the surface equivalence principle (Stratton-Chu formulation~\cite{StrattonChu1939}) implies that a surface of both electric and magnetic currents can perform a reflectionless transformation between any two sets of electric and magnetic fields on opposing sides.  
In essence, any optical operation becomes possible if we can impose an arbitrary polarisation and magnetisation distribution; this being a functionality that artificial nanostructured media can aim to provide.  
During the initial surge of metamaterials research, metallic structures were designed to support loop currents in response to light, and create the effective magnetic response to imprint a magnetisation distribution.  
Perhaps the most iconic of these geometries was the split ring resonator, which permitted a magnetic response using a split in a ring resonator  (loop antenna) to reduce the symmetry and permit coupling into oscillating circulating current from a normally incident plane wave~\cite{Pendry1999}. 
See illustration in Figure~\ref{fig:SRR}.

However, a persisting challenge for this approach was the inherent Ohmic losses of metals leading to unavoidable dissipation of light.
While reflection-based operation could be viable and support efficiencies in excess of 50\% with metals~\cite{Pors2013}, the maximum operational transmission efficiency, even with metasurfaces, was at~\cite{Zhao2013} or below~\cite{Ni2013} 50\%. 
Such limitations from dissipative losses were largely resolved by the predictions~\cite{Evlyukhin2010, Nieto-Vesperinas2010} and realisations~\cite{Garcia-Etxarri2011,Ginn2012, Kuznetsov2012, Evlyukhin2012} that simple nanoparticles made of low-loss and high-refractive-index dielectric materials, such as silicon or germanium, would inherently support both  electric and \textsl{magnetic} dipolar optical resonances, with comparable magnitude to the electric dipole resonances of gold or silver nanoparticles.
Furthermore, as shown nicely in the Supplementary Information of~\cite{Kuznetsov2012}, the magnetic dipole resonance becomes the lowest-frequency resonance as the refractive index of a dielectric sphere increases, with the second resonance being an electric dipole.
As such, it wasn't necessary to fabricate loops or other complicated geometries: high-index dielectric nanoparticles would support resonant circulation of polarisation current with very simple geometries such as arrays of spheres, disks, or bars.  
Subsequent control over both electric and magnetic resonant responses then enabled the metasurface concept presented by Pfeiffer and Grbic~\cite{Pfeiffer2013} to create arbitrary polarisation and magnetisation distributions on a surface, but now with low losses and twin resonances to obtain full $2\pi$ phase control~\cite{Decker2015}.  
Current dielectric metasurfaces can now function at above 90\% transmission efficiency for focusing~\cite{Arbabi2015, Khorasaninejad2016}, holography~\cite{Arbabi2015, Chong2016}, polarisation control and exotic beam forming~\cite{Chong2015, Arbabi2015, Kruk2016}.

While the original premise of metamaterials was to replicate continuous media~\cite{Pendry2006} with constituent elements much smaller than the operational wavelength, this is rarely the case: the majority of metamaterials and metasurfaces have lattice periods on the order of a wavelength, and therefore have some semblance to photonic crystals~\cite{Joannopoulos2008book}.  
Here I will instead make the claim that the key conceptual distinction  of optical metamaterials is that their constituent elements are resonant at the operational wavelength in isolation, being now referred to as {\it nanoantennas}. 
The isolated resonances are what allows almost arbitrary spatial contrast in the phase, amplitude and orientation of imprinted polarisation and magnetisation distributions, and what allows optically thin elements to alter macroscopic forms of light. 
Indeed, I would argue that the most distinctive physical freedom of metamaterials, the one that supposes to provide a functionality beyond the existing photonic crystals, is the resonant properties of the nanoantennas themselves.  
My studies have focused on presenting and developing analyses for operational principles of nanoantennas, and particularly those that support multiple interacting resonances.
I have particularly been interested in the formation of resonances and collective optical responses arising from coupling between nanoparticles, as the analogy and precursor of artificial materials being formed from constituent nanoantennas.
Specifically, I focus on groups of several nanoparticles arranged in closely packed clusters, so-called {\it nanoparticle oligomers}.  
The concept of nanoparticle oligomers was introduced for metal nanoparticles~\cite{Hentschel2010}, where they could allow more complicated optical responses, while simultaneously offering simplification in fabrication, which still typically favours basic nanoparticle geometries such as spheres and other primitive shapes.  
The subsequent transition to incorporate high-index dielectric nanoparticles that support electric and magnetic dipole moments into nanoparticle oligomers, provides an analogue to the local resonant sources of polarisation and magnetisation that nanoantennas represent for artificial materials.
This is particularly interesting given nanoparticle oligomers operate between isolated nanoparticle response and collective media response; collective resonances exist in nanoparticle oligomers, but the resonant properties of any individual nanoparticle remain.
Indeed, oligomers provide a window to explore the evolution between isolated and collective resonant responses that we utilise in the pursuit of artificial media.
This thesis will not emphasize any specific implementation, instead the investigation aims to provide relevant insight on the principles of collective resonances in nanoantennas and artificial nanostructured materials generally.

\newpage
\section{Optical scattering from nanoparticles \label{firstPrinciples}}
%
Here I briefly discuss how distributions of currents and charges presented by the classical electromagnetism model can be parametrised to resemble material phenomena like polarisation and magnetisation.
This aims to give a contextual precursor to models for the optical scattering from nanoparticles presented in Chapter~\ref{chapterModels}, and the subsequent parametrisation of currents into the resonant eigenmodes considered in Chapter~\ref{chapterEigen}.

We must appropriately begin with Maxwell's Equations, which describe the evolution of real electric $\boldsymbol{E}$ and magnetic $\boldsymbol{B}$ fields at the position $\mathbf r$ and time $t$, existing in a homogeneous background medium with permittivity $\epsilon_0$ and permeability $\mu_0$, prescribing a speed of light $c_0 = (\epsilon_0\mu_0)^{-\frac{1}{2}}$, with some charge $\rho$ and current $\boldsymbol J$.
\begin{align}
\boldsymbol{\nabla}\cdot \boldsymbol{B} &= 0 \label{eq:divB}\\
\boldsymbol{\nabla} \times \boldsymbol{E} + \frac{\partial}{\partial t}\boldsymbol{B} &= \mathbf 0  \label{eq:curlE}\\
\boldsymbol{\nabla} \cdot \boldsymbol{E} &= \frac{\rho}{\epsilon_0}  \label{eq:divE}\\
\boldsymbol{\nabla} \times \boldsymbol{B}- {c_0}^{-2} \frac{\partial}{\partial t}\boldsymbol E &= \mu_0  \boldsymbol{J}  \label{eq:curlB}
\end{align}
Note that these equations implicitly assume the continuity of charge $\boldsymbol{\nabla}\cdot \boldsymbol J = -\frac{\partial \rho}{\partial t}$, seen by taking the divergence of (\ref{eq:curlB}), and then substituting (\ref{eq:divE}).
To define interaction with matter, we consider the  Lorentz force imparted on any given charge $q$ moving with a velocity $\boldsymbol v_q$ at a location $\mathbf{r}$ and at time $t$.
\begin{align}
\boldsymbol F_q = q \boldsymbol E(\mathbf r, t) + q \big(\boldsymbol v_q \times \boldsymbol B(\mathbf r, t) \big) \label{eq:force}
\end{align}
Notably, the magnetic field isn't needed to describe the electromagnetic force on $q$ in the inertial reference frame where it is stationary.
The electric field $\boldsymbol E_{q'}$ at position $\mathbf{r}$ and time $t$, produced by some arbitrarily moving charge $q'$, located at $\mathbf{r'}$ at the retardation-adjusted time $t' = t- \frac{R}{c_0}$, with $R = |\mathbf{r}-\mathbf r'|$, was presented by Feynman, \S28~\cite{FeynmanI}, but is equivalent~\cite{Gitman2016} to fields described by Li\'enard-Wiechert potentials~\cite{Jackson}.
\begin{align}
\boldsymbol E_{q' }(\mathbf{r},t)&= \frac{q'}{4 \pi \epsilon_0}\bigg(\frac{\mathbf{\hat n}}{R^2}+ \frac{R}{c_0}\frac{\mathrm d}{\mathrm dt}\Big( \frac{\mathbf{\hat n}}{R^2}\Big )+\frac{1}{{c_0}^2}\frac{\mathrm d^2}{\mathrm dt^2}\mathbf{\hat n}\bigg)
\label{eq:Feynman}
\end{align}
Here $\mathbf{\hat n}$ is the unit vector pointing from $\mathbf{r'} $ to $\mathbf r$, meaning: $\mathbf r - \mathbf{r'} = R\, \mathbf{\hat n}$.
By then writing the \textsl{total} electric field at $(\mathbf r, t)$ as a sum over the fields generated by an arbitrary number of $q'$, $\boldsymbol E(\mathbf r, t) = \sum_{q'} \boldsymbol{E}_{q'}(\mathbf r, t) $, the physical force on $q$ is given by $\boldsymbol F_q = q \boldsymbol E$.
Albeit not remotely practical, we could correctly model all electromagnetic interaction using only electric fields and different reference frames for each $q$.
However, the need for both electric and magnetic fields emerges by defining materials with polarisation $\boldsymbol P$, but also magnetisation $\boldsymbol M$ that interacts directly with $\boldsymbol B$, both of which can be presented as the density of electric dipole moments and magnetic dipole moments per unit volume~\cite{Jackson}.  
In this regard, while the assumption of point dipoles distributed within some background volume is physically reasonable, given existence of atoms and the like, we are concerned with artificial media constructed from nanoparticles that have nonzero volume.
This suggests we should make the distinction of introducing polarisation and magnetisation in terms of currents and charge. 
We therefore consider {density distributions} of charge and current, $\rho(\mathbf r) \rightarrow \mathrm d \rho \equiv \rho(\mathbf r)\mathrm{dr}^3$, $\boldsymbol J(\mathbf r) \rightarrow \mathrm d \boldsymbol J \equiv \boldsymbol J(\mathbf r)\mathrm{dr}^3$, to define the electric dipole $\boldsymbol p$ and magnetic dipole $\boldsymbol m$ of some volume $V$ centred about a point $\mathbf r$.
\begin{align}
\boldsymbol p =& \int _{\mathbf r' \in V}\!\! (\mathbf r' - \mathbf r) \rho(\mathbf r') \mathrm{d r'}^3  ,\quad 
\boldsymbol m = 
\frac{1}{2}
 \int _{\mathbf r' \in V}\!\! (\mathbf r' - \mathbf r) \times \boldsymbol J(\mathbf r') \mathrm{d r'}^3    \label{eq:magdip0}
\end{align}
The equivalent density of point dipoles per unit volume at a point $\mathbf r$, will then be the density of $\boldsymbol p$ or $\boldsymbol m$  in a volume $V\rightarrow 0$ about $\mathbf r$.
\begin{align}
\boldsymbol{P}(\mathbf r) = \underset{V \rightarrow 0}{\text{lim}}\;   \int_{\mathbf r' \in V} \! \! (\mathbf r' - \mathbf r) \rho(\mathbf r') \mathrm{d r'}^3\! \big/V ,\quad
\boldsymbol{M}(\mathbf r) &= \underset{V \rightarrow 0}{\text{lim}}\;   
\frac{1}{2}
\int_{\mathbf r' \in V} \! \! (\mathbf r' - \mathbf r) \times \boldsymbol J(\mathbf r') \mathrm{d r'}^3\! \big/V
\label{eq:PM}
\end{align}
Now, a local misalignment of $\boldsymbol E$ with some $\boldsymbol p$ in the limit of $V\rightarrow 0$ constitutes a local \textsl{restoring} torque $ \boldsymbol{\tau_{\! p}}$ acting on  $\boldsymbol p$, and similarly for $\boldsymbol B$ with $\boldsymbol m$. 
\begin{align}
 \boldsymbol{\tau_{\! p}}  =\Big(\underset{V \rightarrow 0}{\text{lim}} \;\boldsymbol{p}\Big) \times \boldsymbol{E}(\mathbf r),\quad
 \boldsymbol{\tau_{\! m}}    =\Big(\underset{V \rightarrow 0}{\text{lim}} \;\boldsymbol{m}\Big) \times {\boldsymbol{B}(\mathbf r)} 
\label{eq:torque}
\end{align}
If the timescale for movement of charge and current is small compared to that for variation in the applied field, we can then assume local alignment of $\boldsymbol P$ with $\boldsymbol E$, and $\boldsymbol M$ with $\boldsymbol B$, because of the torques $ \boldsymbol{\tau_{\! p}} ,\,\boldsymbol{\tau_{\! m}}$.  
More generally, we can consider fields that oscillate harmonically at an angular frequency $\omega$, for which we can expect some steady state average effect of the torque acting to align dipole moments to the corresponding fields.  
This thereby lets us enstate a general tensor relationship between respective phasors, specifically the one of permittivity and permeability: $\mathbf P=(\boldsymbol{\bbar \epsilon}-\epsilon_0)\mathbf E$, and $\mathbf M=(\boldsymbol{\bbar \mu}-\mu_0)\frac{1}{\mu_0}\mathbf B$.
Such relationships are useful because any arbitrary distribution of time-varying fields, $\boldsymbol u$ for generality, can be expressed as a spectral distribution of harmonic phasors $\mathbf u$ in a causal Fourier representation.
\begin{align}
\boldsymbol{u}(\mathbf r, t) &= \int\limits _{- \infty}^{\infty} \mathbf{u}(\mathbf r, \omega ) e^{- i \omega t} \mathrm d \omega, \quad
\mathbf{u}(\mathbf r, \omega) = \int\limits_{-\infty}^{t_{0}} \boldsymbol{u}(\mathbf r, t ) e^{i \omega t} \mathrm d t \label{eq:Fourier}
\end{align}
I have chosen to end the time integral at a time $t_0$, rather than covering the full $(-\infty, \infty)$ interval, to represent the absence of $\boldsymbol{u}$ beyond the current time $t_0$ in any causal measurement. 
This allows us to consider $\mathbf{u}$ as being derived from an observed $\boldsymbol{u}$, though it does mean that $\mathbf{u}$ treats $\boldsymbol{u}(t)=0$ for $t > t_0$.
The absence of italics will be used herein to denote complex phasors in the frequency domain.

We now simply recognise that the tensor relationship of $\mathbf P,\,\mathbf M$ with $\mathbf E,\,\mathbf B$, which defines materials, can arise from the charge and current distributions in (\ref{eq:PM}), provided the restoring torques $\boldsymbol{\tau_p}$ and $\boldsymbol{\tau_m}$ in (\ref{eq:torque}).
The expression for $\boldsymbol{\tau_{\!p}}$ in (\ref{eq:torque}) follows by substituting $\boldsymbol{F} = \rho \boldsymbol E$ from (\ref{eq:force}) and $\boldsymbol p$ from (\ref{eq:magdip0}) into $\boldsymbol p \times \boldsymbol E$, and assuming $V\rightarrow 0$ for the electric field to be considered as uniform $\boldsymbol{E}(\mathbf r')= \boldsymbol{E}(\mathbf r)$. 
The torque $\boldsymbol{\tau_{\!m}}$ in (\ref{eq:torque}) can be derived for point magnetic dipoles~\cite{Kholmetskii2014}, or from $\boldsymbol{\tau_{\!p}}$ using an electromagnetic duality transformation $(\boldsymbol{E},c_0\boldsymbol{B},\boldsymbol{p},c_0\boldsymbol{m})\rightarrow(c_0\boldsymbol{B},-\boldsymbol{E},c_0\boldsymbol{m},-\boldsymbol{p})$, though this effectively treats $\boldsymbol m$ as being due to (non-physical) magnetic charge~\cite{Figueroa-O'Farrill1998}.  
Otherwise the derivation of $\boldsymbol{\tau_{\!m}}$ can be considered using just currents, see box.
In all cases, the physical notion from the tensor relationship of $\mathbf P$ with $\mathbf E$, and $\mathbf M$ with $\mathbf B$, which defines material, is now that the electric and magnetic fields are heralds of linear and rotational forces between distributed charges.
This will have been at least historically convenient, because it corresponds to the way natural materials respond: atoms and other neutral compositions of charged matter dominantly behave as electric and magnetic dipoles. 
\begin{myfloat}[!h]
\begin{mdframed}[style=BenFrame]
\small
{\bf Magnetic dipoles from a current distribution.}
{Let us define the current  $\boldsymbol{J_m}$, associated with an $\boldsymbol m$, such that it only contains components that contribute to $\boldsymbol m$ from (\ref{eq:magdip0}).  
This means that: $ \boldsymbol{J_m}(\mathbf r') \cdot (\mathbf r'-\mathbf r)\! =\!0$ and $ \boldsymbol{J_m}(\mathbf r')\cdot \boldsymbol m \!=\! 0$.
Consider the quantity $\boldsymbol m \times \boldsymbol B(\mathbf r)$ with (\ref{eq:magdip0}), where $\boldsymbol B$ is the \textsl{total} magnetic field, not just that due to $\boldsymbol{J_m}$.  
\begin{align}
{\boldsymbol{m} \times \boldsymbol{B}(\mathbf r)} 
&= 
\int _{\mathbf r' \in V} \!\! \big[ \underset{\rightarrow \boldsymbol{\tau_m}}{\underbrace{(\mathbf r' - \mathbf r) \!\times\! \big (
{{
 \boldsymbol{J_m}(\mathbf r') \!\times\! \boldsymbol B(\mathbf r)  }}
\big) }}+   \boldsymbol{J_m}(\mathbf r')\!\times\! \big ({{ \boldsymbol B(\mathbf r) \!\times\! (\mathbf r' - \mathbf r)   }} \big)\big] \mathrm{d r'}^3 
\label{eq:mtimesB}
\end{align}
Here we have applied the vector identity  
$(\mathbf a\times \mathbf b)\times \mathbf c = \mathbf a \times (\mathbf b\times \mathbf c) + \mathbf  b \times (\mathbf c\times \mathbf  a)$.
If we take the limit $V\rightarrow0$ to assume $\boldsymbol B(\mathbf{r}')=\boldsymbol B(\mathbf{r})$ and perform a substitution for the local force $\boldsymbol F =  \boldsymbol{J_m}\times \boldsymbol B$ from (\ref{eq:force}), the first term of (\ref{eq:mtimesB}) is the torque $\boldsymbol{\tau_m}$.
The second term of (\ref{eq:mtimesB})  can be expanded with a vector triple product $\mathbf a\times (\mathbf b\times \mathbf c) = \mathbf b  (\mathbf a \cdot \mathbf c)  -  \mathbf  c (\mathbf a\cdot \mathbf  b)$ to make use the orthogonality of $ \boldsymbol{J_m}$ and $(\mathbf r' - \mathbf r)$.
\begin{align}
  \boldsymbol{J_m}(\mathbf r')\times \big ({{ \boldsymbol B(\mathbf r) \times (\mathbf r' - \mathbf r)   }} \big) = - (\mathbf r' - \mathbf r)   \big (  \boldsymbol{J_m}(\mathbf r')\cdot \boldsymbol B(\mathbf r)\big) \label{eq:residual}
\end{align} 
This is zero if $ \boldsymbol{J_m}$ is perpendicular to $\boldsymbol B$, however $\boldsymbol B$ is able to be arbitrarily oriented, such as when it is dominated by external sources.
The second term of (\ref{eq:mtimesB}) is therefore \textsl{not} forbidden by our existing constraints on $\boldsymbol{J_m}$.
The expression for torque $\boldsymbol{\tau_m}$ in (\ref{eq:torque}) appears to require we make a further assumption about $ \boldsymbol{J_m}$ and/or $V$ to make the second term of (\ref{eq:mtimesB}) zero, while keeping the first term nonzero. 
At the very least, if $\boldsymbol{J_m}$ is invariant under arbitrary coordinate rotations about $m$, we can visualise  $\boldsymbol{J_m}$ in uniform rings of circulation about $\boldsymbol m$, {\it i.e.} where $\boldsymbol{J_m}(\mathbf r')$ at fixed $|\mathbf r' - \mathbf r|$ is oriented with uniform magnitude in the direction of $(\mathbf r' - \mathbf r) \times \boldsymbol m$.
The contribution to the torque from (\ref{eq:residual}) integrated over any such ring is then orthogonal to $\boldsymbol B$ and $\boldsymbol m$, hence at least parallel to $\boldsymbol{\tau_{\!m}}$ in (\ref{eq:torque}).
}
\end{mdframed}
\end{myfloat}
But let us now consider the artificial case of a given nanoparticle, occupying a volume $V$, which too only supports an electric and a magnetic dipole moment, and we can write these dipole moments in the frequency domain, {\it i.e.} $\mathbf p$ and $\mathbf m$, using (\ref{eq:Fourier}).
\begin{align}
\mathbf p = \int_V (\mathbf{r}-\mathbf{r}') \rho(\mathbf{r}') \mathrm{dr'}^3 , \qquad \mathbf m = \frac{1}{2}\int_V (\mathbf{r}-\mathbf{r}') \times \mathbf J(\mathbf r') \mathrm{dr'}^3 \label{eq:CartesianDipoles}
\end{align}
We can place a hypothetical bounding sphere around the nanoparticle and use the result of Devaney and Wolf~\cite{Devaney1974}, which states that all fields external to any sphere that bounds an arbitrary object will be precisely described by the radiation of the complete set of {\it spherical harmonic multipoles}.
Moreover, the radiation of the given dipole moments  $\mathbf p$ and $\mathbf m$  in (\ref{eq:CartesianDipoles}) can be described by some corresponding $a_1$ and $b_1$ scattering coefficients~\cite{Grahn2012}:
\begin{subequations}\label{eq:pandmintermsofa1andb1}
\begin{align}
\mathbf p &= \frac{6 \pi i \epsilon_0}{ k^3}\, \Big[\frac{1}{\sqrt{2}} (a_{11}- a_{1-1})\mathbf{\hat x} + \frac{i}{\sqrt{2}} (a_{11}+ a_{1-1})\mathbf{\hat y}+   a_{10}\mathbf{\hat z}\Big] \\
\mathbf m &= \frac{6 \pi i}{k^3} \sqrt{\frac{\epsilon_0}{\mu_0}}\, \Big[ \frac{1}{\sqrt{2}} (b_{11}- b_{1-1})\mathbf{\hat x} + \frac{i}{\sqrt{2}} (b_{11}+ b_{1-1})\mathbf{\hat y}+  b_{10}\mathbf{\hat z}\Big]
\end{align}
\end{subequations}
Further details on the spherical multipole decomposition, and spherical nanoparticles, are provided in the second box.
It is also worth mentioning that, unlike (\ref{eq:pandmintermsofa1andb1}), the general relationship between the dipole moments $\mathbf p$ and $\mathbf m$ in (\ref{eq:CartesianDipoles}) and the {\sl total} $a_1$ and $b_1$ coefficients, will require a series of additional correction terms to the dipole moments, where each correction term radiates identically to the $a_1$ or $b_1$ coefficient~\cite{Miroshnichenko2015}.  
These correction terms generally become more relevant as the volume $V$, of the physical system in (\ref{eq:CartesianDipoles}), approaches the order of the wavelength of light.
However, if we instead have prescribed knowledge of the $a_1$ and $b_1$ coefficients, then we can choose to define {\sl effective} dipole moments $\mathbf p$ and $\mathbf m$ of point dipoles that will radiate identically to the known $a_1$ and $b_1$ coefficients.  
This is what is done for the case of spheres in (\ref{eq:spheresonwards}).
Such effective dipole moments can be different to those defined in (\ref{eq:CartesianDipoles}), given they will account also for the additional correction factors.   
In any case, we return to our earlier discussion: the mentioned result of Devaney and Wolf~\cite{Devaney1974}, combined now with the equivalence of radiation from $\mathbf p$ and $\mathbf m$ to that described by $a_1$ and $b_1$, allows us to conclude that all fields external to the smallest bounding sphere around the given dipolar nanoparticle are the same as the fields radiated by point dipoles with moments $\mathbf p$ and $\mathbf m$.  
This means that an oscillating or circulating current in a nanoparticle is indistinguishable, at points external to a bounding sphere, to a point of ``true" polarisation or magnetisation.
\begin{myfloat}[!h]
\begin{mdframed}[style=BenFrame]
\small
{\bf The spherical multipole decomposition.}
Details here on the spherical multipole decomposition are from \S9~of~\cite{Jackson}, and also from~\cite{GrahnThesis}, details on the scattering from a sphere are from \S4~of~\cite{Bohren1983}. 
The {\it vector spherical harmonics} $\mathbf X_{lm}$ are origin-dependent vector fields defined from the {\it scalar spherical harmonics} $Y_{lm}$ as per:
\begin{align}
\mathbf X_{lm}  = \frac{1}{\sqrt{l(l+1)}} ( \mathbf r \times \boldsymbol \nabla) Y_{lm},   \quad Y_{lm} = \sqrt{\frac{2l + 1}{4 \pi}\frac{(l-m)!}{(l+m)!}}P^m_l(\cos \theta) \, e^{i m \phi}
\end{align}
Here $\theta$ and $\phi$ are the azimuthal and polar angles of $\mathbf r$, and $P^m_l$ is the {\it associated Legendre function}.
 The scattering coefficients $a_{lm}$ and $b_{lm}$ denote complex amplitudes for the decomposition of a given $\mathbf E$-field distribution in terms of the vector spherical harmonics:
\begin{align}
\mathbf E (\mathbf r) = \sum \limits_{l=0}^{\infty}\sum \limits_{m=-l}^l  \frac{a_{lm}}{k} \boldsymbol \nabla \times \Big(f_l (k|\mathbf r|) \mathbf X_{lm}\Big) + b_{lm} g_l(k|\mathbf r|) \mathbf X_{lm}
\end{align}
The functions $f_{l}(k |\mathbf r|)$ and $g_{l}(k |\mathbf r|)$ are normalisation factors for the $a_{lm}$ and $b_{lm}$ coefficients of the form:  $f_{l}(k |\mathbf r|) \!=\! A_l \,h^{\!(1)}_l\!(k |\mathbf r|) + B_l \,h^{\!(2)}_l\! (k |\mathbf r|)$, where $A_l$ and $B_l$ are free $l$-dependent coefficients, and $h^{(1)}_l$ and $h^{(2)}_l$ are the $l$-th order {\it spherical Hankel functions} of the first and {second kind}.
The specific normalisation I use corresponds to that for homogeneous spheres in \S4 of~\cite{Bohren1983}, which provides analytic solutions for $a_{lm}$ and $b_{lm}$ under plane wave illumination, known as {\it Mie Theory}~\cite{Mie1908}.
Being spheres, the orientation of the plane wave polarisation can be neglected, subsuming the $m$ dependence to define scalar $a_l$ and $b_l$ coefficients.
Specifically, for a sphere of radius $r$ and refractive index $n$, the $a_l$ and $b_l$ coefficients due to a plane wave with unit amplitude at the origin are expressed in terms of {\it Ricatti-Bessel functions}: $\psi_l(x) = x j_l(x)$ and $\xi_l(x) = x h_l^{(1)}(x)$, and their derivatives with respect to $x$ (denoted here as $\psi'$ and $\xi'$), where $j_l$ is the $l$-th order {\it spherical Bessel function of the first kind}.  
\begin{align}
a_l &= \frac{n \psi_l(nr) \psi'_l(r)-\psi_l(r) \psi'_l(nr)}{n \psi_l(nr) \xi'_l(r)-\xi_l(r) \psi'_l(nr)} \\
b_l &= \frac{\psi_l(nr) \psi'_l(r)-n\psi_l(r) \psi'_l(nr)}{\psi_l(nr) \xi'_l(r)-n\xi_l(r) \psi'_l(nr)}
\end{align}
For the specific case of $l=1$, we can use the following substitutions: 
\begin{align}
\psi_1(x) &= \frac{\sin x}{x}-\cos x, \quad 
\psi'_1(x) = \frac{\cos x}{x} - \frac{\sin x}{x^2}   + \sin x \\
\xi_1(x) &= 
e^{i x}\big (\!-1 - \frac{i}{x} \big)
,   \quad \xi'_1(x) = 
e^{i x} \big(\!-i  + \frac{1}{x}  +\frac{i}{x^2})
\end{align}
We then obtain a simple relationship between the dipole moments of spheres and the spherical scattering coefficients for a plane wave with field $\mathbf{E_0},\mathbf{H_0}$ at the origin.    
\begin{align}
\mathbf p &= \frac{6 \pi i \epsilon_0}{k^3} a_1 \mathbf{E_0}, \quad \mathbf m = \frac{6 \pi i}{k^3} b_1 \mathbf{H_0} \label{eq:spheresonwards}
\end{align}  
\end{mdframed}
\end{myfloat}
This is a conclusion at the heart of artificial optical materials: it explains why sufficiently small nanoparticles are able to serve, at least optically, as building blocks of artificial materials in a manner analogous to atoms in conventional materials.  
There are some residual distinctions, such as effective magnetisation from a nanoparticle will be induced by the anti-symmetric component of the electric field over a volume $V\!\not \rightarrow 0$ of $\boldsymbol m$ in (\ref{eq:magdip0}), and not the magnetic field, which is relevant to $V \!\rightarrow 0$.
However, this distinction is also a reason for the access to strong magnetic responses when using nanoparticles: the larger displacements $(\mathbf r - \mathbf r')$ of circulating currents in (\ref{eq:CartesianDipoles}), relevant to $\mathbf m$. 
Yet, we must ultimately recognise that the models of dipoles, polarisation and magnetisation still inherently correspond to material as found in nature, and we have no particular guarantee that assemblies of arbitrarily shaped subwavelength nanoantennas will respond to fields as per linear or rotational movement of current.
Indeed, the macroscopic description of polarisation and magnetisation has already been recognised to not align with even layered media~\cite{Sheinfux2014}, and particularly as we reduce symmetry, the resonant optical responses of even single nanoparticles cease to align with purely electric or magnetic dipoles~\cite{Evlyukhin2011}.  
As such, it is not necessarily appropriate to use a homogenised polarisation and magnetisation description for nanostructured media, indeed different models and analyses are necessary to model the scattering of light from assemblies of nanoparticles. 
This is the reason why alternate modelling approaches are presented in Chapter~\ref{chapterModels}, and it also illustrates a motivation to parametrise optical responses instead according to eigenmodes in Chapter~\ref{chapterEigen} onwards.  

\clearpage
\section{Outline of context statement}

\begin{figure}[H]
\centerline{\includegraphics[width=\textwidth]{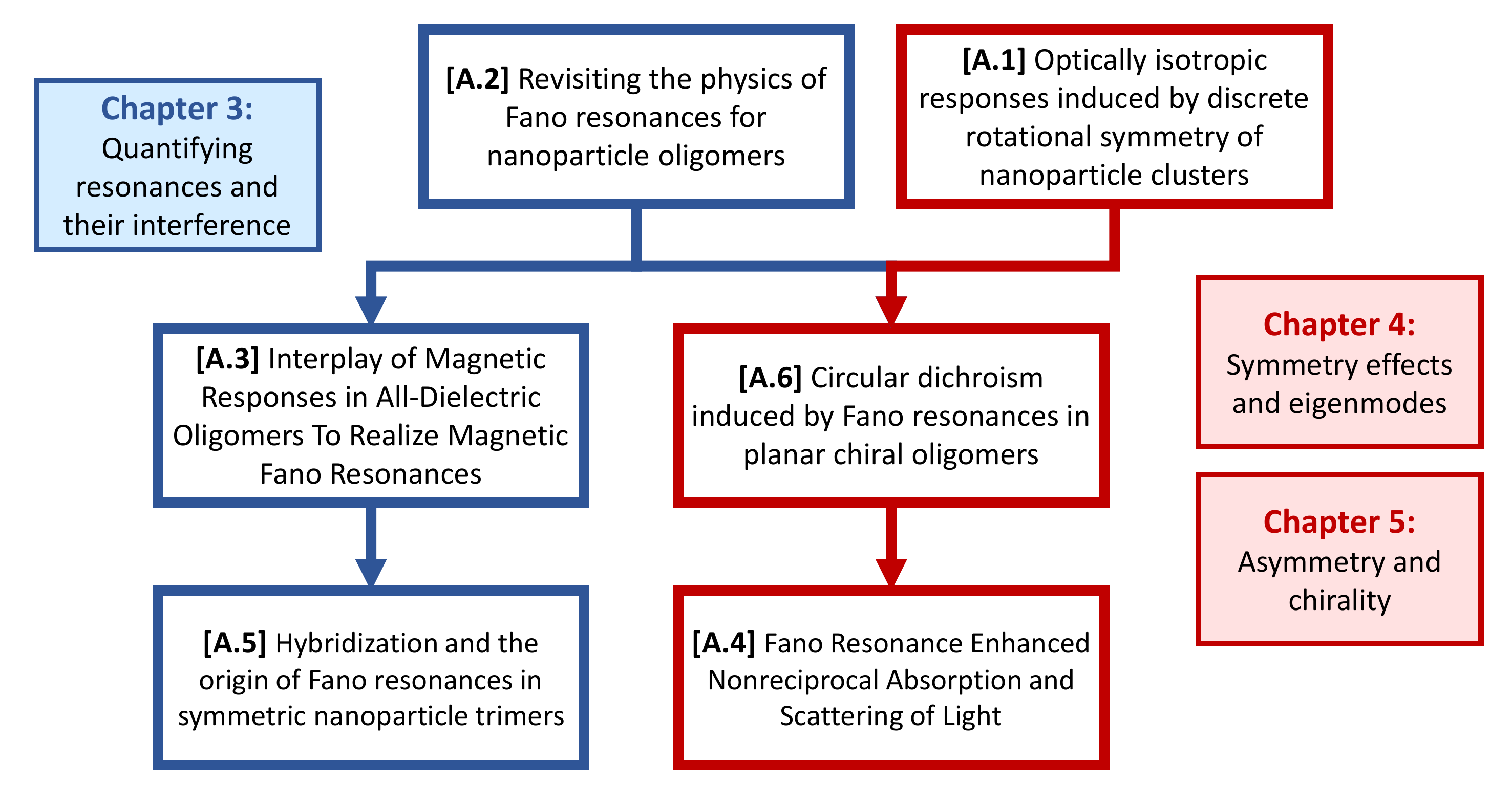}}
\caption{The six papers from Appendix~\ref{pubs} with their relationships denoted by arrows, and the corresponding Chapters denoted by colour.}
\label{fig:resentment}
\end{figure}

This thesis presents a set of analysis tools that were developed to rigorously quantify the collective optical resonances of coupled nanoparticles in oligomer arrangements, before then presenting arguments that use these tools to explore {\it a priori} the properties of geometry and resonances that impose specific optical effects.  
The three specific optical effects I emphasize are: Fano resonances in dielectric nanoparticle oligomers (Chapter~\ref{chapterEigen}), polarisation-independent scattering and absorption (Chapter~\ref{chapterSymmetry}), and circular dichroism in absorption (Chapter~\ref{chapterChiral}), which are the key results presented over the six works in Appendix~\ref{pubs}. 
This thesis ultimately aims to collate and contextualise these six works, in addition to introducing retrospective insights.
The relations between each paper and the Chapters are illustrated in Figure~\ref{fig:resentment}.
Regarding the structure of the thesis itself, Chapter~\ref{chapterModels} firstly presents to models that were used in my studies to analytically quantify, and numerically simulate, optical scattering in nanoparticle systems. 
This is essentially providing a means for retrospective analysis of a fixed nanoparticle system, from which Chapter~\ref{chapterEigen} then shows that the eigenmodes of these models provide a useful basis to quantify any given geometry.
This particular chapter then also presents the argument as to why nonorthogonal eigenmodes are necessary for interference phenomena to exist, particularly Fano resonances.
In the following two Chapters~\ref{chapterSymmetry} and \ref{chapterChiral}, I am able to consider the optical properties of an undefined geometry in terms of a generic set of eigenmodes. 
This serves as an attempt to relate realistic geometric design considerations to a designated optical effect by investigating correspoding necessary properties of the generic geometry's eigenmodes. 
Chapter~\ref{chapterSymmetry} uses this approach to relate geometric symmetry to the degeneracies of eigenmodes, and a subsequent derivation that discrete rotational symmetry leads to polarization-independent scattering and absorption properties. 
Chapter~\ref{chapterChiral} then illustrates a different form of prospective geometric relation, starting instead from an assumption of nonorthogonal eigenmodes, from which a new form of circular dichroism in absorption is shown to be possible.
Here the knowledge that Fano resonances imply nonorthogonal eigenmodes, from Chapter~\ref{chapterEigen}, allows us to repurpose empirical knowledge of oligomer geometries that support Fano resonances to realise circular dichroism in absorption with planar chiral oligomers.   
This then completes the thesis; the chapters having served to trace successive steps that illustrate a manner in which one can impose specified scattering quantities from geometric design freedoms by analysing the necessary properties of eigenmodes.   
The specific summary of each chapter is now listed. 
\begin{itemize}
\item{
Chapter~\ref{chapterModels} defines the induced current model and the coupled dipole model, which I use for derivations, analysis and numerical simulation of optical scattering in the subsequent chapters. 
These two models are amended versions of the models presented in the works of Appendix~\ref{pubs}, each having the benefit of cumulative and retrospective developments encountered during my studies.  
This Chapter is therefore intended to serve as a reference for similar modelling, and also for justifying my choices in utilising these specific models.     
}

\item{
Chapter~\ref{chapterEigen} presents the analysis technique for describing Fano resonances as interference between the nonorthogonal eigenmodes, and its implementation for nanoparticle oligomer systems.
This contrasts plasmonic and high-index dielectric nanoparticle oligomers, culminating in a demonstration of Fano interference occurring between multiple magnetic dipolar resonances, specific to dielectric nanoparticle oligomers.
}

\item{
Chapter~\ref{chapterSymmetry} discusses the effects of geometric symmetry and reciprocity on the optical properties of resonant nanostructures. 
This revisits the derivation of polarisation-independent scattering and absorption losses due to discrete rotational symmetry in~[\ref{Nanoscale}], using retrospective knowledge of reciprocal eigenmode degeneracy in~[\ref{LPOR}]. 
}

\item{
Chapter~\ref{chapterChiral} considers instead the absence of symmetry and discusses both geometric and optical chirality, relating to their influence on scattering properties of nanostructures and their resonances.
It reviews circular dichroism effects from the perspective of symmetries, before presenting a new form of circular dichroism in the material absorption that is attributed to interaction between nonorthogonal resonances.
}
\end{itemize}

\chapter{Models for optical scattering \label{chapterModels}}

%

%
For the context of investigating scattering from nanoparticle systems, it is necessary to have models to describe scattering systems that resemble neither homogenised media nor simple Rayleigh scatterers.
Here I compile the two modelling approaches used in the papers of [\ref{pubs}], where I remove variations between papers while also introducing some retrospective  insights.
The first section presents the description of scattering in terms of combined free currents and polarisation currents, which allows us to directly consider the physical source of fields within any nanostructured optical system, thereby providing a largely unapproximated model for optical scattering.
We then present the coupled dipole model as a practical simplification that allows direct investigation and straightforward simulation for the dominant resonances of compact nanoparticle systems.
This model is tailored specifically for describing nanoparticle oligomer geometries, and particularly those consisting of high-index dielectric nanoparticles.

\section{Induced current model}
We begin our analysis of linear optical scattering systems by acknowledging there is no need to recognise the distinction between oscillating free current and polarisation current.
Any tensor conductivity $ {\boldsymbol{\bbar{\sigma}}}$ and susceptibility $ {\boldsymbol{\bbar{\chi}}}$ can be incorporated into an effective permittivity ${\boldsymbol{\bbar{\epsilon}}}$ that relates electric field $\mathbf{E}$ to a \textsl{total} electric current $\mathbf{J}$ containing both free and polarisation currents.
\begin{align}
{\boldsymbol{\bbar{\epsilon}}}\equiv ( {\boldsymbol{\bbar{\chi}}}+1)\epsilon_0 - \frac{ {\boldsymbol{\bbar{\sigma}}}} {i \omega}  \quad \Rightarrow \quad \mathbf{J} (\mathbf{r},\omega) e^{-i\omega t} = - i \omega  \left [ {\boldsymbol{\bbar{\epsilon}}}(\mathbf{r})- \epsilon_0 \right ] \cdot \mathbf{E}(\mathbf{r},\omega) e^{-i\omega t} \label{eq:induced current}
\end{align}
There is also the simplification that most optical materials have a negligible permeability difference to the background medium, allowing us to neglect the radiation from any magnetisation current.
It is also notationally convenient to use the $\mathbf H$-field, which is now defined relative to the uniform background permeability $\mathbf B = \mu_0 \mathbf H$. 
To relate the currents $\mathbf J$ to fields $\mathbf {E_s}$, $\mathbf{H_s}$ they radiate, we can make use of the dyadic Green's function  $\mathbf{\bbar G_0}$.
\begin{align}
 {\mathbf{\bbar G_0}}(\mathbf{r},\mathbf{r'})
 &=    \left [\mathbf{\bbar{I}}+ \frac{1}{k^2} \boldsymbol{\nabla \nabla}\right] \frac{e^{i k R}}{4 \pi   R} \\
& = \frac{ e^{i k R }}{4 \pi  R}\bigg[ \Big(1 + \frac{i}{k R}  - \frac{1}{k^2 R^2} \Big)\, \mathbf{\bbar I} - \Big(1 + \frac{3 i}{k R}  - \frac{3}{k^2 R^2}\Big) \mathbf{\hat n}\mathbf{\hat n}^\mathrm{T} \bigg]   \label{eq:G0_1} 
\end{align}
where $R = \left| \mathbf{r}-\mathbf{r'}\right|$ and $\mathbf{\hat n}$ is the unit vector pointing from $\mathbf{r'}$ to $\mathbf{r}$, in other words: \mbox{$R \, \mathbf{\hat n}= \mathbf{r}-\mathbf{r'}$}.
This is a solution for the dyadic wave equation: 
\begin{align}
\boldsymbol{\nabla} \times \boldsymbol{\nabla} \times   {\mathbf{\bbar G_0}}(\mathbf{r},\mathbf{r'})  - k^2   {\mathbf{\bbar G_0}}(\mathbf{r},\mathbf{r'}) =  \delta(\mathbf{r} -\mathbf{r'})\mathbf{\bbar I}
\end{align}
where $\delta$ is a Dirac delta function and $\mathbf{\bbar I}$ is the identity matrix.  
This is notably relevant for the wave equation for the electric field shown in (\ref{eq:wavEfield}), which is obtained from substituting (\ref{eq:curlE}) into  (\ref{eq:curlB}) while assuming harmonic $e^{-i \omega t}$ time dependence.
\begin{align}
\boldsymbol{\nabla} \times \boldsymbol{\nabla} \times   \mathbf{E}(\mathbf r)  - k^2   \mathbf{E}(\mathbf r)  =  -{i \omega \mu_0}\mathbf{J}(\mathbf r)  \label{eq:wavEfield}
\end{align}
As such, the dyadic Green's function specifies the radiation from a Dirac delta point source of electric current at $\mathbf r'$.  
The electric field $\mathbf{E_s}$ radiated by an arbitrary  distribution of electric current can therefore be expressed using the dyadic Green's function by integrating the electric fields generated from each point of electric current~\cite{Yaghjian1980}.
\begin{align}
\mathbf{E_s}(\mathbf{r})  &= i \omega \mu_0\;  \bigg(  \text{\scriptsize P.V.} \! \! \int    \Big [\mathbf {\bbar{G}_{0}}(\mathbf{r},\mathbf{r'}) - \mathbf{\bbar{\,L}} \frac{\delta (\mathbf{r}-\mathbf{r'})}{k^2 } \Big] \! \cdot  \mathbf{J}(\mathbf{r'}) \;{{\mathrm{dr'}}^{3}}\bigg)  \label{eq:scatt}\\
\mu_0 \mathbf{H_s}(\mathbf{r}) &=\frac{1} {i \omega} \boldsymbol{\nabla}\times \mathbf{E_s}(\mathbf{r}) \;,\qquad \text{from (\ref{eq:curlE})}\label{eq:scattH}
\end{align}
Here $k$ is the wavenumber, $\omega$ is the angular frequency, $\epsilon_0$ and $\mu_0$ are the permittivity and permeability of the background medium, and  the volume of the scattering object is assumed to be finite.
The $\text{\small P.V.}$ implies a principal value exclusion of $\mathbf{r'}=\mathbf{r}$ when performing the integration, and $ \mathbf{\bbar{\,L}}$ is the source dyadic necessary to account for the shape of the infinitesimal  volume that forms this exclusion~\cite{Yaghjian1980}.
Source dyadics have been derived for different shaped exclusions, the simplest being spheres or cubes:~$ \mathbf{\bbar{\,L}}= \frac{1}{3}$, but more complicated expressions have also been derived for ellipsoids, cylinders, rectangular parallelepiped  and others~\cite{Yaghjian1980}.  
The different source dyadics are necessary to ensure the same electric field is obtained, irrespective to the shape of the exclusion. 
This point becomes necessary in numerical simulations, which consider discretised meshes of continuous objects and thereby exclude self-interaction in a single mesh element, which requires a source dyadic that can account for the shape of the exclusion volume.  
From the expression (\ref{eq:scatt}), and referring to  (\ref{eq:G0_1}), can now take the opportunity to define explicitly the {\it near-field} as the component of $\mathbf{E_s}$ that scales with  $R^{-2}$ or  $R^{-3}$, and the {\it far-field} as the component scaling with $R^{-1}$.
The time-average power flux density $\frac{1}{2}(\mathbf{E_s}^*\times \mathbf{H_s})$ of the far-field scales as $R^{-2}$, while its total flux area expands with $R^2$, meaning the total energy of these scattered fields do not decay: they propagate and can be observed at distances very far from the source.
By the same reasoning, the near-field will decay the further from the source they get: they do not propagate, but will dominate the total field at very small $R$.
We can now solve for the current induced by an external electric field $\mathbf{E_0}$.
The total internal field  $\mathbf{E} = \mathbf{E_0}+\mathbf{E_s}$ is related to the current through the effective permittivity in (\ref{eq:induced current}), therefore we can combine this with $\mathbf{E_s}$ in (\ref{eq:scatt}), to relate the induced currents and the external field.
\begin{align}
 - i \omega [ \boldsymbol{\bbar \epsilon} (\mathbf{r}) \!-\! \epsilon_0] \! \cdot  \mathbf{E}_{0}(\mathbf{r})   =&\,{\mathbf{J}(\mathbf{r})} \! -\! \frac{k^2}{\epsilon_0} [  {\boldsymbol{\bbar\epsilon}(\mathbf{r})  \!-\! \epsilon_0}]  \bigg( \! \text{\scriptsize P.V.} \! \! \int \!\!\Big [ \mathbf{\bbar G_0}(\mathbf{r},\mathbf{r'})\! -\! \mathbf{\bbar{\,L}} \frac{\delta (\mathbf{r}\!-\!\mathbf{r'})}{k^2 } \Big] \!\cdot  \mathbf{J}(\mathbf{r'})\,{\mathrm{dr'}^3}\bigg)   \label{eq:current equation}
\end{align}
In the absence of magnetisation, (\ref{eq:current equation})  will determine the currents induced in any finite object under any arbitrary excitation, then (\ref{eq:scatt}) will describe the fields radiated by this current.
It is, however, highly nontrivial to obtain general solutions for (\ref{eq:current equation}), even with very simple geometries.
Therefore, I use (\ref{eq:current equation}) primarily as an analytical model for investigating general principles of scattering systems.  
In Chapter~\ref{chapterEigen} the current model is used to relate far-field interference features to the nonorthogonality of different resonant distributions of current, and in Chapters~\ref{chapterSymmetry} and \ref{chapterChiral} it is used to explore consequences of symmetry and asymmetry.
In these pursuits, I focus on how the energy within any given current distribution is lost.
Specifically, we consider two broad loss channels: power transported elsewhere as electromagnetic radiation, or power transported elsewhere as anything that isn't electromagnetic radiation.
The former is radiative losses, which will represent the power of scattering when the currents  are induced by externally applied electromagnetic fields, and the latter is dissipative losses, which we will call absorption, and encompasses heat generation, photocurrents, and other linear loss mechanisms\footnote{
I will neglect nonlinear loss mechanisms, which can represent multi-photon absorption, frequency mixing, and other effects that only become relevant at high electric field intensities. 
}.  
Assuming a lossless background medium, the total radiated power can be calculated by considering any surface encompassing the current distribution, which is finite by our initial definitions, and performing a surface integral of the normal component of time-averaged Poynting vector for the scattered fields $\mathbf{S_s} =\frac{1}{2} \mathrm{Re}\{\mathbf{E_s}^*\times \mathbf{H_s}\}$, as per (\ref{eq:scatt}) and  (\ref{eq:scattH}).
Such a calculation is relatively straightforward to implement numerically, but analytically we can rely on the derivations of Markel~\cite{Markel1995}, who calculated the total radiation by an arbitrary arrangement of electric dipoles bound by an infinite spherical surface.
\begin{align}
P_{scat} =&  \frac{1}{2}\sqrt{\frac{\epsilon_0}{\mu_0}} (4\pi {\epsilon_0})^{-2}\bigg( \sum \limits_i  \frac{8 \pi k^4} {3} |\mathbf{p}_i|^2  + 4 \pi k \sum \limits_{j \neq i}   \mathrm{Im} \Big \{ \mathbf{p}_i^* \! \cdot\big(4 \pi k^2 \mathbf {\bbar{G}_{0}}(\mathbf{r}_i,\mathbf{r}_j) \big)   \cdot    \mathbf{p}_j \Big\} \bigg) \label{eq:Pscat_dip}
\end{align}
Here I have added coefficients as necessary adjustment factors to express Markel's result in SI units for $\mathbf{p}$.
We also need to note that the right hand term $\mathbf {\bbar{G}_{0}}(\mathbf{r}_i,\mathbf{r}_j) \big)   \cdot    \mathbf{p}_j $ is related to the radiated electric field at $\mathbf r_i$ by the dipole at $\mathbf r_j$, the expression for which I will later cover in (\ref{eq:dipEs}).
The next step simply requires translating ($\ref{eq:Pscat_dip}$) to apply to continuous systems.
We first write a current distribution $\mathbf{J}(\mathbf r)$ as an equivalent polarization distribution $\mathbf P(\mathbf r) = - i \omega \mathbf J (\mathbf r) $.  
Each infinitesimal volume $\mathrm {dr}^3$, over which $\mathbf P $ can be considered constant, will thereby have a dipole moment defined as $\mathbf{p}_{\scriptscriptstyle (\mathbf r)}= \mathbf P(\mathbf r) \mathrm {dr}^3$.  
Our continuous current representation can then be represented with the set of dipole moments $\{\mathbf{p}_{\scriptscriptstyle (\mathbf r)}\!\}$, which we can substitute into (\ref{eq:Pscat_dip}), and noting that the summations in (\ref{eq:Pscat_dip}) will become integrals.
\begin{align}
P_{scat} =&  \frac{1}{2}\sqrt{\frac{\epsilon_0}{\mu_0}} (4\pi {\epsilon_0})^{-2}\int \!\! \Big[ \frac{8 \pi k^4} {3} |\mathbf P(\mathbf{r}_i)|^2 {\mathrm{dr}_i}^3
 \nonumber 
\\
 & + (4 \pi)^2 k^3 \bigg ( \text{\scriptsize P.V.} \! \!  \int   \mathrm{Im} \Big \{\mathbf P^* (\mathbf{r}_i) \! \cdot \! \Big( \mathbf {\bbar{G}_{0}}(\mathbf{r}_i,\mathbf{r}_j) \! -\! \mathbf{\bbar{\,L}}  \frac{\delta (\mathbf{r}_i-\mathbf{r}_j)}{k^2 } \Big)   \!\cdot\!    \mathbf P (\mathbf{r}_j) \Big\}{\mathrm{dr}_j}^3 \bigg ) \Big]{\mathrm{dr}_i}^3 \label{eq:meanReviewer}
\end{align}
We can now neglect the first term in (\ref{eq:meanReviewer}) because it is proportional to $\mathrm{dr}_i^3$ {\sl after} volume integration.  
Meaning this term will get arbitrarily small as the voxel gets arbitrarily small, and it is physically equivalent to the radiation of the single voxel in isolation: it is the only term that remains if we define $\mathbf P(\mathbf r_j) = \mathbf 0$ at all $\mathbf r_j \neq \mathbf r_i$.  
The expression for scattered power after neglecting such terms can now be written in terms of currents given: $\mathbf J = - i \omega \mathbf P$.    
\begin{align}
P_{scat} =&  \frac{\omega \mu_0  }{2} \bigg( \text{\scriptsize P.V.} \! \! \iint \mathrm{Im}\Big \{ \mathbf{J}^* (\mathbf{r}) \cdot \Big( \mathbf {\bbar{G}_{0}}(\mathbf{r},\mathbf{r'}) - \mathbf{\bbar{\,L}}  \frac{\delta (\mathbf{r}-\mathbf{r'})}{k^2 }  \Big) \cdot \mathbf{J}(\mathbf{r'}) \Big \}   \mathrm{dr}^3 \mathrm{dr'}^3 \bigg)  \label{eq:Pscat} \\
= &  - \frac{1}{2 }\int  \mathrm{Re}\{ \mathbf{J}^*(\mathbf{r}) \cdot \mathbf{E_s}(\mathbf r) \}\mathrm{dr}^3 \;, \qquad \text{by substituting (\ref{eq:scatt})}
\end{align}
To now calculate the total {\sl absorbed} power there are two immediate options: perform a power balance calculation globally, or perform a power balance locally (at each point) and integrate over the whole space.  
The global power balance refers to calculating absorption as the difference between the net input power $P_{in}$ entering the system and the net electromagnetic power leaving the system, which is a surface integral of the \textsl{total} time-averaged Poynting vector leaving any bounding surface $\Omega$ around the scattering object.
\begin{align}
P_{abs} = P_{in}- \oint_\Omega  \frac{1}{2} \mathrm{Re}\{\mathbf{E}^*(\mathbf r)\times \mathbf{H}(\mathbf r)\} \cdot \mathrm{d}\mathbf{A} \label{eq:pbalance}
\end{align}
Here $\mathrm d \mathbf A$ is the area differential element represented by a surface-normal vector.  
This calculation is straightforward to implement numerically for scattering of propagating electromagnetic waves, because $P_{in}=0$, but otherwise it will require the value of $P_{in}$ to calculate absorption.  
The general advantage of this calculation approach in  (\ref{eq:pbalance}) is that it doesn't require knowledge of either the scattering object's material or specific geometry, other than requiring a bounding surface.
The disadvantage is that we need to know $P_{in}$, which becomes challenging whenever the excitation source is complicated, as might be the case for modelling nonlinear mixing or harmonic generation processes.  
It is then often instead better to use knowledge of the material and geometry to calculate the local power balance within the material.
Draine~\cite{Draine1988} derived an expression on the total absorption in arbitrary systems of electric dipole moments, by calculating the local difference between the radiative and total losses of each individual dipole.  
\begin{align}
P_{abs} =& -\frac{1}{2}\sqrt{\frac{\epsilon_0}{\mu_0}}  \frac{k}{{\epsilon_0}^2}  
\bigg( \sum \limits _i   \frac{k^3}{6 \pi} |\mathbf{p}_i|^2
+ {\mathbf{p}_i}^* \! \cdot \, \mathrm{Im} \{  {\boldsymbol{\bbar \alpha}_i}^{-1}\} \cdot  \mathbf{p}_i  \bigg) \label{eq:Pabs_dip}
\end{align}
We can again translate this expression to consider a continuous current system using the discretisation previously considered for scattered power, which amounted to \mbox{$\mathbf{J}(\mathbf r) \mathrm{dr}^3 \equiv - i \omega \mathbf{p}_{\scriptscriptstyle (\mathbf r)}$}, and implies that \mbox{$\boldsymbol{\bbar \alpha }_{\scriptscriptstyle (\mathbf r)}\epsilon_0  =(\boldsymbol{\bbar \epsilon}(\mathbf{r})-\epsilon_0) \mathrm{dr}^3$}.
When performing this substitution from $\mathbf p$ to $\mathbf J$, the term $|\mathbf p_{i}|^2\! \rightarrow\! |\mathbf p_{\scriptscriptstyle (\mathbf r)}|^2\! =\!\Big | \frac{\mathbf J}{\omega}\Big|^2 \mathrm{dr}^6$ will notably remain proportional to $\mathrm{dr}^3$ {\sl after} the volume integration.  
As such, this term becomes arbitrarily small and can be neglected, meaning we neglect the contribution to absorption from each voxel in isolation, which is reasonable given each voxel is infinitesimally small.   
\begin{align}
P_{abs} =& -\frac{1}{2 \omega }\int \mathbf{J}^*\!(\mathbf r)  \cdot \mathrm{Im} \{  (\boldsymbol{\bbar \epsilon}(\mathbf r)- \epsilon_0)^{-1}\} \cdot  \mathbf{J}(\mathbf r)\;  \mathrm{dr}^3   \label{eq:Pabs}
\end{align}
Note that, by writing in terms of the total electric field $\mathbf E$ using $\mathbf{J}  = -i \omega (\boldsymbol{\bbar \epsilon}(\mathbf r)- \epsilon_0) \mathbf E$, the expression in (\ref{eq:Pabs}) becomes:
\begin{align}
P_{abs} =& \frac{1}{2 }\int  \mathrm{Re} \{ \mathbf{E}^*\!(\mathbf r)  \cdot  \mathbf{J}(\mathbf r) \}  \; \mathrm{dr}^3 
\end{align}
The final way to quantify optical response is the total amount of power that interacts with the given object; this quantity is the sum of $P_{scat}$ and $P_{abs}$, and it is known as the extinction.  
\begin{align}
P_{ext}&= P_{scat}+ P_{abs} \nonumber \\ & = - \int \Big(  \frac{1}{2 } \mathrm{Re}\{ \mathbf{J}^* (\mathbf{r}) \cdot\mathbf{E_s}(\mathbf r) \}  +\frac{1}{2 \omega} \mathbf{J}^* (\mathbf{r})  \cdot\; \mathrm{Im} \{  (\boldsymbol{\bbar \epsilon}(\mathbf{r}) - \epsilon_0)^{-1}\} \cdot \mathbf{J}(\mathbf{r}) \Big) \;  \mathrm{dr}^3 \label{eq:prelimExt}
\end{align}
If we now consider the input power of our system to be written in terms of an externally applied field $\mathbf{E_0}$, it is then possible to write the scattered field $\mathbf{E_s}$ as the difference of total and incident fields $\mathbf{E_s} = \mathbf{E}-\mathbf{E_0}$, and use (\ref{eq:induced current}) to equate $\mathbf{E} = \frac{1}{- i \omega}(\boldsymbol{\bbar \epsilon}(\mathbf{r}) - \epsilon_0)^{-1}  \cdot  \mathbf{J}$.
If we now substitute $\mathbf{E_s} = \frac{1}{- i \omega}(\boldsymbol{\bbar \epsilon}(\mathbf{r}) - \epsilon_0)^{-1}  \cdot  \mathbf{J} - \mathbf{E_0}$ into (\ref{eq:prelimExt}) we are then able to obtain a simplified expression for extinction.
\begin{align}
P_{ext}&=   \frac{1}{2 }  \int  \mathrm{Re}\{ \mathbf{E_0}^* (\mathbf r) \cdot \mathbf{J}(\mathbf{r}) \}\;  \mathrm{dr}^3  \label{eq:Pext}
\end{align}
This is now precisely a volume integral of the time-averaged power imparted locally by the fields $\mathbf{E_0}$ on the currents $\mathbf{J}$~\cite{LandauLifshitzVol5}, and therefore shows that extinction is the total amount of power removed from the excitation fields $\mathbf{E_0}$.
Extinction is often therefore considered to be equal to the difference of input and output power ($1 - \text{transmission}$) through surfaces or media, but this should be treated carefully as it neglects the power of the scattered field that is co-propagating with the incident field.  
An ideal $\lambda/2$ waveplate has 100\% transmission when rotating the linear polarisation of an incident plane wave by $\pi/2$, but thereby produces 100\% extinction, because none of the original plane wave with original polarisation remains. 
Similarly, reflectionless metasurfaces have also been shown to allow near 100\% co-polarized transmission, while simultaneously altering the phase of this transmission anywhere over the complete $2\pi$ range of phase shifts, defined relative to the original illumination~\cite{Decker2015}.
That the resonators in the metasurface are able to change the phase of a wavefront implies they must be the source of significant scattering, and hence extinction, yet transmission remains near 100\%.
As such, one generally cannot use the extinction calculated through equations such as (\ref{eq:prelimExt}) or  (\ref{eq:Pext}) to infer transmission.

Before concluding, it is worth recognising that the quantities of scattering, absorption and extinction, are regularly considered in terms of their corresponding {\it cross-section} $\sigma$, which refers to the cross-sectional area of the illuminating plane wave that contains the power $P$.  
In other words, a plane wave with intensity $I_0 =  |\mathbf{E_0}|^2$ will have a power flux density of $\frac{1}{2}\sqrt{\frac{\epsilon_0}{\mu_0}}\, I_0 $, and the area of a cross-section $\sigma$ is therefore related to the corresponding power $P$ as per:
\begin{align}
P =\frac{1}{2}\sqrt{\frac{\epsilon_0}{\mu_0}}\, I_0 \, \sigma \label{eq:Psigma}
\end{align}
The use of cross-sections is practical because a realistic light source that has a finite beamwidth.  
A cross-section therefore prescribes a limit on the maximum power of scattering, absorption or extinction for any fixed input light intensity.
From now onwards, we will consider cross-sections rather than total power loss.
However, the current model remains very nontrivial to find general continuous solutions for simple geometries.
In the next section, I will therefore outline the use of the dipole model, as a way of imposing discretisation to replace the integrals here with finite sums, and thereby allow explicit solutions for scattering response to be found as matrix equations.
More specifically, we will turn our attention to arrangements of simple nanoparticles in oligomer geometries.

\section{Coupled dipole model}

Here we consider the optical responses of nanoparticles placed in oligomer arrangements as to produce more complex optical features. 
The constituent nanoparticles of oligomers are typically both sufficiently simple and subwavelength in size, for their lowest-energy resonances to resemble dipoles at optical or near-infrared wavelengths.
Given electric $\mathbf p$ and magnetic $\mathbf m$ dipole moments radiate identically to the $a_1$ and $b_1$ spherical scattering coefficients, see (\ref{eq:pandmintermsofa1andb1}) and (\ref{eq:spheresonwards}), we can again make use of the conclusion from Devaney and Wolf~\cite{Devaney1974}. 
Specifically,  the scattered field of a dipolar nanoparticle with only the moments $\mathbf p$ and $\mathbf m$ will be described by only $a_1$ and $b_1$ coefficients, hence the near fields external to the smallest bounding sphere around the nanoparticle itself can be exactly modelled by that of $\mathbf p$ and $\mathbf m$.  
We can therefore use a dipole model as a simplified alternative to the current model for nanoparticle oligomers with only minor limitations in accuracy, for even closely spaced nanoparticles\footnote{
An example of the validity of near-field predicted by the dipole model can be found in the Supplemental Material of [\ref{ACS}].
}.
In essence, the dipole model enables us to analyse the dominant optical properties of a collective nanoparticle oligomer by considering only the dominant optical resonances of the individual nanoparticles.

\begin{figure}[!ht]
\centerline{\includegraphics[width=\textwidth]{{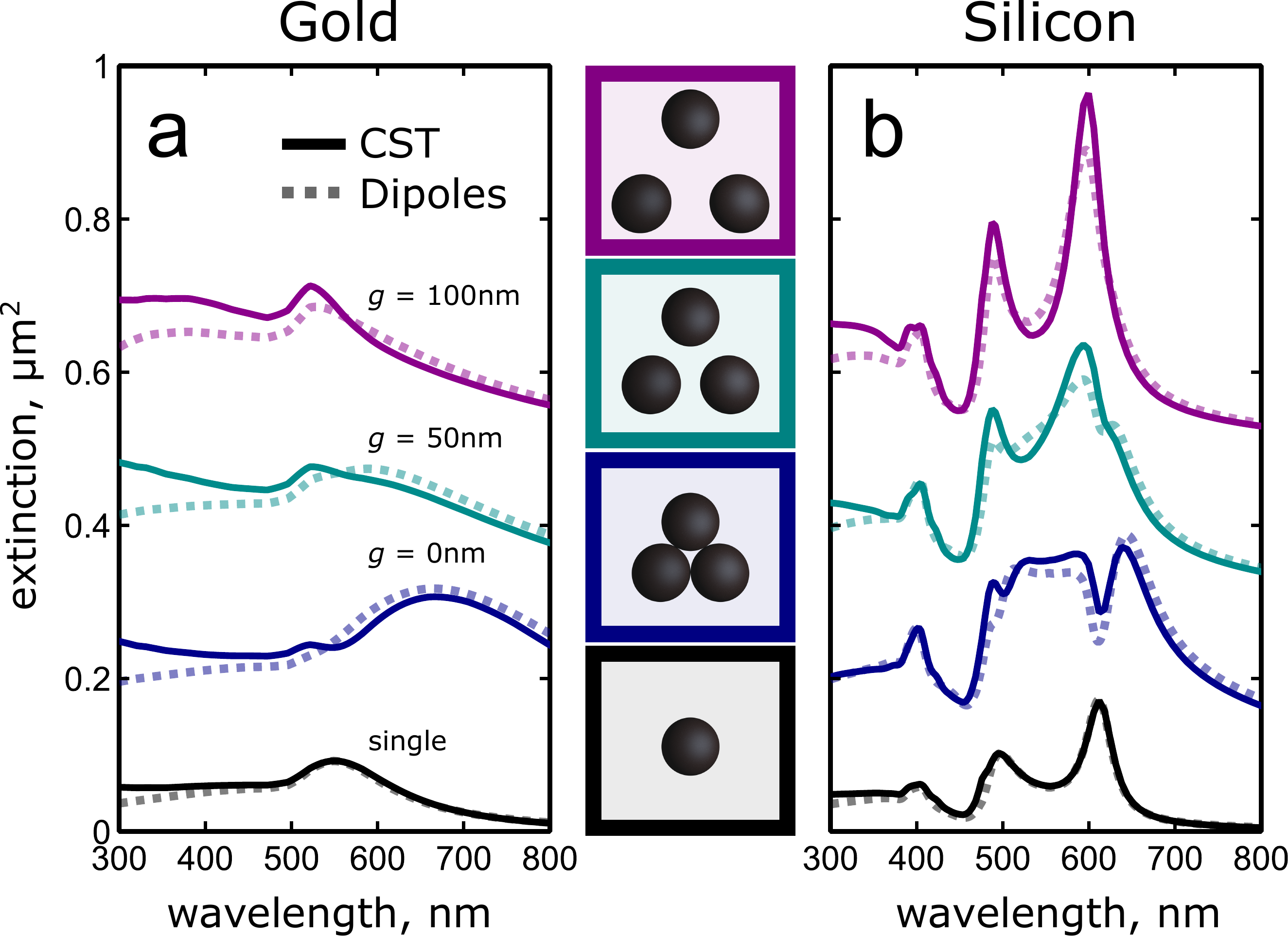}}}
\caption{A comparison of the extinction cross-section calculated using the dipole model in (\ref{eq:Dmodel}) with that calculated using  CST Microwave Studio.
Calculations are for 150~nm nanospheres made of (a) gold and (b) silicon, when arranged as symmetric trimers with varying separation $g$ between nanoparticles.
Figure taken from Chapter~10 of~\cite{Agrawal2017}.   }
\label{fig:trimerComparison}
\end{figure}

We now first consider the case of small plasmonic nanoparticles, where the individual nanoparticle response is dominantly an electric dipole, and we can therefore use the dipole approximation\cite{Draine1988, Draine1994} to describe the optical properties of the oligomer.
Moreover, analogous to (\ref{eq:scatt}) and (\ref{eq:scattH}) for currents, we can write the scattered electric and magnetic fields from a system of electric dipole moments $\{ \mathbf p_i \}$ in terms of dyadic Green's functions.
\begin{align}
\mathbf{E_s}(\mathbf{r})  &= \frac{k^2}{\epsilon_0}  \sum \limits_i \mathbf {\bbar{G}_{0}}(\mathbf{r},\mathbf{r}_i) \cdot  \mathbf{p}_i \label{eq:dipEs} \\
\mathbf{H_s}(\mathbf{r}) &= \frac{1}{i \omega \mu_0} \boldsymbol{\nabla}\times \mathbf{E_s}(\mathbf{r}) =  \frac{ k c_0}{i} \sum \limits_i  \boldsymbol{\nabla}\times\mathbf {\bbar{G}_{0}}(\mathbf{r},\mathbf{r}_i) \cdot  \mathbf{p}_i  \label{eq:magfieldelecdip}
\end{align}
Here the Green's function is the same as that defined for currents in (\ref{eq:G0_1}), but we can also write the effect of $\mathbf{\bbar G_0}$ acting on arbitary vector $\mathbf u$:
\begin{align}
\mathbf{\bbar G_0}(\mathbf{r},\mathbf{r}') \cdot \mathbf{u} &= \frac{ e^{i k R }}{4 \pi  R}\bigg[ \Big(1 + \frac{i}{k R}  - \frac{1}{k^2 R^2} \Big)\, \mathbf{u} - \Big(1 + \frac{3 i}{k R}  - \frac{3}{k^2 R^2}\Big)( \mathbf{\hat n}\cdot\mathbf{u})\,\mathbf{\hat n}\bigg]  \label{eq:EEcoupling}
\\
\boldsymbol{\nabla}\!\times \!\mathbf{\bbar G_0} (\mathbf{r},\mathbf{r}') \cdot \mathbf{u}
&=i k \frac{ e^{i k R }}{4 \pi  R}\left(1 + \frac{i}{k R}  \right) \mathbf{\hat n}\times\mathbf{u} \label{eq:EMcoupling}
\end{align}
Here $\mathbf{\hat n}$ is the unit vector pointing from $\mathbf{r}$ to $\mathbf{r'}$, so that: \mbox{$\mathbf{r} - \mathbf{r'} = R \mathbf{\hat n}$}.
The dipole moment of the $i^{\text{th}}$ point dipole (located at $\mathbf r_i$) can generally be related to the total electric field at $\mathbf r_i$, being the sum of $\mathbf{E_0}$ and $\mathbf{E_s}$, using a tensor electric dipole polarisability, {\it i.e.} \mbox{$\mathbf p_i = \boldsymbol{\bbar\alpha}_{{\scriptscriptstyle E}}^{ (i)}[\mathbf{E_0}(\mathbf r_i) + \mathbf{E_s}(\mathbf r_i)]$}. 
By substituting $\mathbf{E_s}$ from (\ref{eq:dipEs}), we then obtain an expression for each electric dipole moment $\mathbf{p}_{i}$ in an arbitrary dipole system as a function of the externally applied electric field distribution $\mathbf{E_0}$:
\begin{align}
\mathbf{p}_{i}  = \boldsymbol{\bbar\alpha}_{{\scriptscriptstyle E}}^{(i)}\epsilon_{0}\mathbf{E_{0}}(\mathbf{r}_{i})&+\boldsymbol{\bbar\alpha}_{\scriptscriptstyle E}^{(i)}k^{2}
\sum \limits_{j\neq i} \mathbf{\bbar G_0}(\mathbf{r}_{i},\mathbf{r}_{j}) \cdot \mathbf{p}_{j}
\label{eq:ED model}
\end{align}
For a system of $N$ dipoles, the expression in (\ref{eq:ED model}) forms a matrix equation of rank $3N$, which we can then solved for any arbitrary excitation as per an ordinary matrix equation.
Note that, unlike the current model in (\ref{eq:current equation}), the source dyadic can be neglected here given we are able to exclude a single point, rather than shaped volumes, to avoid the $R^{-3}$ singularity of $\mathbf{\bbar G_0}$ when $\mathbf r_i = \mathbf r_j$. 
In Figure~\ref{fig:trimerComparison}a, we present the validity of this model for a symmetric trimer arrangement of gold nanospheres.
For the spherical nanoparticles, the electric and magnetic dipole polarisabilities are scalars and defined in terms of the $a_1$ and $b_1$ scattering coefficients from Mie theory~\cite{Mie1908, Bohren1983}:
\begin{align}
\alpha_{\scriptscriptstyle E} = \frac{6 i \pi a_1}{k^3} \quad
\alpha_{\scriptscriptstyle H} = \frac{6 i \pi b_1}{k^3} \label{eq:polarisabilities} 
\end{align} 
Note that I use the convention for electric and magnetic dipole polarizabilities: $\mathbf p = \alpha_{\scriptscriptstyle E} \epsilon_0 \mathbf E$, $\mathbf m = \alpha_{\scriptscriptstyle H} \mathbf H$.
The extinction spectra calculated from the dipole model are compared to those calculated using commercial electromagnetic simulation software, {\it CST Microwave Studio}, which uses a frequency domain solver that is based on the finite element method~\cite{Jin_FEM}.  
The details of finite element methods are beyond the scope of this thesis, but I will at least outline that they aim to solve a boundary value problem for the general class of wave equations where a differential operator $\hat { \mathcal L}$ acts on an unknown vector field $\boldsymbol \psi$, and is equal to some imposed driving vector field $\mathbf f$, that is: $\hat{\mathcal L} \boldsymbol \psi  = \mathbf f$.  
One relevant manner by which such equations can be solved, is to minimise a corresponding {\it functional} defined in terms of an unknown $\boldsymbol \psi$.  
Moreover, by defining a so-called {\it test} or {\it trial} function for $\boldsymbol \psi$ as a linear combination of some known basis functions, the minimisation of the functional can be equated to a matrix equation for the coefficients of each basis function, and is solvable by matrix inversion.
A minimised trial function will then approach the correct solution for $\boldsymbol \psi$.
Details of the above steps, which correspond to the {\it Ritz Method}, can be found in \S2 of~\cite{Jin_FEM}.
However, the distinguishing feature of the finite element method, is that it subdivides the full simulation volume into a volumetric mesh of smaller domains, and defines different trial functions in each voxel. 
This allows a simpler set with only a handful of basis functions to be used for each voxel's trial function.
Furthermore, it makes the calculation method more easily translatable to complex geometries, as one can simply reduce the size of each voxel until the fixed basis functions can correctly represent the minimised trial function.    
In this sense, CST recreates a given scattering geometry as a finite volumetric mesh, then superimposes an external electric and magnetic field distribution as the driving term $\mathbf f$, much like $\mathbf{E_0}$ in (\ref{eq:current equation}).  
It then performs its own version of a finite element method calculation, which has been able to provide quantitative  agreement with experiment, such as in Fig.~6 of [\ref{ACS}].

~

Even with very small gaps between the spheres, the dipole model offers an accurate prediction of the trimer's response in Figure~\ref{fig:trimerComparison}a, with the exception of that coming from the single particle electric quadrupole response.
To next use the dipole model to consider high-index dielectric nanoparticles, we must also account for the potential magnetic dipole resonance in addition to the electric dipole resonance.
To this end, the dipole model in (\ref{eq:ED model}) can then be extended to include both electric and magnetic dipoles.
It is straightforward to define magnetic dipoles induced by the total magnetic field with magnetic polarizabilities, such as denoted in (\ref{eq:polarisabilities}), but we must also recognize that the total magnetic field will now include magnetic field radiated by electric dipoles (\ref{eq:magfieldelecdip}), and also {\it vice versa}~\cite{Mulholland1994}.
As such, we can write electric and magnetic fields radiated by a system of electric and magnetic dipoles: 
\begin{align}
\mathbf{E_s} (\mathbf r)
&= \frac{k^{2}}{\epsilon_0}
  \bigg(\sum \limits_i \mathbf{\bbar G_0}(\mathbf{r},\mathbf{r}_{i})\cdot  \mathbf{p}^{(i)}
-\frac{1}{ikc_0}\,\boldsymbol{\nabla}\times \mathbf{\bbar G_0}(\mathbf{r},\mathbf{r}_{i}) \cdot \mathbf{m}^{(i)}\bigg)
 \\
 \mathbf{H_s} (\mathbf r)
&= k^{2}  \bigg(\sum \limits_i  \mathbf{\bbar G_0}(\mathbf{r},\mathbf{r}_{i}) \cdot \mathbf{m}^{(i)}
+\frac{c_0}{ik}\,\boldsymbol{\nabla}\times \mathbf{\bbar G_0}(\mathbf{r},\mathbf{r}_{i}) \cdot \mathbf{p}^{(i)}\bigg) 
\end{align}
This leads to the two equations in (\ref{eq:Dmodel}) for the electric and magnetic dipole moments induced by external electric and magnetic field distributions: (\ref{eq:dualP}) from the total electric field and (\ref{eq:dualM}) from the total magnetic field.
\begin{subequations}\label{eq:Dmodel}
\begin{align}
\mathbf{p}^{(i)}  = \boldsymbol{\bbar\alpha}_{E}^{(i)}\epsilon_{0}\mathbf{E_{0}}(\mathbf{r}_{i})&
+\boldsymbol{\bbar\alpha}_{E}^{(i)}k^{2}
  \bigg(\underset{j\neq i}{{\sum}}\mathbf{\bbar G_0}(\mathbf{r}_{i},\mathbf{r}_{j})\cdot  \mathbf{p}^{(j)}
-\frac{1}{{ik}c_0}\,\boldsymbol{\nabla}\times \mathbf{\bbar G_0}(\mathbf{r}_{i},\mathbf{r}_{j}) \cdot \mathbf{m}^{(j)}\bigg) \label{eq:dualP} \\
 \mathbf{m}^{(i)} = \boldsymbol{\bbar\alpha}_{H}^{(i)}\mathbf{H_{0}}(\mathbf{r}_{i})&
 +\boldsymbol{\bbar\alpha}_{H}^{(i)}k^{2}  \bigg(\underset{j\neq i}{{\sum}}\mathbf{\bbar G_0}(\mathbf{r}_{i},\mathbf{r}_{j}) \cdot \mathbf{m}^{(j)}
+\frac{c_0}{{ik}}\,\boldsymbol{\nabla}\times \mathbf{\bbar G_0}(\mathbf{r}_{i},\mathbf{r}_{j}) \cdot \mathbf{p}^{(j)}\bigg) \label{eq:dualM}
\end{align}
\end{subequations}
To summarize notation: $\mathbf{p}^{(i)}$ ($\mathbf{m}^{(i)}$) is the electric (magnetic) dipole moment of the $i^{\mathrm{th}}$ nanoparticle, $\mathbf{\bbar G_0}(\mathbf{r}_{i},\mathbf{r}_{j}) $ is the free space dyadic Green's function between the locations of the $i^{\mathrm{th}}$ and $j^{\mathrm{th}}$ dipoles, $\boldsymbol{\bbar\alpha}_{E}^{(i)}$ ($\boldsymbol{\bbar\alpha}_{H}^{(i)}$) is the tensor electric (magnetic) dipole polarisability of the $i^{\mathrm{th}}$ particle, $c_0 = {1}/\!{\sqrt{\epsilon_0\mu_0}}$ is the speed of light in the background medium and $k$ is the background wavenumber.
As can be seen in Figure~\ref{fig:trimerComparison}b, this coupled electric and magnetic dipole model is able to accurately predict the extinction of a trimer, with the exception of the single nanoparticle's magnetic quadrupole response.  
Once again, the quantities we consider from the dipole model will be the cross-sections. 
We can again define the scattering cross-section from the integral of the far-field scattered power, now following the derivations by Merchiers {\it et al.}~\cite{Merchiers2007}.
\begin{align}
\sigma_s = & \frac{k}{{\epsilon_0}^2 I_0} \mathrm{Im}\bigg\{ \sum \limits_i  \epsilon_0\mathbf{E_0}^*(\mathbf{r}_i) \cdot  \mathbf{p}_i + {\epsilon_0}{\mu_0}\mathbf{H_0}^*(\mathbf{r}_i) \cdot  \mathbf{m}_i \nonumber \\[-1ex]
&+\mathbf{p}_i^*\cdot\big (\frac{ i k^3}{6 \pi}  + (\boldsymbol{\bbar\alpha}_{\scriptscriptstyle E}^{(i)})^{-1}\big)  \cdot \mathbf{p}_i 
+
{\epsilon_0}{\mu_0} \mathbf{m}_i^* \! \cdot (\frac{2}{3} i k^3  + \big(\boldsymbol{\bbar\alpha}_{\scriptscriptstyle H}^{(i)})^{-1}\big)  \cdot  \mathbf{m}_i \bigg\}
\end{align}
Here $I_0$ is the average intensity of a plane wave excitation to relate the area of a cross-section $\sigma$ to the total power $P$ as per (\ref{eq:Psigma}).
The absorption cross-section can next be calculated from the local losses of the internal electric and magnetic field~\cite{Draine1988}.
\begin{align}
\sigma_a =& \frac{-k}{{\epsilon_0}^2 I_0} \mathrm{Im}\bigg\{ \mathbf{p}_i^*\cdot\big (\frac{ i k^3}{6 \pi}  + (\boldsymbol{\bbar\alpha}_{\scriptscriptstyle E}^{(i)})^{-1}\big)  \cdot \mathbf{p}_i +{\epsilon_0}{\mu_0} \mathbf{m}_i^* \! \cdot (\frac{ i k^3}{6 \pi}  + \big(\boldsymbol{\bbar\alpha}_{\scriptscriptstyle H}^{(i)})^{-1}\big)   \cdot \mathbf{m}_i \bigg\}
\end{align}
The extinction cross-section is again written as the sum of absorption and scattering.
\begin{align}
\sigma_e =& \frac{k}{\epsilon_0 I_0} \mathrm{Im}\bigg\{ \sum \limits_i  \mathbf{E_0}^*(\mathbf{r}_i) \cdot  \mathbf{p}_i + \mu_0\mathbf{H_0}^*(\mathbf{r}_i) \cdot  \mathbf{m}_i \bigg\}
\label{eq:extinctionDip}
\end{align}
We now conclude with a pair of comparisons between the current and dipole models.
Firstly, (\ref{eq:ED model}) is the discrete equivalent to current model in (\ref{eq:current equation}), and the extension to include magnetic dipoles in (\ref{eq:Dmodel}) is therefore analogous to accounting for magnetisation currents.
This also means that (\ref{eq:Dmodel}) becomes invariant under duality transformations: $(\mathbf{E}, c_0 \mathbf{B}) \rightarrow (c_0 \mathbf{B}, -\mathbf{E}) $ with $(\mathbf{p}, c_0 \mathbf{m}) \rightarrow (c_0 \mathbf{m}, -\mathbf{p}) $, and any electric and magnetic dipoles satisfing (\ref{eq:dualP}) necessarily also satisfy (\ref{eq:dualM}), or vice versa, as discussed in [\ref{ACS}].
Secondly, we can note that the scattering and absorption from each dipole in isolation is accounted for in the dipole model, (\ref{eq:Pscat_dip}) and (\ref{eq:Pabs_dip}), but can be neglected for a point in the current model,  (\ref{eq:Pscat}) and   (\ref{eq:Pabs_dip}), given it is an infinitesimally small voxel of current. 
The influence of an individual source in the current model is only felt on the surrounding media, and so the current model explicitly describes a purely collective response.
On the other hand, the dipole model still accounts for losses from individual dipoles, in recognition that they can represent a significant resonant object. 
In a nanoparticle oligomer, we therefore account for the isolated response of each nanoparticle, yet their coupling ensures that no nanoparticle can be considered independently, and so the overall optical response remains a collective property of the oligomer.
Nanoparticle oligomers exist on the boundary between collective and isolated optical responses, and subsequently present a unique challenge to understand their collective optical properties.
In the next Chapter, I present the use of eigenmodes to model the collective resonances of oligomers directly.

\chapter{Quantifying resonances and their interference \label{chapterEigen}}

This Chapter presents the approach developed during my studies for modelling resonances using the eigenmodes of nanoparticle systems, and the subsequent conclusion that interference between resonances, and particularly Fano resonances, can be described by the overlap of nonorthogonal eigenmodes.
I also take some time to contrast the derivation of eigenmodes and Fano resonances between high-refractive-index dielectric nanoparticle oligomers and plasmonic nanoparticle oligomers.  
The aim is to combine the key results of the works [\ref{PRA},\ref{ACS},\ref{PRB}] in Appendix~\ref{pubs}, which have explored and developed these particular areas for nanoparticle oligomers.  
This culminates in the demonstration of Fano interference between multiple magnetic dipoles when using dielectric nanoparticle oligomers in [\ref{ACS}] .

\section{Resonances in terms of eigenmodes}
One of the key properties of nanoantennas for artificial materials is their capacity to support a \textsl{resonant} optical response.
Understanding the characteristics of a resonance is therefore important for tailoring it toward the given application, be it electric or magnetic field localisation, directional scattering, or any number of other potential properties.
In this regard, the most common way to  characterise a resonance quantitatively is with a multipolar decomposition~\cite{Bohren1983, Raab2005}.
This is obtained by projecting the scattered field onto vector spherical harmonics to obtain a spherical multipole decompositions, or by projecting the internal current distributions onto Cartesian multipoles, and potentially also their various correction factors, to obtain a Cartesian multipole decomposition~\cite{Chen2011, Grahn2012, Miroshnichenko2015}.
Yet these multipolar decompositions ultimately remain a choice of basis for the scattering responses, and one where multipoles depend on the choice of origin; a multipole expansion is performed about an origin to describe the fields in the surrounding space.
There may be an intuitive choice of origin for simple nanostructures, but it can be more ambiguous for complex nanostructures, or arrangements of nanoparticles. 
Additionally, there is the practical problem that any given multipole is not necessarily going to align with a given resonance of a considered nanoantenna, which can lead to a large number of multipoles being necessary to describe a single resonance.
In this context, it was desirable to quantify resonances for a fixed object in a unique and origin-independent manner, particularly for later quantifying the interaction between resonances.  We therefore began to consider the eigenmodes of the current model in the presence of a driving field.
Moreover, (\ref{eq:current equation}) has an associated eigenmode equation, where an eigenmode $|v \rangle$ has a current distribution ${\mathbf{j}_{v}}$ and eigenvalue $\lambda_v$ that satisfies:
\begin{align}
i \omega   \lambda_v {\mathbf{j}_{v}}(\mathbf{r})    =&  -[ {{\boldsymbol{\bbar {\epsilon}}}(\mathbf{r}) - \epsilon_0}]^{-1} \cdot {{\mathbf{j}_{v}}(\mathbf{r})} +\frac{1}{\epsilon_0}\int   {\mathbf{\bbar G_0}}(\mathbf{r},\mathbf{r'}) \cdot  {\mathbf{j}_{v}}(\mathbf{r'}) \;{\mathrm{dr'}}^3 \label{eq:eigenmode equation}
\end{align}
These eigenvalues represent scalar impedances (or susceptibilities $\lambda^{-1}$) and the eigenmodes represent the associated origin-independent basis of stable current distributions.  These have a number of desirable properties:
\begin{itemize}
\item{By nature of being eigenmodes, the set of eigenmodes also represents the \textsl{only} basis for the optical response where each basis vector represents a current distribution that is subject to energy conservation in isolation.
In a formal sense, the real component of the eigenvalue must be greater than zero to be passive, or less than zero to be active, where {\it active} means inputting net energy into the system and {\it passive} refers to being not active.
This follows from the sign of the extinction losses (\ref{eq:Pext}), when using eigenmodes as solutions to (\ref{eq:current equation}): $\mathbf{E_0}\rightarrow \lambda_v  \mathbf j_v$, $\mathbf{J}\rightarrow \mathbf j_v$.}
\item{The complex frequency where an eigenvalue becomes zero corresponds to a self-sustaining field distribution, which is a formal way to define a resonance~\cite{Stratton}.
In fact, given the eigenmodes form almost always\footnote{See discussion on exceptional points following Figure~\ref{fig:heptamers}.} a complete and linearly independent basis, \textsl{every} such self-sustaining resonance must be associated with at least one zero eigenvalue.  
Note that 'self-sustaining' refers to a distribution of currents and fields whose magnitude does not decay in time, and requires no external input of energy.  
That is: a solution to (\ref{eq:current equation}) with $\mathbf{E_0}=\mathbf 0$ and a nontrivial current distribution $\mathbf J \neq \mathbf 0$. 
}
\end{itemize}
An eigenmode decomposition therefore uniquely maps to the complete set of resonances at \textsl{complex} frequencies, while also providing a complete set of necessarily passive basis vectors (assuming the absence of gain media) at \textsl{real} frequencies, and further connecting these two physical attributes together in a consistent and origin-independent modal framework.
However, it does introduce an issue in that the current model (\ref{eq:current equation}) contains loss and is thereby generally non-Hermitian, meaning its eigenmodes are not necessarily orthogonal.
The specific excitation of each eigenmode in the current model can still be found through the impact of reciprocity, or time-reversal symmetry of (\ref{eq:divB})-(\ref{eq:curlB}), on the eigenmodes of any arbitrary system. 
This is discussed further in Chapters~\ref{chapterSymmetry} and {\ref{chapterChiral}, but for now it suffices that Onsager's arguments~\cite{Onsager1930,Onsager1931} or the {\it Fluctuation Dissipation Theorem}~\cite{Callen1951}, require that the dyadic Green's function and permittivity tensor must be symmetric, although complex and not necessarily Hermitian.
\begin{align}
\mathbf{\bbar G_0}(\boldsymbol{x},\boldsymbol{x}')  = \mathbf{\bbar G_0}(\boldsymbol{x}',\boldsymbol{x}),  \quad\mathbf{\bbar G}= \mathbf{\bbar G_0}^{T},\quad {{\boldsymbol{\bbar {\epsilon}}}} = {{\boldsymbol{\bbar {\epsilon}}}}^{T}
\label{eq:onsager}
\end{align}
The overall operator of the eigenvalue equation (\ref{eq:eigenmode equation}) then represents a complex symmetric matrix, and there are a number of ways to show that this makes any two nondegenerate eigenmodes $\mathbf{j}_{v},\mathbf{j}_{w}$ orthogonal under unconjugated complex projections, see Chapter~7 of~\cite{Bai2000}.  
For an example, one can write the matrix in Gantmacher's normal form~\cite{Gantmacher1959}, which enforces such orthogonality between nondegenerate eigenmodes~\cite{Craven1969}.
\begin{align}
\int  {\mathbf{j}_{v} (\mathbf{r})} \cdot {\mathbf{j}_{w} (\mathbf{r})}   \; {\mathrm{dr}}^3= 0\,, \quad \mathrm{when}\;\,\lambda_v \neq \lambda_w\:. \label{eq:this is responsible for orthogonality}
\end{align}
The excitation $a_v$  of any eigenmode $\mathbf{j}_{v}$ can then be determined through unconjugated dot products between the eigenmode and  the driving field distribution, analogous to the more familiar use of true complex projections for the excitation of orthogonal eigenmodes.
\begin{align}
\lambda_v a_v = \frac{\int  {\mathbf{j}_{v} (\mathbf{r})} \cdot {\mathbf{E_0}(\mathbf{r})}   \; {\mathrm{dr}}^3}{\int  {\mathbf{j}_{v} (\mathbf{r})} \cdot {\mathbf{j}_{v} (\mathbf{r})}   \; {\mathrm{dr}}^3} \label{eq:excitationCoefficient}
\end{align}
Therefore, despite the eigenmodes being nonorthogonal, any given eigenmode's excitation is determined entirely by the given eigenmode's current distribution and the driving field.
The only exception to (\ref{eq:excitationCoefficient}) is $\int \mathbf j_{v} \cdot \mathbf j_{v} = 0$, which can generally be disregarded as accidental, although in Chapter~\ref{chapterChiral}, and specifically (\ref{eq:newexcitation}), I discuss eigenmodes whose symmetry enforces $\int \mathbf j_{v} \cdot \mathbf j_{v} = 0$, and require a different calculation of $a_v$. 
However, turning now to our consideration of nanoparticle oligomers, we can consider analogous eigenmodes of systems made from purely electric dipoles.
 An eigenmode $|v \rangle$, now having electric dipoles $\mathbf{p}_v$, will satisfy (\ref{eq:ED model}) as:
\begin{align}
\mathbf{p}_v^{(i)} = \boldsymbol{\bbar \alpha}_{\scriptscriptstyle E}^{(i)} \epsilon_0 \lambda_v \mathbf{p}_v^{(i)}  +  \boldsymbol{\bbar \alpha}_{\scriptscriptstyle E}^{(i)} \sum \limits_{j\neq i} k^2 \mathbf{\bbar G_0}(\mathbf{r}_i,\mathbf{r}_j)\cdot \mathbf{p}_v^{(j)}
\label{eq:eig equation}
\end{align}
However, when considering the model including magnetic dipoles (\ref{eq:Dmodel}), the eigenmodes need to be simultaneously constructed of both electric dipoles and magnetic dipoles, which  have different units.
To address this difference of units, my initial approach of [\ref{PRA}] was to separate (\ref{eq:Dmodel}) and consider different eigenmode equations for electric and magnetic dipoles, with cross terms to describe driving of electric dipoles by an applied magnetic field, and the driving of magnetic dipoles by an applied electric field.
In effect, this approach finds the eigenmodes of either the electric \textsl{or} the magnetic dipole systems, and their polarisabilities (eigenvalues), irrespective of the effect they have on the other dipole system.
However, while this remains a full description of the dipole system, and it provides information on the resonances of electric and magnetic systems in the presence of each other, it \textsl{does not} describe the simultaneous stable oscillations of both the electric \textsl{and} magnetic dipoles.
To consider the resonances of the collective system, we must consider both electric and magnetic dipoles together for single eigenmodes.
In this regard, it is desirable to introduce relative scaling between the electric and magnetic dipoles, and between the electric and magnetic fields, to maintain fixed units of polarisability for the resulting eigenvalues.
Moreover, the units can be standardised if the magnetic dipoles are scaled by a factor of ${c_0}^{-1}$, and the magnetic field by a factor of $\sqrt{\mu_0/\epsilon_0}$. 
This also makes the eigenmodes (not eigenvalues) independent of $\epsilon_0$ and $\mu_0$, see (\ref{eq:DmodelEig}), as might be expected given polarisabilities are defined relative to an arbitrary background.  
An eigenmode $|v \rangle$, having electric dipoles $\mathbf{p}_v$ and magnetic dipoles $\mathbf{m}_v$, will then satisfy (\ref{eq:Dmodel}) as:\\[-5ex]
\begin{changemargin}{-0.5cm}{0cm}
\begin{subequations}\label{eq:DmodelEig}
\begin{align}
 \lambda_v \, \mathbf{p}_v^{(i)} &=\,{(\boldsymbol{\bbar{\alpha}}_{{\scriptscriptstyle E}}^{(i)} \epsilon_{0} )}^{\!\!-1} \!\!\!\cdot   \mathbf{p}_v^{(i)}
-  \frac{k^{2}}{\epsilon_0}  \Big(\mathbf{\bbar G_0}(\mathbf{r}_{i},\mathbf{r}_{j}) \cdot \mathbf{p}_v^{(j)}  +\frac{1}{ik}\nabla\!\times\! \mathbf{\bbar G_0}(\mathbf{r}_{i},\mathbf{r}_{j}) \cdot [c_0^{-1} \mathbf{m}_v^{(j)}]\Big)\\[2ex]
  \lambda_v \, [c_0^{-1} \mathbf{m}_v^{(i)}] &=\,
   {(\boldsymbol{\bbar{\alpha}}_{{\scriptscriptstyle H}}^{(i)} \epsilon_0)}^{\!\!-1} \!\!\! \cdot  [c_0^{-1}   \mathbf{m}_v^{(i)}]
-  \frac{k^{2}}{\epsilon_0}  \Big(\mathbf{\bbar G_0}(\mathbf{r}_{i},\mathbf{r}_{j}) \cdot [c_0^{-1} \mathbf{m}_v^{(j)}]
-\frac{1}{ik}\nabla\!\times\!\mathbf{\bbar G_0}(\mathbf{r}_{i},\mathbf{r}_{j}) \cdot\mathbf{p}_v^{(j)} \Big)
\end{align}%
\end{subequations}%
\end{changemargin}
This expression describes a  matrix equation for eigenmodes of the electric and magnetic dipole system describing $N$ nanoparticles, however the associated $6N\times 6N$ matrix will, notably, not be symmetric when there is non-negligible coupling between the electric and magnetic dipoles.
Therefore the corresponding eigenmodes will not maintain orthogonality analogous to that in (\ref{eq:this is responsible for orthogonality}) for currents.

I now want to conclude with some discussion related to [\ref{PRB}], and consider the derivation of eigenmodes for nanoparticle oligomers using the eigenmodes of its different resonant subsystems. 
The aim of this particular work was to try and reconcile our eigenmode model with {\it plasmonic hybridisation theory}~\cite{Prodan2003}, which is an existing method to determine the collective optical response of metallic nanoparticle systems that support localised plasmon resonances.
While the underlying model in plasmonic hybridisation theory treats individual metallic nanoparticles as electron gas density distributions and is generally quite different to the models presented in Chapter~\ref{chapterModels}, it has an intuitive and conceptual goal.
By dividing a given metal nanoparticle system into two or more resonant subsystems with sufficiently simple properties, the collective properties can be deduced from how these subsystems combine. 
The theory itself originally neglected retardation of interaction between charges, meaning it focused on smaller nanoparticle systems than I consider, but an amended version was later proposed to account for retardation~\cite{Turner2010}, therefore justifying its use on systems including nanoparticle oligomers.
In [\ref{PRB}], I aimed to parallel the construction procedure of plasmonic hybridisation theory, but instead being relevant to our models for optical scattering in Chapter~\ref{chapterModels}: deriving the collective eigenmodes of nanoparticle oligomers from the eigenmodes of their interacting subsystems.
In doing this, we wanted to provide alternate and simplified commentary for the derivation of resonances without requiring the same complexity of plasmonic hybridisation theory. 
Moreover, the plasmonic hybridisation calculation is nontrivial for even simple systems, such as concentric spheres~\cite{Prodan2004} and two-particle dimers~\cite{Nordlander2004}, which has led to it becoming more regularly used as a conceptual tool to designate experimental and numerical observations of scattering that does not resemble that of the constituent nanoparticles in isolation.
Our intention was to provide an alternate avenue to allow simplified derivations of collective resonance formation, but also to extend this approach to apply to high-index dielectric nanoparticles.
In this regard, symmetric nanoparticle trimers were used as an example geometry, given these did not have a general solution in plasmonic hybridisation theory at the time.
We also later used asymmetric nanoparticle dimers as a second example in \mbox{\S10.3.3} of~\cite{Agrawal2017}, because this particular geometry was becoming of particular interest for high-index dielectric nanoparticles~\cite{Albella2015,Zywietz2015,Yan2015, Shibanuma2016}.
Our proposed method used the eigenmodes of each isolated subsystem as basis vectors for the collective optical response, and used the radiation from each of these basis vectors to quantify coupling channels between the basis vectors.
This allowed us to re-express the eigenvalue problem in (\ref{eq:DmodelEig}) with sets of a few coupled equations.
General expressions for the collective eigenmodes and eigenvalues could then be found directly from these coupled equations, and were able to provide quantitative agreement to full-wave numerical simulations, and an experimental measurement of transmission through a dielectric nanoparticle trimer. 
This demonstrated that the eigenmode model presented here could replicate the dominant collective resonance formation of plasmonic hybridisation theory, while simultaneously offering dramatic simplifications with quantitatively reliable modelling.
Given we are able encapsulate much of the effect of plasmonic hybridisation through a simple dipole model, I will turn our attention to the formation of specific resonant interference features known as {\it Fano resonances} in the coming section.
Moreover, Fano resonances in plasmonic nanoparticle oligomers were regularly attributed to the presence of plasmonic hybridisation, largely because it enabled non-radiative coupling between resonances~\cite{Giannini2011, Francescato2012, Lovera2013}.
The use of eigenmodes now offers an unambiguous way to quantify resonances, and is therefore a platform on which to quantify and understand interactions between resonances. 
In the coming section, I present my work toward understanding both Fano resonances and modal interference generally, and particularly the implementation of nanoparticle oligomers to realize such features.

\section{Eigenmode interference and Fano resonances}

One particular area that has garnered significant attention in recent years is the study of Fano resonances  in nanoparticle oligomers and other cluster structures~\cite{Hentschel2010, Fan2010_NL}.
For the context of nanoparticle scattering systems, Fano resonances have come to refer to a resonant interference in the total scattered power that is typically observed as a spectrally sharp, asymmetric lineshape in the extinction.
The name itself owes to an asymmetric lineshape in the energy spectrum of atomic photoionisation explained by Fano~\cite{Fano1961} to be due to constructive and destructive two-channel  interference between photoionisation from a broad ground state and from a discrete autoionised state of an atom.
Yet analogous asymmetric spectra appear in a wide range of optical, atomic and mechanical systems that support at least two-channel interference~\cite{Miroshnichenko2010}.
To this background, Fano resonances in plasmonic nanoparticle scattering systems became predominantly described by the interference between a strongly scattering ``bright resonance'' and a weakly scattering ``dark resonance"~\cite{Fan2010_NL, Francescato2012}.  
By nature of being a poor radiator, a dark resonance is expected to be less damped, and thereby spectrally sharp, while a bright resonance is heavily damped by radiation losses and spectrally broad.
Notably, this depicts a coupled oscillator model~\cite{Joe2006, Gallinet2011, Gallinet2012}, where a harmonically driven (bright) oscillator $m_1$ is damped by coupling $\gamma$ to an undriven (dark) oscillator $m_2$, such as the illustration in Figure~\ref{fig:coupledOscillator}a.
Here the feedback acting on $m_1$ due to a resonance in $m_2$ creates interference in the energy dissipation of the driving force provided by $m_1$ (the extinction). 
\begin{figure}[!ht]
\centerline{\includegraphics[width =0.9\textwidth]{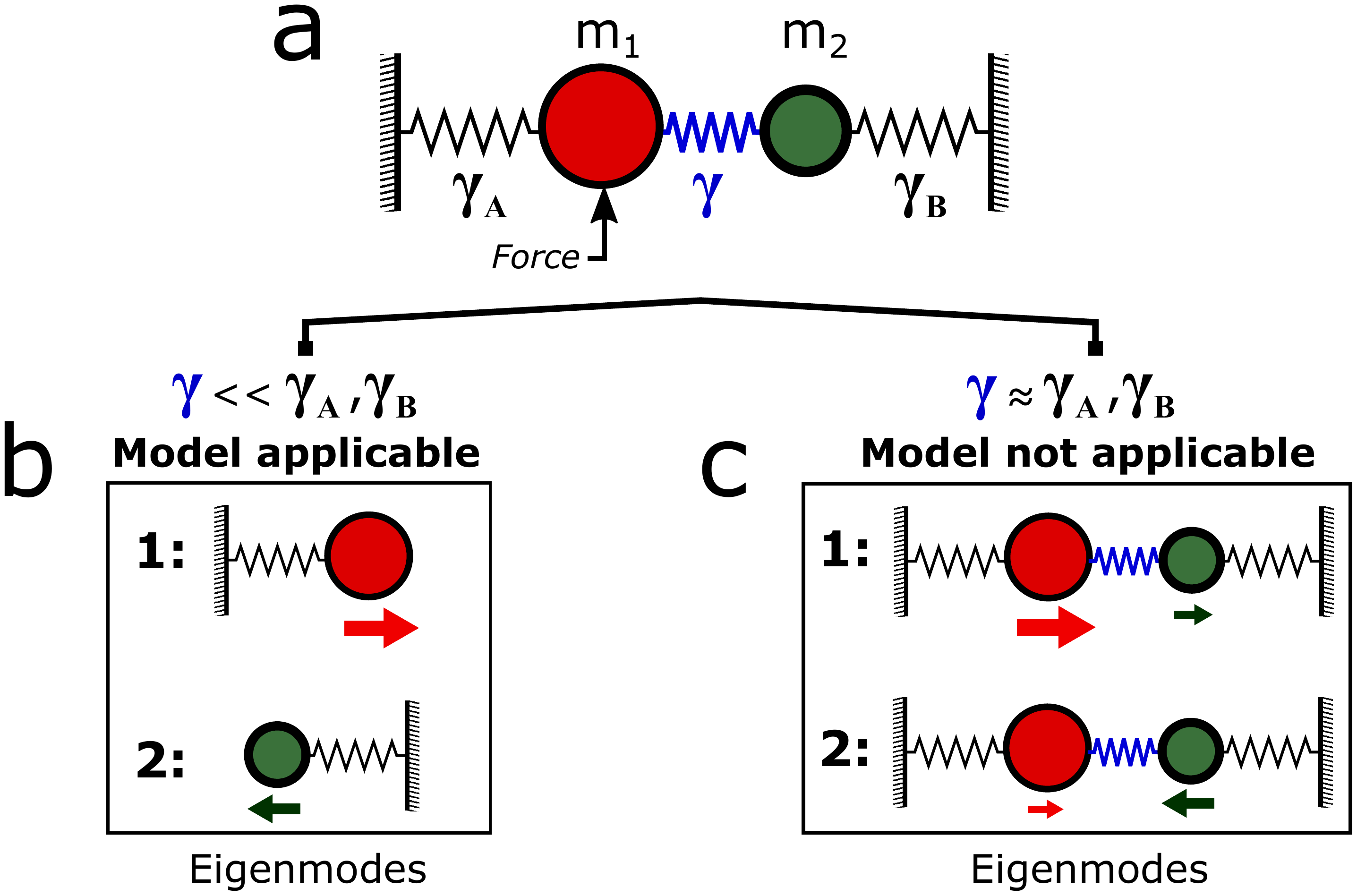}}
\caption{(a) The coupled oscillator model for two-channel Fano resonances; only the mass $m_1$ receives a direct external driving force, but it is coupled by $\gamma$ to a second mass $m_2$.  Qualitative illustration of two regimes where the interfering resonances will correspond to: (b) isolated resonances of $m_1$ and $m_2$, and (c) distributed resonances of the collective system.}
\label{fig:coupledOscillator}
\end{figure}
This interference can be either constructive or destructive, given the expected $\pi$ change in relative phase acquired over the resonance of $m_2$, enabling a characteristic asymmetric Fano lineshape in extinction.
However, it is less clear how to apply the coupled oscillator model if we can't identify distinct resonant subsystems for $m_1$ and $m_2$, or if the choice becomes arbitrary or ambiguous.  
Furthermore, there is an implicit requirement that the  coupling $\gamma$ is considered as small for individual oscillators $m_1$ and $m_2$ to resemble the resonances that one can physically observe in the total channel system, Figure~\ref{fig:coupledOscillator}b.
If the identified resonant subsystems for $m_1$ and $m_2$ are instead strongly coupled, then the physical resonances of the two-channel system will not resemble the isolated resonances of $m_1$ and $m_2$, Figure~\ref{fig:coupledOscillator}c.
However, there is still a two-channel system that allows interference between resonances, hence a continued propensity for Fano resonances.
The bright and dark mode depiction is simply no longer representative of the strongly coupled system, indeed both eigenmodes are bright, meaning they both couple directly to the applied force.
Notably, a case where we can't clearly identify distinct resonant subsystems for $m_1$ and $m_2$  is likely also a case where they are strongly coupled, which is generally the situation arising in my consideration of nanoparticle oligomers.  
At least when utilising coupled plasmonic nanoparticles, there are indeed a host of oligomers, and nanoparticle arrangements generally, from which Fano resonances can arise~\cite{Rahmani2013_lpor}.  
However, it was subsequently predicted that Fano resonances should also occur in symmetric oligomers made of silicon nanoparticles~\cite{Miroshnichenko2012}, where there is no plasmon hybridisation between nanoparticles, previously considered responsible for providing the coupling $\gamma$ in Figure~\ref{fig:coupledOscillator}a.
To this background, the formal treatment proposed in [\ref{PRA}] was developed to model Fano resonances directly from the resonances of the collective system \textsl{without requiring separation into distinct resonant subsystems}.  
This model identified a common theme underlying the physics of Fano resonances in both plasmonic and all-dielectric oligomers: the existence of Fano resonances can be attributed to the fact that the eigenmodes of these scattering systems are not orthogonal and, therefore: {they can directly interfere with each other in extinction}.
As I will present herein, this is a general and rigorous conclusion applicable to all interference between resonances that impacts the extinction.

We begin by noting that both the current model (\ref{eq:current equation}) and the dipole model (\ref{eq:Dmodel}) describe open, radiative systems.
As such, even in the absence of material loss, the system does still exhibit radiation losses and is generally, therefore, {\it non-Hermitian}.
The immediate consequence of this non-Hermicity is that the eigenmodes we have defined in (\ref{eq:eigenmode equation}), (\ref{eq:eig equation}) and (\ref{eq:DmodelEig}) are not necessarily orthogonal.
To recognise the effect of this nonorthogonality, and its relation to the Fano resonances, we can refer to the extinction cross-section by combining (\ref{eq:Pext}) and (\ref{eq:Psigma}).
\begin{align}
\sigma_{\mathrm{ext}}  &= \frac{1}{I_0}\sqrt{\frac{\mu_0}{\epsilon_0}} \; \mathrm{Re}\Big [ \int \mathbf{E}_0^* \! \cdot \mathbf{J}  \; \mathrm{dr}^3 \Big ]\:. \label{eq:extinction definition}
\end{align}
We now recognise that an arbitrary applied field and its induced currents can be defined in terms of a linear superposition of the eigenmodes.
\begin{align}
 \mathbf{E}_{0} &= \sum \limits _v a_v \lambda_v {\mathbf{j}_{v}}  \quad \Rightarrow \quad \mathbf{J} = \sum \limits _v a_v {\mathbf{j}_{v}}\:. \label{eq:first J}
\end{align}
We are then able to rewrite the total extinction (\ref{eq:extinction definition}) in terms of eigenmodes and eigenvalues.
Moreover, we can divide the extinction into two contributions: {\it direct terms} that provide contributions to extinction from individual eigenmodes, and also {\it interference terms} coming from the overlap between different eigenmodes.
\begin{align}
\sigma_{\mathrm{ext}}  &=  \frac{1}{I_0} \sqrt{\frac{\mu_0}{\epsilon_0}}\sum \limits_{v} \Bigg ( \; \underset{\mathrm{direct\;terms}}{\underbrace{\mathrm{Re}[\lambda_v]  \int |a_v|^{2}|\mathbf{j}_{v}|^{2}  \; \mathrm{dr}^3 }}\; + \sum \limits_{w \neq v}\;\underset{\mathrm{interference\;terms}}{\underbrace{\mathrm{Re}\Big[ a_v^{*} a_w \lambda_v^*  \int \mathbf{j}_{v}^{*}\cdot \mathbf{j}_{w} \; \mathrm{dr}^3 \Big]}}\Bigg) \:.
\label{eq:decomp}
\end{align}
We firstly recognise that the direct terms must always be greater than zero if we assume the system is passive; an eigenmode is an isolated solution to (\ref{eq:current equation}) and so must always produce positive extinction in the absence of gain media: it can't generate power.
Secondly, each given eigenmode's excitation, being the $a_v$ coefficients of (\ref{eq:first J}), is independent from the excitations of other eigenmodes as shown in (\ref{eq:excitationCoefficient}).
Therefore, the only way any interaction between two or more eigenmodes can affect the extinction cross-section is through interference terms.  
The existence of nonzero interference terms, is therefore required for Fano resonances to exist  in our model, meaning Fano resonances can only be described here by the nonorthogonality of eigenmodes.  
Given eigenmodes map uniquely to resonances, this also coincidentally shows that the largely accepted condition for one resonance being dark (orthogonal to the incident field) is \textsl{not} a requisite for Fano resonances.
These are the conclusions of [\ref{PRA}] for nanoparticle oligomers, and analogous conclusions were also reached in parallel by a separate work~\cite{ Forestiere2013} using models for plasmons in metallic nanoparticles systems.
The absence of dark resonances in Fano resonances was further in agreement with other recent works where Fano resonances were proposed and observed to occur between resonances that were driven directly by the incident field~\cite{Frimmer2012, Lovera2013}.
However, even beyond the conclusion that dark resonances are not necessary requisites for Fano resonances,  our model required that Fano resonances be due to nonorthogonal eigenmodes, and we could explicitly define requisites for nonorthogonal eigenmodes, as presented in [\ref{LPOR}]. 
For eigenmodes to be nonorthogonal, we must require either: (i) non-negligible retardation in coupling, $k \not \rightarrow 0$ or $k R \not \rightarrow 0$, to prevent  $\mathbf{\bbar G_0}$ becoming real and symmetric, hence Hermitian with orthogonal eigenmodes; or (ii) multiple materials following the argument of (31)-(35) in  [\ref{LPOR}].  

The eigenmode  decomposition (\ref{eq:decomp}) is also not specific to plasmonic nanoparticle systems, it simply requires that we can calculate eigenmodes, which meant we could equally explore interference that appeared in high-index dielectric nanoparticle oligomers, as was done in \mbox{[\ref{PRA}, \ref{ACS}, \ref{PRB}]}.  
To this extent, a comparison between plasmonic and dielectric Fano resonant oligomers, heptamers, is given in  Figure~\ref{fig:heptamers}. 
Here I show an eigenmode decomposition of extinction of each heptamer, in which I plot the direct terms of extinction (\ref{eq:decomp}) from the dominant eigenmodes, overlaid with the total extinction.
The difference between total extinction and the sum of direct terms is then the sum of interference terms due to nonorthogonal eigenmode overlap.
For the case of the gold nanoparticle heptamer in Figure~\ref{fig:heptamers}a, we have a typical scenario for a Fano resonance: the overlap of a broad resonance and a sharp resonance, which leads to destructive interference.
This gold heptamer is modelled after the investigation in~\cite{Hentschel2010}, and shows a classical example of a two-channel Fano resonance, albeit with interference between the resonances being due to eigenmode nonorthogonality.
\begin{figure}[!ht]
\centerline{\includegraphics[width=0.9\textwidth]{{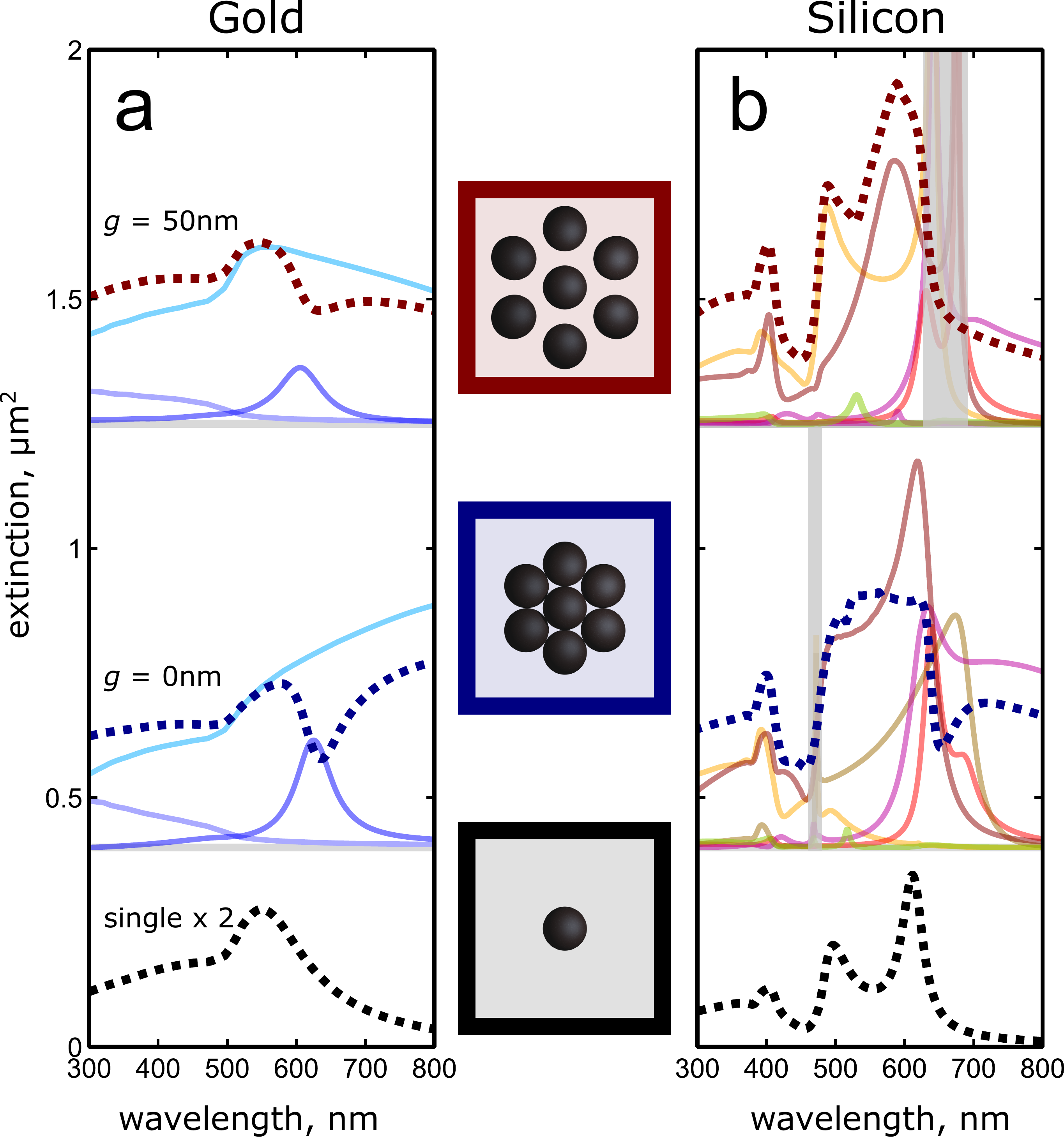}}}
\caption{(Dashed lines) Extinction spectra of (a) gold and (b) silicon heptamers, simulated using the dipole model (\ref{eq:Dmodel}) and showing the role of eigenmode interference in producing Fano resonances.
(Solid lines) Overlaid direct terms to extinction, as per (\ref{eq:decomp}), for all three excited eigenmodes of the gold heptamers, and the six most dominant eigenmodes for the silicon heptamers.
The grey regions hide wavelength bands near exceptional points.  
Both gold and silicon nanospheres are 150~nm in diameter.
Figure taken from Chapter~10 of~\cite{Agrawal2017}.}
\label{fig:heptamers}
\end{figure}
However, as seen in Figure~\ref{fig:heptamers}b, the situation becomes dramatically more complicated for a silicon nanoparticle heptamer.
The number of eigenmodes of this system is much higher.
More formally, the number of eigenmodes of the gold heptamer that can be excited by a plane wave is limited by symmetry to three pairs of polarisation-degenerate eigenmodes following the argument of [\ref{PRA}], equations~(2)-(10).  
This number increases to six, due to extra magnetic dipoles in the silicon quadrumer when repeating the same argument and neglecting the electric-magnetic dipole coupling (\ref{eq:EMcoupling}).  
It then increases beyond six with electric-magnetic dipole coupling, because such coupling allows dipoles to be oriented parallel to the propagation direction; see (\ref{eq:EMcoupling}) regarding cross-coupling in (\ref{eq:Dmodel}).
The first consequence of more eigenmodes seen in Figure~\ref{fig:heptamers}b is many more signatures of interference occurring across the extinction spectra.
The second consequence is that we have to deal with the formation of  {\it exceptional points}~\cite{Heiss2001, Dembowski2001, Heiss2012}.
An exceptional point refers to a point degeneracy of two or more eigenvalues when the corresponding eigenmodes also become linearly dependent.
The linear dependence of two or more eigenmodes subsequently implies that the span of the complete set of eigenmodes has reduced, indicating that the eigenmodes cease to be a complete basis for the response space.   
One can instead recover the complete basis by defining generalised eigenvectors, see Chapter~6 of~\cite{Bronson2009}.
Ultimately, however, the eigenmodes are a bad basis in the vicinity of an exceptional point: there is a component of the response space that is becoming orthogonal to the eigenspace.
In Figure~\ref{fig:heptamers}b we observe that the excitation magnitude of coalescing eigenmodes can diverge in the vicinity of an exceptional point, given there is a component of the excitation field becoming orthogonal to the eigenmodes while the still in the span of the eigenspace ({\it i.e.} when not precisely at the exceptional point). 
Exceptional points can exist even in simple plasmonic and dielectric oligomer systems, as found when deriving eigenmodes in [\ref{PRB}] and in \S10.3.3 of~\cite{Agrawal2017} with complex frequencies, but appear more regularly when there are more interacting eigenmodes.
I have therefore attempted to divert attention away from the direct extinction terms of individual coalescing eigenmodes in Figure~\ref{fig:heptamers}b, given they diverge as the wavelength nears an exceptional point.
This nonetheless illustrates an unexpected level of complexity of interactions that arises there are many intercoupled resonances, the example being Figure~\ref{fig:heptamers}a vs. \ref{fig:heptamers}b.

Yet [\ref{ACS}] instead presents a controlled use of four silicon nanoparticles in a symmetric quadrumer geometry for the purposes of interfering two collective magnetic resonances. 
Referring to the illustration in Figure~1b of [\ref{ACS}], the global circulation of electric dipoles resembles a magnetic dipole analogous to a circulation of current (\ref{eq:magdip0}), while the parallel aligned magnetic dipoles of each nanoparticle will also radiate like a magnetic dipole analogous to a volume integral of magnetisation $\mathbf {\bar m} = \int_{\!\mathbf r \in V} \!\mathbf M(\mathbf r) \mathrm{dr}^3$.
If we were to treat this quadrumer as a point source, the net magnetisation of this point would have contributions from the circulating electric dipoles resembling circulating current (\ref{eq:PM}), but also a second channel from aligned internal magnetic dipoles, which would more resemble a fictitious oscillating magnetic charge.
While this is ultimately a consequence of incorrectly perceiving the quadrumer as a point source, there are tangible effects of the presence of these two channels for the quadrumer's collective magnetic response: the two channels are coupled through internal electric-magnetic dipole coupling, and thereby enable two-channel interference and Fano resonances.
This is precisely what is realized in [\ref{ACS}].
A symmetric silicon quadrumer, and its microwave analogue, were both demonstrated to support magnetic-magnetic Fano interference when perceiving the response of the quadrumer as a single collective object. 
This magnetic Fano resonance was then later used to enhance the local internal magnetic dipole moments of silicon quadrumer arrays for the purpose of enhancing their third harmonic generation~\cite{Shorokhov2016}, using interference to build on previous demonstrations of third harmonic generation due to magnetic resonances of individual silicon disks~\cite{Shcherbakov2014}.
As such, this demonstrates how considered use of high-index dielectric nanoparticles in oligomers can enable both unconventional optical properties and functional outcomes.

\chapter{Symmetry effects and eigenmodes \label{chapterSymmetry}}

Symmetry is one of very few tools where simple analytical arguments can provide quantifiable constraints on the otherwise nontrivial dependency shared between the geometry of a nanoparticle system and its collective optical response.  
Section~\ref{sec:geoSym} will provide a short introduction to the decomposition of optical scattering responses based on their properties under geometric symmetry transformations, and then \ref{sec:tSym} discusses the consequences of time-reversal symmetry.
These ideas will then be combined in Section~\ref{sec:rotSym} to present analysis on the eigenmodes of geometries that have at least 3-fold discrete rotational symmetry, and the subsequent manifestation of polarisation-independent scattering and absorption.

\section{Geometric symmetry \label{sec:geoSym}}

In this section, I provide a short summary on the use of group theory to quantify geometric symmetry in the models from Chapter~\ref{chapterModels}, and also the eigenmodes in Chapter~\ref{chapterEigen}.
The content will largely follow from more comprehensive works on the application of group theory to condensed matter systems~\cite{Harris1989, Cotton1989}, and particularly~\cite{Dresselhaus2008}, though presented here as relevant to specifically nanoparticle scattering systems.  
In this regard, I first restrict our consideration to specifically {\it  symmetry point groups}, which consider geometric transformations acting on coordinates (points) in a given space.  
A symmetry point group $G$ refers to a particular set of transformation operations that form a {\it group} because each operation $\hat{R}$ in  $G$ also has an inverse operation $\hat{R}^{-1}$ in  $G$.
The relevant symmetry for the models in Chapter~\ref{chapterModels} is going to be the invariance under a given transformation $\hat R$ of either: a permittivity distribution $\boldsymbol{\bbar \epsilon}(\mathbf{r})$, or an arrangement of discrete dipole polarisabilities $\boldsymbol{\bbar \alpha}_{\scriptscriptstyle E}$, $\boldsymbol{\bbar \alpha}_{\scriptscriptstyle H}$ located at points in space $\mathbf{r}$.
The application of a symmetry operation $\hat{R}$ on the corresponding geometric space can then be described by transforming each $\mathbf{r} \rightarrow \hat{R}\mathbf{r}$.
However, to quantify symmetry operations on vector fields ({\it e.g.} $\mathbf E$, $\mathbf B$, $\mathbf J$, {\it etc.}), it is often intuitive to consider the equivalent transformation of the coordinate system from the inverse symmetry operation $\hat R^{-1}$, which is also necessarily in $G$.
When $\hat R^{-1}$ acts on the coordinate system of a vector field $\mathbf{u}(\mathbf{r})$, which is distributed over our geometric space, the vector field will be transformed by $\hat R$ as \mbox{$\mathbf{u}(\mathbf{r}) \rightarrow \hat{R}\mathbf{u}(\hat{R}\mathbf{r})$}.
I should now emphasize that $\hat{R}$ here denotes an operation, while the explicit transformation will be enacted by some matrix $\mathbf{\bbar D}^{(\!\hat R)}\!$.
The form of $\mathbf{\bbar D}^{(\!\hat R)}$ will depend on the properties of the space on which $\hat R$ acts, that is: the specific quantity a given $\mathbf{u}$ represents.
As an example, parallel components of electric and magnetic fields have opposite parity under a reflection of the coordinate system $\hat R = \hat \sigma$.  
This follows because magnetic field is defined from the curl of electric field in (\ref{eq:curlE}); the magnetic field is related to local rotation of electric field at the given location, discussed in Section~\ref{firstPrinciples}.   
As such, if $\hat \sigma (\mathbf E\! \cdot\! \mathbf{\hat x}) \mathbf{\hat x} = (\mathbf E\! \cdot\! \mathbf{\hat x}) \mathbf{\hat x}$,  for some direction $\mathbf{\hat x}$, then $\hat \sigma ( \mathbf B\! \cdot\! \mathbf{\hat x}) \mathbf{\hat x} = -  ( \mathbf B\! \cdot\! \mathbf{\hat x})  \mathbf{\hat x}$.  

Returning now to our given symmetry group $G$, but now also presuming to have a quantity $\mathbf u$, the set of symmetry operations $\hat R$ in $G$ will be described for $\mathbf u$ by some set of transformation matrices $\mathbf{\bbar D}^{(\!\hat R)}$.  
This set of matrices is called a {\it representation} of $G$.
Yet, much like our ability to consider $\mathbf E$ and $\mathbf H$ separately owing to their different transformation properties under reflection ({\it e.g.} rather than considering a concatenated 6-vector), we can choose to decompose our desired vector fields according to the transformation properties under symmetry operations.  
This is where the limiting concept of {\it irreducible representations} becomes useful.
A given representation is {\it reducible} if there is a single unitary basis transformation that can express all matrices $\mathbf{\bbar D}^{(\!\hat R)}$ from the representation in the same diagonal matrix block form, as it indicates each block is a smaller dimensional representation of a subspace of the given $\mathbf u$.
For example, if there is a single unitary matrix $\mathbf{\bbar U}$ that allows a representation of matrices $\mathbf{\bbar D}^{(\!\hat R)}$ to be expressed in a block diagonal form seen below for all $\hat R$ in $G$, then the original representation can be reduced into the two representations having matrices $\mathbf{\bbar D'}^{(\!\hat R)}$ and $\mathbf{\bbar D''}^{(\!\hat R)}$.
\begin{align}
\mathbf{\bbar U}^\dagger \cdot \mathbf{\bbar D}^{(\!\hat R)} \cdot \mathbf{\bbar U}  
=
\left(
\begin{array}{c|c}
& \\[-2ex]
\quad \mathbf{\bbar D'}^{(\!\hat R)}\quad
  & \\[1ex] \hline
  & \mathbf{\bbar D''}^{(\!\hat R)}
\end{array} \right )
\end{align}
An irreducible representation $\Gamma$ is a representation for $G$, consisting a set of matrices  $\mathbf{\bbar D}_\Gamma^{(\!\hat R)}$ for all $\hat R$ in $G$, which cannot be rewritten in a diagonal block form of lower dimensional representations.
Conversely, this means that any symmetry operation can be decomposed into a matrix that is equivalent to a block diagonal form, where each block is a matrix in the irreducible representations of the corresponding operation.
The specific $\mathbf{\bbar D}_\Gamma$ matrices of irreducible representations are described by character tables, which list the character (trace) of each matrix $\mathbf{\bbar D}_\Gamma^{(\!\hat R)}$ in the given symmetry group.
In Table~\ref{table:charTabs}, I write down a set of character tables for a few symmetries that will be used in this and the next section, and have corresponding example geometries illustrated in Figure~\ref{fig:localChiralFields}a.
\begin{table}[!h]
\begin{changemargin}{-0.85cm}{0cm}
\begin{center}
{\large $\boldsymbol{C_4}:\quad\!\!\!$}
\begin{tabular}{|c|c|c|c|c||c|}
\hline & & & & & \\[-2ex]
& \large$\hat E$ &\large $\hat C_2$ &\large $\hat C_4$ &\large $(\hat C_4)^3$ & {\small examples for} $[x,y,z]$\\
\hline & & & &  & \\[-2ex]
\large $A$ & 1 & 1& 1& 1& $z$\\[1ex]
\large $B$ & 1 & 1& -1& -1& $x^2-y^2$\\[1ex]
\large $E$ & $\bigg\{
\begin{array}{c} 1\\ 1\end{array}$ &
 $\begin{array}{c}-1 \\ -1\end{array}$ &
 $\begin{array}{c}i \\ -i \end{array}$ & 
$\begin{array}{c} -i \\ i \end{array}$ & 
$\begin{array}{c}x + i y\\ x - i y\end{array}$
\\[1.5ex]\hline
\end{tabular}

{\large $\boldsymbol{S_4}:\quad\!\!$}
\begin{tabular}{|c|c|c|c|c||c|}
\hline & & & & & \\[-2ex]
& \large$\hat E$ &\large $\hat C_2$ &\large $\hat S_4$ &\large $(\hat S_4)^3$ & {\small examples for} $[x,y,z]$\\
\hline & & & &  & \\[-2ex]
\large $A$ & 1 & 1& 1& 1& $z$\\[1ex]
\large $B$ & 1 & 1& -1& -1& $x^2-y^2$\\[1ex]
\large $E$ & $\bigg\{
\begin{array}{c} 1\\ 1\end{array}$ &
 $\begin{array}{c}-1 \\ -1\end{array}$ &
 $\begin{array}{c}i \\ -i \end{array}$ & 
$\begin{array}{c} -i \\ i \end{array}$ & 
$\begin{array}{c}x + i y\\ x - i y\end{array}$
\\[1.5ex]\hline
\end{tabular}

{\large $\boldsymbol{C_{4v}}:\;\,\!\!$}
\begin{tabular}{|c|c|c|c|c|c||c|}
\hline & & & & & & \\[-2ex]
& \large$\,\hat E\,$ &\large $\hat C_2$ & \large $2\hat C_4$ & \large $2 \hat \sigma_v$ & \large $2  \hat \sigma_d$ & {\small examples for} $[x,y,z]$\\
\hline & & & & & & \\[-2ex]
\large $A_1$ & 1 & 1& 1& 1& 1&  $z$\\[1ex]
\large $A_2$ & 1 & 1& 1& -1& -1&  $ \frac{\mathrm dx}{\mathrm dy} - \frac{\mathrm  dy}{ \mathrm dx} $\\[1ex]
\large $B_1$ & 1 & 1& -1& 1& -1& $x^2- y^2$ \\[1ex]
\large $B_2$ & 1 & 1& -1& -1& 1& $xy$\\[1ex]
\large $E$ & 2 & -2 & 0& 0& 0 & $[x,y]$ \\[1ex]\hline
\end{tabular}

\mbox{{\large $\boldsymbol{C_{4h}}:\;\,\!\!$}
\begin{tabular}{|c|c|c|c|c|c|c|c|c||c|}
\hline & & & & & & & & & \\[-2ex]
& \large$\hat E$ &\large $\hat C_2$ & \large $\hat C_4$ & \large $\!(\hat C_4)^3\!$ & \large $\hat i$
& \large $\hat S_4$  & \large $\!(\hat S_4)^3\!$& \large $ \hat \sigma_h$ & {\small examples for} $[x,y,z]$\\
\hline & & & & &  & & & &  \\[-2ex]
\large $A_g$ & 1 & 1& 1& 1& 1& 
 1& 1& 1& $x^2 + y^2$\\[1ex]
\large $A_u$ & 1 & 1& 1& 1& -1&
  -1& -1& -1&  $ z $\\[1ex]
\large $B_g$ & 1 & 1& -1& -1& 1&  
 -1& -1& 1& $x^2- y^2$ \\[1ex]
\large $B_u$ & 1 & 1& -1& -1& -1&  
1& 1& -1&  \\[1ex]
\large $E_g$ & $\bigg\{
\begin{array}{c} 1\\ 1\end{array}$ &
 $\begin{array}{c}-1 \\ -1\end{array}$ &
 $\begin{array}{c}i \\ -i \end{array}$ & 
$\begin{array}{c} -i \\ i \end{array}$ & 
$\begin{array}{c} 1 \\ 1 \end{array}$ & 
$\begin{array}{c} -i \\ i \end{array}$ & 
$\begin{array}{c} i \\ -i \end{array}$ & 
$\begin{array}{c} -1 \\ -1 \end{array}$ & 
$\!\!\!\!\!\!\begin{array}{c}\frac{\mathrm d z}{\mathrm d y}\!-\!\frac{\mathrm d y}{\mathrm d z} \!+\! i (\frac{\mathrm d x}{\mathrm d z}\!-\!\frac{\mathrm d z}{\mathrm d x}) \\ \frac{\mathrm d z}{\mathrm d y}\!-\!\frac{\mathrm d y}{\mathrm d z} \!-\! i (\frac{\mathrm d x}{\mathrm d z}\!-\!\frac{\mathrm d z}{\mathrm d x}) \end{array}\!\!\!\!\!\!$
\\[3ex]
\large $E_u$ & $\bigg\{
\begin{array}{c} 1\\ 1\end{array}$ &
 $\begin{array}{c}-1 \\ -1\end{array}$ &
 $\begin{array}{c}i \\ -i \end{array}$ & 
$\begin{array}{c} -i \\ i \end{array}$ & 
$\begin{array}{c} -1 \\ -1 \end{array}$ & 
$\begin{array}{c} i \\ -i \end{array}$ & 
$\begin{array}{c} -i \\ i \end{array}$ & 
$\begin{array}{c} 1 \\ 1 \end{array}$ & 
$\begin{array}{c}x + i y\\ x - i y\end{array}$
\\[1.5ex]\hline
\end{tabular}}

\end{center}
\end{changemargin}

\caption{Character tables for the $C_4$,  $S_4$, $C_{4v}$ and $C_{4h}$ symmetry groups (see also the example illustrations in Figure~\ref{fig:localChiralFields}a).  Rows denote irreducible representations, columns denote symmetry operations: $\hat E$ is the identity, $\hat i$ is a point inversion, $\hat C_n$ and $\hat S_n$ are rotation and improper rotation by $\frac{2\pi}{n}$, $\hat \sigma$ is a reflection plane parallel ($\hat \sigma_v$, $\hat \sigma_d$) or perpendicular ($\hat \sigma_h$) to the principle rotation axis.  Table indices are the trace of the matrix representation for the corresponding symmetry operation.}
\label{table:charTabs}

\end{table}

For the remainder of this Chapter, we will require only that the given $\mathbf{\bbar D}_\Gamma$ exists, or consider exclusively one-dimensional irreducible representations, where each $\mathbf{\bbar D}_\Gamma$ is just a scalar ${ D}_\Gamma$, and equal to its character. 
The main use of irreducible representations in this Chapter will be to separate a vector field or distribution $\mathbf u$ into a finite set of orthogonal distributions $\mathbf{u}_\Gamma$ that each transform according to a single irreducible representation $\Gamma$.  
Here each $\Gamma$ is an irreducible representation for a symmetry group $G$, which consists only symmetry operations the given geometry is invariant under. 
In essence, we will show that one can always write a given $\mathbf{u}$ as a linear combination of some set of $\mathbf{u}_\Gamma$:
\begin{align}
\mathbf u(\mathbf r) = \sum \limits _{\Gamma} a_\Gamma \mathbf u_\Gamma(\mathbf r)
\label{eq:symmetrydecomp}
\end{align}
Here $a_\Gamma$ are complex coefficients.
The ability to write this decomposition isn't immediately obvious, so we can start from the definition that the irreducible representations describe all possible transformations under the symmetry operations of $G$, following from their definition as a basis for representations for the group $G$~\cite{Dresselhaus2008}.
This is equivalent to saying that there is no portion of $\mathbf{u}$ that doesn't lie in the span of some given $\mathbf{u}_\Gamma$, because it would indicate a component of $\mathbf u$ doesn't transform under symmetry operations according to any irreducible representation in $G$.  
If it can now be shown that $\mathbf{u}_{\Gamma_{\!\scriptstyle i}}$ and $\mathbf{u}_{\Gamma_{\!\scriptstyle j}}$ are orthogonal when $\Gamma_i \neq \Gamma_j$, we infer linear independence of each $\mathbf{u}_{\Gamma}$ and therefore the ability to uniquely decompose $\mathbf{u}$ into $\mathbf{u}_\Gamma$ as per (\ref{eq:symmetrydecomp}). 
This is precisely what is shown in the following box.
\begin{myfloat}[H]
\begin{mdframed}[style=BenFrame]
\small
{\bf Orthogonality of basis vectors from different irreducible representations.}
When applying symmetry operations $\hat{R}^{-1}$ to the coordinate system,  the inner product between $\mathbf{u}_{\Gamma_{\!\scriptstyle i}}$ and $\mathbf{u}_{\Gamma_{\!\scriptstyle j}}$ will be conserved because symmetry operations $\hat{R}$ are necessarily unitary (e.g. to conserve geometry).
\begin{align}
\int [\mathbf{\bbar D}_{\Gamma_{\!\scriptstyle i}}^{\!(\hat R)}\!\! \cdot \mathbf{u}_{\Gamma_{\!\scriptstyle i}}(\mathbf{r})]^* \! \cdot [\mathbf{\bbar D}_{\Gamma_{\!\scriptstyle j}}^{\!(\hat R)} \!\!\cdot \mathbf{u}_{\Gamma_{\!\scriptstyle j}}(\mathbf{r})] \mathrm{dr}^3 = \int [\mathbf{u}_{\Gamma_{\!\scriptstyle i}}  (\mathbf r)]^* \! \cdot \mathbf{u}_{\Gamma_{\!\scriptstyle j}} (\mathbf r)  \mathrm{dr}^3
\label{eq:conservedInnerProduct}
\end{align}
Here we have also used the fact that $\mathbf{ \hat{x}}$ and $\hat{R}\mathbf{ \hat{r}}$ are equivalent integration variables because $\hat{R}$ preserves the geometric integration volume, allowing us to describe the effect of any $\hat{R}$ entirely by the matrix representations $\mathbf{\bbar D}_{\Gamma_{\!\scriptstyle i}}^{(\!\hat R)}$ and $\mathbf{\bbar D}_{\Gamma_{\!\scriptstyle j}}^{(\!\hat R)}$ acting on $\mathbf{u}_{\Gamma_{\!\scriptstyle i}}$ and $\mathbf{u}_{\Gamma_{\!\scriptstyle j}}$, respectively.
We now utilise the {\it Wonderful Orthogonality Theorem}~\cite{Dresselhaus2008} and unitary $\hat R$.
\begin{align}
\sum_{\hat R} [\mathbf{\bbar D}_{\Gamma_{\!\scriptstyle i}}^{(\!\hat R^{\,{\text{\mbox{\tiny -1\!}}}})}]_{ab} \, [\mathbf{\bbar D}_{\Gamma_{\!\scriptstyle j}}^{(\!\hat R)}]_{a' b'}= \frac{N_{G}}{L}\delta_{\Gamma_{\!\scriptstyle i},\Gamma_{\!\! \scriptstyle j}} \delta_{a,a'}\delta_{b,b'} \Rightarrow \sum_{\hat R} [\mathbf{\bbar D}_{\Gamma_{\!\scriptstyle i}}^{(\!\hat R)}]^* \! \cdot [\mathbf{\bbar D}_{\Gamma_{\!\scriptstyle j}}^{(\!\hat R)}]= 0
\label{eq:wonderfulorthogonality}
\end{align}
Here  $a,a',b,b'$ denote matrix indices, $N_{G}$ is the number of $\hat R$ in the group $G$, $\delta$ denotes a Kronecker delta function, and $L$ is the dimension of both ${\Gamma_j}$ and ${\Gamma_j}$, where we can neglect $L_{\Gamma i}\neq L_{\Gamma j}$ due to the $\delta_{\Gamma_{\!\scriptstyle i},\Gamma_{\!\! \scriptstyle j}} $ term.
The right hand side of (\ref{eq:wonderfulorthogonality}) also assumes $\Gamma_i\neq\Gamma_j$. 
In any case, we can then consider a sum of $N_{G}$ identical inner products of $  \mathbf{u}_{\Gamma_{\!\scriptstyle i}} $ and $\mathbf{u}_{\Gamma_{\!\scriptstyle j}}$.
\begin{align}
N_{G} \int   \mathbf{u}_{\Gamma_{\!\scriptstyle i}} ^* \! \cdot \mathbf{u}_{\Gamma_{\!\scriptstyle j}} \mathrm{dr}^3& = \sum \limits_{\hat R} \int  [\mathbf{\bbar D}_{\Gamma_{\!\scriptstyle i}}^{(\!\hat R)}\mathbf{u}_{\Gamma_{\!\scriptstyle i}}]^* \! \cdot [\mathbf{\bbar D}_{\Gamma_{\!\scriptstyle j}}^{\!(\hat R)} \mathbf{u}_{\Gamma_{\!\scriptstyle j}} ] \mathrm{dr}^3
\,, \;\; \text{using }(\ref{eq:conservedInnerProduct}) \nonumber \\ 
&= 0\,, \;\; \text{using }(\ref{eq:wonderfulorthogonality})
\label{eq:provenorthogonality}
\end{align}
Given there is always at least an identity operation in any $G$, we know $N_{G}\geq 1$, and therefore (\ref{eq:provenorthogonality}) shows $\mathbf{u}_{\Gamma_{\!\scriptstyle i}}$ and $\mathbf{u}_{\Gamma_{\!\scriptstyle j}}$ must be orthogonal if $\Gamma_i\neq\Gamma_j$.
\end{mdframed}
\end{myfloat}

Let us now consider the more explicit case of some scattering object whose geometry is invariant under operations of some symmetry group $G$, which has a set of irreducible representations $\{\Gamma\}$.  
Using (\ref{eq:symmetrydecomp}), we can separate any external field distribution $\mathbf{E_0}$  into the individual components $ \boldsymbol{\scriptstyle\mathcal{E}}_{\Gamma}$ that transform according to the corresponding $\Gamma$.
\begin{align}
\mathbf{E_0}(\mathbf{r})=\sum \limits_\Gamma a_\Gamma \boldsymbol{\scriptstyle\mathcal{E}}_{\Gamma}(\mathbf{r})
\end{align}
The current distribution $\boldsymbol{\scriptstyle\mathcal J}_{\Gamma}$, induced by $ \boldsymbol{\scriptstyle\mathcal{E}}_{\Gamma}$, can now be written as a linear combination of eigenmode current distributions $\mathbf{j}_{v}$, following (\ref{eq:eigenmode equation}). 
\begin{align}
 \boldsymbol{\scriptstyle\mathcal{E}}_{\Gamma}  &= \sum \limits _v b_{{\scriptscriptstyle \Gamma},v} \lambda_v {\mathbf{j}_{v}}  \quad \Rightarrow \quad  \boldsymbol{\scriptstyle\mathcal J}_{\Gamma} = \sum \limits _v b_{{\scriptscriptstyle \Gamma},v} {\mathbf{j}_{v}}\:. 
 \label{eq:eigensymmetry}
\end{align}
Importantly, we can show that each eigenmode ${\mathbf{j}_{v}}$ of the geometry will, once excluding accidental degeneracies, transform according to only a single irreducible representation $\Gamma$.
If this were not true, a single ${\mathbf{j}_{v}}$ could be written as a linear combination of $\mathbf{u}_\Gamma$ that each transform according to different irreducible representations $\Gamma$ using (\ref{eq:symmetrydecomp}). 
Because each $\Gamma$ has a unique set of transformation properties, the number of linearly independent $\mathbf{u}_\Gamma$ in the given $\mathbf j_v $ specifies the number of $\hat R$ that can act on ${\mathbf{j}_{v}}$ to create a new, linearly independent eigenmode ${\mathbf{j}'_v}$, which must be degenerate with ${\mathbf{j}_{v}}$ (having the same eigenvalue $\lambda_v$) given $\hat R$ is equivalent to a coordinate transformation. 
In other words, the number of  $\mathbf{u}_\Gamma$ is equal to the degeneracy order of ${\mathbf{j}_{v}}$. 
We can subsequently construct a set of ${\mathbf{j}_{v}}$ and all ${\mathbf{j}'_{v}}$  whose size is equal to, and thereby spans, the set of $\mathbf{u}_\Gamma$.
This means, conversely, that we can write each  $\mathbf{u}_\Gamma$ as a linear combination of the set of ${\mathbf{j}_{v}}$ and ${\mathbf{j}'_{v}}$, which means each $\mathbf{u}_\Gamma$ is an eigenmode with eigenvalue $\lambda_v$.
Hence we have reached our original statement that each eigenmode transforms according to a single $\Gamma$, once excluding accidental degeneracies.
We can also conclude that $ \boldsymbol{\scriptstyle\mathcal J}_{\Gamma}$ in (\ref{eq:eigensymmetry}) transforms according to $\Gamma$, given it is a linear combination of  ${\mathbf{j}_{{\scriptscriptstyle \Gamma},v}}$. 
These results will be used in the derivations of the next section to constrain the scattering and absorption properties of rotationally symmetric nanostructures.

\section{Reciprocity and time-reversal symmetry  \label{sec:tSym}}

In this section, I want to explore a specific conclusion necessary for arguments I consider in the remainder of this thesis: any applied electric field distribution $\mathbf{E_0}$ experiences exactly the same rate of energy loss (extinction) as the distribution $\mathbf{E_0}^*\!$. 
When $\mathbf {E_0}$ is a plane wave with wavevector $\mathbf k$, $\mathbf{E_0}^*$ is a co-polarised plane wave with wavevector $-\mathbf k$, and this invariance of extinction from plane waves propagating in opposite directions is one basic conclusion attributed to {\it reciprocity} in optics.
So I begin with reciprocity, the most common definition of which for optics is an equivalence of some measurement when interchanging source and detector.
While clearly dependent on the chosen definition of the detector's measurement, there are a number relevant theorems to this extent in optics, of which Potton~\cite{Potton2004} provides an extensive review.
The most commonly utilised version of reciprocity, and relevant to optical far-field scattering, can be paraphrased in a manner analogous to that presented in \S4~of \cite{Potton2004} when summarising such reciprocal theorems.
\begin{itemize}\item \small The amplitude of a scattered plane wave propagating along direction $\mathbf{\hat s}_{\boldsymbol\beta}$ with polarization state $ \mathbf {\hat E}_{\boldsymbol\beta}$, due to an incident plane wave propagating along $\mathbf{\hat s}_{\boldsymbol\alpha}$ with polarization $\mathbf {\hat E}_{\boldsymbol\alpha}$.
\end{itemize}~\\[-4.5ex]
 {\sl \quad Equal to:}\\[-4.5ex]
\begin{itemize}
\item \small The amplitude a scattered plane wave propagating along $-\mathbf{\hat s}_{\boldsymbol\alpha}$ with polarization $\mathbf {\hat E}_{\boldsymbol\alpha}$, due to incident plane wave propagating along $-\mathbf{\hat s}_{\boldsymbol\beta}$ with polarization $\mathbf {\hat E}_{\boldsymbol\beta}$.
\end{itemize}
However, the models for scattering used in this thesis align instead with volumetric field distributions, so I instead focus entirely on the {\it Lorentz Reciprocity Theorem} seen in (\ref{eq:LorentzReciprocity}), a derivation of which can be found in \S89 of~\cite{LandauLifshitzVol8}.    
\begin{align}
\int_V \mathbf E_{\boldsymbol \alpha} \cdot \mathbf J _{\boldsymbol \beta} - \mathbf E_{\boldsymbol \beta} \cdot \mathbf J _{\boldsymbol \alpha} \mathrm{d}V = \int_{\Omega_V} \mathbf E_{\boldsymbol \alpha} \times \mathbf H_{\boldsymbol \beta} - \mathbf E_{\boldsymbol \beta} \times \mathbf H_{\boldsymbol \alpha} \mathrm{d}\Omega
\label{eq:LorentzReciprocity}
\end{align}
This defines a relationship between any two solutions of Maxwell's Equations denoted by subscript $\alpha$, $\beta$: the fields $\mathbf E_{\boldsymbol \alpha},\mathbf H_{\boldsymbol \alpha}$ and $\mathbf{E}_{\boldsymbol \beta}$,$\mathbf{H}_{\boldsymbol \beta}$, with corresponding current distributions $\mathbf J_{\boldsymbol \alpha}$ and $\boldsymbol{J_{\beta}}$.
For compact source currents with finite energy, the surface integral on the right hand side of this equation will go to zero whenever we can assume one of two conditions.
In the case of a background with any dissipative loss \mbox{$\mathrm {Im}\{ \epsilon_0, \mu_0\} < 0$}, one can simply take the volume $V$ as infinite, such that any fields at the boundary $\Omega_V$ have propagated infinitely far from their finite energy sources, and hence the fields are zero.  
Alternatively, we can assume a homogeneous background exists at an infinite distance from our compact sources with finite energy, which means the only fields on the boundary of an infinite spherical volume $V$ will be both plane waves, and propagating radially outwards (along $\mathbf{\hat r}$).
As such, on the boundary $\Omega_V$, we can substitute $\mathbf H$ as per a plane wave $\mathbf E = \sqrt{\frac{\mu_0}{\epsilon_0}}\mathbf{\hat r}\times \mathbf H$, when using the scalar triple product relation $\mathbf a\cdot (\mathbf b\times \mathbf c) = \mathbf b  \cdot (\mathbf a \times \mathbf c) $.  
This then equates the surface integral to zero:
\begin{align}
\int_{\Omega_V} \mathbf E_{\boldsymbol \alpha} \times \mathbf H_{\boldsymbol \beta} - \mathbf E_{\boldsymbol \beta} \times \mathbf H_{\boldsymbol \alpha} \mathrm{d}\Omega &= \int_{\Omega_V} \mathbf{\hat r} \cdot (\mathbf E_{\boldsymbol \alpha} \times \mathbf H_{\boldsymbol \beta} - \mathbf E_{\boldsymbol \beta} \times \mathbf H_{\boldsymbol \alpha}) \mathrm{d}\Omega \nonumber 
\\
&  = \int_{\Omega_V} \sqrt{\frac{\epsilon_0}{\mu_0}} (\mathbf E_{\boldsymbol \alpha} \cdot \mathbf E_{\boldsymbol \beta} - \mathbf E_{\boldsymbol \beta} \cdot \mathbf E_{\boldsymbol \alpha}) \mathrm{d}\Omega \;=\;  0
\end{align}
The two assumptions we just considered will generally account for all radiative systems of interest, hence we can reliably use a simpler reciprocity equation without the surface integral:
\begin{align}
\int_{V} \mathbf E_{\boldsymbol \alpha} \cdot \mathbf J _{\boldsymbol \beta} - \mathbf E_{\boldsymbol \beta} \cdot \mathbf J _{\boldsymbol \alpha} \mathrm{d}V = 0\label{eq:LorentzReciprocityWith0}
\end{align}

To reach the desired conclusion, being the equal extinction experienced by applied fields $\mathbf{E_0}$ and $\mathbf{E_0}^*$, let  us define $\mathbf J_{\boldsymbol \alpha}$ as the currents induced by some $\mathbf {E_0}$, and  $\mathbf J_{\boldsymbol \beta}$ as the currents induced by $\mathbf {E_0}^*$.  
We can then consider four solutions to Maxwell's Equations grouped into two sets: 
\begin{itemize}
\item[(i)] $\mathbf {E_0}$ and $\mathbf {E_0}^*$ are imposed separately, and respectively induce currents $\mathbf J_{\boldsymbol \alpha}$, $\mathbf J_{\boldsymbol \beta}$, with total fields  $\mathbf E_{\boldsymbol \alpha}$, $\mathbf E_{\boldsymbol \beta}$.
\item[(ii)] The two currents $\mathbf J_{\boldsymbol \alpha}$ and $\mathbf J_{\boldsymbol \beta}$ are imposed separately in the background medium of (i), respectively producing scattered fields $\mathbf E_{\boldsymbol \alpha}-\mathbf {E_0}$ and $\mathbf E_{\boldsymbol \beta}-\mathbf {E_0}^*$.
\end{itemize}
Both (i) and (ii) share the same current distributions $\mathbf J_{\boldsymbol \alpha}$ and  $\mathbf J_{\boldsymbol \beta}$, and we can therefore take the difference of (\ref{eq:LorentzReciprocityWith0}) due to (i) and (ii), which gives the result we are interested in:
\begin{align}
\int_{V} \mathbf{E_0} \cdot \mathbf J _{\boldsymbol \beta} \; \mathrm{d}V  = \int_{V} \mathbf{E_0}^*\! \cdot \mathbf J _{\boldsymbol \alpha}\; \mathrm{d}V \label{eq:equalLosses}
\end{align}
By referring to the definition of extinction in (\ref{eq:extinction definition}), and remembering that $\mathbf J_{\boldsymbol \alpha}$ is induced by $\mathbf {E_0}$ and $\mathbf J_{\boldsymbol \beta}$ is induced by $\mathbf {E_0}^*$, the conclusion in (\ref{eq:equalLosses}) states that $\mathbf {E_0}$ and  $\mathbf {E_0}^*$ experience precisely the same power loss, as desired.
However, before concluding that reciprocity is therefore responsible for this conclusion, we must recognise that the derivation of the Lorentz Reciprocity Theorem (\ref{eq:LorentzReciprocity}) itself assumes permittivity $\boldsymbol{\bbar \epsilon}$, and permeability $\boldsymbol{\bbar \mu}$, are symmetric matrices.
The symmetry of these matrices is generally considered to be a consequence of {\it Onsager's Reciprocity}~\cite{Onsager1930, Onsager1931}, which in turn was derived from the existence of {\it microscopic reversibility}: the equal probability of any two time-separated random fluctuation events occurring in the opposite order for a system in thermal equilibrium.
More specifically, Onsager showed that, when given a system in thermal equilibrium, a matrix $\boldsymbol{\bbar{\mathcal M}}$ that relates the time evolution of any set of time-averaged displacement quantities $q$ to the rate of change with $q$ of the maximum system entropy $S$ at thermal equilibrium, will be symmetric.  
That is: \mbox{$\frac{\mathrm d q_i}{\mathrm d t} = \sum_j \mathcal M_{ij} \frac{\partial S}{\partial q_j} $}, and $\mathcal M_{ij}=\mathcal M_{ji}$. 
The relevant application to electric displacement $\mathbf D$ seems to arise by recognising ${\partial S}/{\partial D_j}$ acts as a force on $\mathbf D$ and is thereby proportional to the local electric field, hence the reciprocal relation enforces symmetry on permittivity defined through $\mathbf D = \boldsymbol{\bbar \epsilon} \mathbf{E}$.
However, the later application of the {\it Fluctuation Dissipation Theorem}~\cite{Callen1951} to generalised susceptibilities, being frequency domain tensors that relate physical quantities to the generalized forces they experience, gets the same matrix symmetry requirements more directly, as discussed in \S125 of~\cite{LandauLifshitzVol5}, or indeed in \S96 of~\cite{LandauLifshitzVol8} for specifically permittivity and permeability. 

I will now define time-reversal symmetry in electromagnetism, and provide a generic argument as to why time-reversal symmetry implies microscopic reversibility.
The reversal of time transformation $t\rightarrow -t$ will leave Maxwell's Equations in (\ref{eq:divB})-(\ref{eq:curlE}) unchanged, provided the current and magnetic field are time-antisymmetric, {\it i.e.}  $\boldsymbol E(t) \rightarrow \boldsymbol E(-t)$, $\boldsymbol B(t) \rightarrow -\boldsymbol B(-t)$, $\rho(t)\rightarrow \rho(-t)$, $\boldsymbol J(t)\rightarrow-\boldsymbol J(-t)$.
From a physical perspective, $\boldsymbol J\rightarrow -\boldsymbol J$ is needed to preserve the continuity of charge equation $\boldsymbol{\nabla}\cdot \boldsymbol J = -\frac{\partial \rho}{\partial t}$, and $\boldsymbol{B}\rightarrow -\boldsymbol B$  is needed to preserve the Lorentz force (\ref{eq:force})\footnote{
Given the magnetic field can be attributed to relative velocity of different inertial reference frames, see \S13~\cite{FeynmanII}, and that this velocity changes sign if $t\rightarrow -t$, so the time-antisymmetry of $\boldsymbol B$ should be expected.
}. 
From the ability to reverse time without changing Maxwell's Equations, we have an implicit freedom  in any electromagnetic system to define the causal progression of time as either $t$ increasing or decreasing. 
When now presented with a steady state solution to Maxwell's Equations that is distributed over all time, this solution must simultaneously be valid for both directions of causal time.  
As such, the evolution of this system from its state at time 0 to $t_0$ is identical to and that describing evolution from the same state at time 0 to  $-t_0$.
The conditions that specify a probability for any random fluctuation of fields or currents to occur in this steady state system at $t_0$ must therefore be precisely the same as at time $-t_0$.    
Given that the definition of time 0 is actually arbitrary in a steady state solution, our previous argument starting from a time origin of $t_0$ applies to the probability of random fluctuations at time 0 and time $2t_0$.  
This is the microscopic reversibility required by Onsager's arguments, and it therefore suggests that time-reversal symmetry is necessary for Lorentz reciprocity, and our subsequent conclusion that the applied field distributions $\mathbf{E_0}$ and $\mathbf{E_0}^*$ experience equal extinction.  
To go even further, in the box I show that the equal extinction of $\mathbf{E_0}$ and $\mathbf{E_0}^*$ can even follow directly from time-reversal, without introducing permittivity or permeability, let alone their symmetry.  
This is significant given the earlier review by Potton~\cite{Potton2004}, because it was able to conclude that reciprocity is distinct from time-reversal symmetry. 
I will continue to refer to the equal extinction of $\mathbf{E_0}$ and $\mathbf{E_0}^*$ as being due to reciprocity, not time-reversal symmetry, in recognition that I needed to assume linearity and include the Lorentz force experienced by charges, which is not fundamentally different to the assumptions used in the derivation of the Lorentz Reciprocity Theorem.

\begin{myfloat}[H]
\begin{mdframed}[style=BenFrame]
\small
{\bf Derivation from time-reversal.}  
Here I will derive two statements that collectively show that $\mathbf {E_0}$ and $\mathbf {E_0}^*$ experience the same extinction.
\begin{itemize}
\item[(i)] 
The difference in extinction experienced by $\mathbf {E_0}$ and $\mathbf {E_0}^*$ is equal to the difference in causal extinction experienced by a single $\mathbf {E_0}$, but between the two directions of causal time. 
\item[(ii)]
Any applied field distribution $\mathbf {E_0}$, when imposed on a given physical system, must experience the same causal extinction for both directions of causal time evolution.  
\end{itemize}

\quad {\sl Derivation of (i)}.  By restricting ourselves to systems whose optical response depends only on electric currents and fields, time-reversal will act as: $\boldsymbol E(t) \rightarrow \boldsymbol E(-t)$, $\boldsymbol J(t)\rightarrow-\boldsymbol J(-t)$.
We again set fields and currents at times proceeding the current time $t_0$ as zero, and the corresponding casual time interval also changes under time-reversal $(-\infty, t_0]\rightarrow [-t_0,\infty)$.  
The Fourier transformation (\ref{eq:Fourier}) of the time-reversed electric field $\boldsymbol E'(t)=\boldsymbol E(-t)$ then defines the phasors $\mathbf E'(\omega)$ of the time-reversed system:\\[-4ex]
\begin{align}
\mathbf{E}'(\mathbf r, \omega) 
= \int\limits_{-t_0}^{\infty} \boldsymbol{E}'(\mathbf r, t ) e^{ i \omega t} \mathrm d t  
= \int\limits_{-\infty}^{t_0} \boldsymbol{E}(\mathbf r, t ) e^{- i \omega t} \mathrm d t 
= \mathbf E^*(\mathbf r, \omega)  \label{eq:Fourier2}
\end{align}\\[-3ex]
The same argument shows $\mathbf J'(\omega) = -\mathbf J^*(\omega)$ using $\boldsymbol J'(t)=-\boldsymbol J(-t)$.
In other words, $(\mathbf E^*,-\mathbf J^*)$ is a solution to Maxwell's Equations whenever $(\mathbf E,\mathbf J)$ is a solution, noting the time dependence of these phasors is uniformly $e^{-i \omega t}$ and they are distributed over all time ($-\infty,\infty$).  
Now we consider the application of this to three cases: \\[-3ex]
\begin{itemize}[leftmargin=20pt]
\item{$\mathbf {E_0}$ is the phasor of a solution to Maxwell's Equations (\ref{eq:divB})-(\ref{eq:curlB}) in the absence of charge and currents, so $\mathbf {E_0}^*$ is also a solution.}\\[-3ex]
\item{ $\mathbf {E_0}$ is imposed over a scattering geometry and $(\mathbf E,\mathbf J)$ is the solution of (\ref{eq:current equation}), so $(\mathbf E^*,-\mathbf J^*)$ is also a solution.}\\[-3ex]
\item{ $\mathbf J$ is imposed in the background medium and $(\mathbf E_s,\mathbf J)$ is the solution of (\ref{eq:scatt}), so $(\mathbf{E_s}^*,-\mathbf J^*)$ is also a solution.}\\[-3ex]
\end{itemize}
Assuming a linear relation between currents and fields, we can construct the current induced by imposing $\mathbf{E_0}^*$ over the scattering geometry as the being the currents associated with \mbox{$\mathbf E^*  - \mathbf{E_s}^*\!$}.  
In other words, the external field $\mathbf {E_0}^*$ will induce $-\mathbf{J}^*$. 
The xtinction (\ref{eq:Pext}) of $\mathbf {E_0}^*$ from $-\mathbf{J}^*$ is the negative to that of $\mathbf {E_0}$ from $\mathbf {J}$: the energy lost per unit time changes sign, as largely expected given we reversed causal time. 
Conversely, the {\it causal extinction} defined as energy lost per unit causal time remains unchanged.  
The statement (i) then follows.\\[-1ex]

\quad {\sl Derivation of (ii)}. The current model (\ref{eq:current equation}) describes the coupling between two solutions to Maxwell's Equations: system A as $\mathbf{E_0}$ in the background medium, and system B as a static distribution of charge $q$ in the background medium (as the origin of currents).
System A, by its definition as the fixed field distribution $\mathbf {E_0}e^{-i \omega t}$, does not depend on the choice of causal time.  
The optical properties of system B are governed by the forces $\boldsymbol F_q$ experienced by the charges $q$, which also don't depend on the direction of causal time: using the definition of time-reversal, $\boldsymbol E(t) \rightarrow \boldsymbol E(-t)$, $\boldsymbol B(t) \rightarrow -\boldsymbol B(-t)$, $\rho(t)\rightarrow \rho(-t)$, $\boldsymbol J(t)\rightarrow-\boldsymbol J(-t)$, it follows that $\boldsymbol F_q(t) \rightarrow \boldsymbol F_q(-t) $ in (\ref{eq:force}).   
We now recognise that the extinction of $\mathbf{E_0}$, as in (\ref{eq:Pext}), is the time-averaged rate of energy transfer from system A into system B.  
As neither system A nor system B depend on the direction of causal time in isolation, a difference in energy transferred from system A to system B, which depends only on the choice of causal time, implies a preferred direction of overall time evolution. 
We know such a preference does not exist from our capacity to perform an inconsequential time-reversal transformation on Maxwell's Equations.  
The statement (ii) then follows.

\end{mdframed}
\end{myfloat}

\newpage
\section{Discrete rotational symmetry \label{sec:rotSym}}

Here I use the eigenmode model of Chapter~\ref{chapterEigen} to derive the result presented in [\ref{Nanoscale}]: \mbox{$C_n$ ($n\!\geq\!3$)} discrete rotational symmetry of an object forces the total power it scatters (\ref{eq:Pscat}) and absorbs (\ref{eq:Pabs}) to be independent from the angle of linear polarisation for plane waves propagating parallel to the $n$-fold rotation axis.
This result is interesting because both the far-field scattering pattern and the near-field distribution remain highly dependent on the incident plane wave's polarisation angle, even while the total scattering and total absorption remain independent.
The reason to return to this result is to retrospectively understand what symmetry has actually done to the individual resonances causing this effect.
Of the specific relevance is equations (6) to (12) of [\ref{LPOR}], where it was derived that, because the permittivity $\boldsymbol{\bbar \epsilon}$ is a symmetric matrix required by Onsager's reciprocity, {degeneracy is enforced between pairs of eigenmodes belonging to irreducible representations that are complex conjugates of each other}.
This result then affects geometries whose symmetry is invariant under the operations of the $C_n$ ($n\!\geq\!3$) group, such as the $E$ irreducible representation for $C_3$ in Table~\ref{table:C3}, for which rotations are equated to phase shifts.\footnote{Note that the degeneracy of left- and right-circulating eigenmodes, belonging to the $E_1$ irreducible representation of $C_n$ ($n\!\geq\!3$), has some similarities with a preliminary example considered by Onsager in {\small (4.1)}-{\small (4.11)} of~\cite{Onsager1930}.
There, for hexagonal and tetragonal crystals with $C_3$ and $C_4$ symmetries, Onsager derived the enforced equivalence of clockwise and counter-clockwise heat currents due to microscopic reversibility (defined in Section~\ref{sec:tSym}).  }
We will show here that this same reciprocity induced degeneracy derived in [\ref{LPOR}] implies the polarisation-independent scattering and absorption cross-sections derived in [\ref{Nanoscale}].
\begin{table}[!h]
\centerline{
{\large $\boldsymbol{C_3}$:\quad}
\begin{tabular}{|c|c|c|c||c|}
\hline & & & & \\[-2ex]
& \large$\hat E$ &\large $\hat C_3$ &\large $(\hat C_3)^2$ & {\small examples for} $[x,y,z]$\\
\hline & & & & \\[-2ex]
\large $A$ & 1 & 1& 1& $z$\\[1ex]
\large $E$ & $\bigg\{\begin{array}{c}1\\ 1\end{array}$ & $\begin{array}{c}\phi \\ \phi ^*\end{array}$ & $\begin{array}{c}\phi ^*\\ \phi \end{array}$ & $\begin{array}{c}x + i y\\ x - i y\end{array}$\\[1.5ex]\hline
\end{tabular}\qquad\quad}
\caption{Character table for the $C_3$ symmetry group.  Rows denote irreducible representations, columns denote symmetry operations ($\hat E$ is the identity, $\hat C_3$ is a rotation by $\frac{2\pi}{3}$), indices are the character of the corresponding matrix representation for the symmetry operation ($\phi = e^{\frac{i 2 \pi}{3}}$)}.
\label{table:C3}
\end{table}

To begin, we reiterate that the $E_1$ irreducible representation of $C_n$ can be separated into two, one-dimensional irreducible representations that are complex conjugates of each other, which we will denote as $E_{\raisebox{0.1ex}{\text{\tiny+\hspace{-0.1ex}}}1}$ and $E_{\raisebox{0.1ex}{\text{\,\scriptsize-}}1}$.
Using the aforementioned result from [\ref{LPOR}] that the eigenmodes of $E_1$ will come in degenerate pairs between $E_{\raisebox{0.1ex}{\text{\tiny+\hspace{-0.1ex}}}1}$ and $E_{\raisebox{0.1ex}{\text{\,\scriptsize-}}1}$  due to reciprocity, we can treat $E_1$ as a single two-dimensional irreducible representation.  
Moreover, we consider any excitation field $  \boldsymbol{\scriptstyle\mathcal{E}}_{\Gamma}$ that transforms according to a single $\Gamma = E_1$ (\ref{eq:eigensymmetry}), and write out its corresponding eigenmode decomposition.
\begin{align}
  \boldsymbol{\scriptstyle\mathcal{E}}_{\Gamma}  &= \sum \limits _v b_v \lambda_v {\mathbf{j}_{{\scriptscriptstyle \Gamma},v}}  \quad \Rightarrow \quad  
 \hat{C}_n  \boldsymbol{\scriptstyle\mathcal{E}}_{\Gamma}  = \sum \limits _v b_v \lambda_v [\hat{C}_n {\mathbf{j}_{{\scriptscriptstyle \Gamma},v}} ]
 \label{eq:CnUse1}
\end{align}
Here $\hat{C}_n$ denotes the symmetry operation in $C_n$ for rotation by $\frac{2\pi }{n}$.
The symmetry rotated eigenmode $\hat{C}_n {\mathbf{j}_{{\scriptscriptstyle \Gamma},v}} $ is also generally able to be linearly independent to ${\mathbf{j}_{{\scriptscriptstyle \Gamma},v}} $, given ${\mathbf{j}_{{\scriptscriptstyle \Gamma},v}} $ can contain both $E_{\raisebox{0.1ex}{\text{\tiny+\hspace{-0.1ex}}}1}$ and $E_{\raisebox{0.1ex}{\text{\,\scriptsize-}}1}$ transformation properties. 
A plane wave propagating parallel to the principal axis of $C_n$ transforms according to $E_1$, given LCP and RCP respectively transforming according to $E_{\raisebox{0.1ex}{\text{\,\scriptsize-}}1}$ and $E_{\raisebox{0.1ex}{\text{\tiny+\hspace{-0.1ex}}}1}$.
These plane waves are therefore encompassed by our consideration of a generic $\boldsymbol{\scriptstyle\mathcal{E}}_{\Gamma}$, and we now simply need to derive the consequences of $E_1$ eigenmode degeneracy on the total scattering (\ref{eq:Pscat}) and total absorption (\ref{eq:Pabs}) of $\boldsymbol{\scriptstyle\mathcal{E}}_{\Gamma}$.  
Moreover, for each ${\mathbf{j}_{{\scriptscriptstyle \Gamma},v}} $, we can define a new degenerate eigenmode ${\mathbf{j}^{(\theta)}_{{\scriptscriptstyle \Gamma},v}} $, which is rotated by an angle $\theta$.
\begin{align}
\mathbf{j}^{(\theta)}_{{\scriptscriptstyle \Gamma},v} = (\cos \theta - \sin \theta \cot {\scriptstyle \frac{2\pi}{n}})\mathbf{j}_{{\scriptscriptstyle \Gamma},v} 
+\sin\theta \csc {\scriptstyle \frac{2\pi}{n}} [\hat{C}_n {\mathbf{j}_{{\scriptscriptstyle \Gamma},v}} ]
 \label{eq:CnUse2}
\end{align}
Here ‘rotation’ is defined relative to $\mathbf{j}_{{\scriptscriptstyle \Gamma},v} $ by treating $\hat{C}_n {\mathbf{j}_{{\scriptscriptstyle \Gamma},v}} $ as a second basis vector oriented with a rotated angle of  $2\pi/n$.
We can also define a new excitation field  $ \boldsymbol{\scriptstyle\mathcal{E}}^{(\theta)}_{\Gamma}$ that is rotated by an angle $\theta$ relative to  $ \boldsymbol{\scriptstyle\mathcal{E}}_{\Gamma}$ in (\ref{eq:CnUse1}), with corresponding induced currents  $   \boldsymbol{\scriptstyle\mathcal{J}}^{(\theta)}_{\Gamma}$.
\begin{align}
 \boldsymbol{\scriptstyle\mathcal{E}}^{(\theta)}_{\Gamma}  &= \sum \limits _v b_v \lambda_v {\mathbf{j}^{(\theta)}_{{\scriptscriptstyle \Gamma},v}} \, , 
 \quad  \boldsymbol{\scriptstyle\mathcal{J}}^{(\theta)}_{\Gamma}  = \sum \limits _v b_v {\mathbf{j}^{(\theta)}_{{\scriptscriptstyle \Gamma},v}}  
\label{eq:linearcombos}
\end{align}
We next split each $\mathbf{j}_{{\scriptscriptstyle \Gamma},v}$ into terms of its constituent degenerate eigenmode pairs  $\mathbf{j}_{{\scriptscriptstyle +}v}$ and $\mathbf{j}_{{\scriptscriptstyle -}v}$ that transform according to $E_{\raisebox{0.1ex}{\text{\tiny+\hspace{-0.1ex}}}1}$ and $E_{\raisebox{0.1ex}{\text{\,\scriptsize-}}1}$, respectively. 
The $\hat C_n$ operation used in (\ref{eq:CnUse1}) and (\ref{eq:CnUse2}) is then specifically described by scaling  $\mathbf{j}_{{\scriptscriptstyle +}v}$ and $\mathbf{j}_{{\scriptscriptstyle -}v}$ by $e^{+i2\pi⁄n}$ and $e^{-i2\pi⁄n}$.  
\begin{align}
\mathbf{j}_{{\scriptscriptstyle \Gamma},v}=  A_{v} \mathbf{j}_{{\scriptscriptstyle +}v} + B_{v} \mathbf{j}_{{\scriptscriptstyle -}v}
\quad \Rightarrow \quad 
\hat C_n \mathbf{j}_{{\scriptscriptstyle \Gamma},v}=  A_{v} \mathbf{j}_{{\scriptscriptstyle +}v} e^{i2\pi⁄n}+ B_{v} \mathbf{j}_{{\scriptscriptstyle -}v} e^{-i2\pi⁄n}
\label{eq:Cndecomp}
\end{align}
 Here $A$ and $B$ are complex scalars, and (\ref{eq:Cndecomp}) also shows that the arbitrary rotation by $\theta$, defined in (\ref{eq:CnUse2}), is a unitary operation.
\begin{align}
\Big | (\cos \theta - \sin \theta \cot {\scriptstyle \frac{2\pi}{n}}) 
+\sin\theta \csc {\scriptstyle \frac{2\pi}{n}}e^{\pm i 2 \pi /n}\Big|^2 =1
\label{eq:unitaryrotation}
\end{align}
To consider the expressions for scattering (\ref{eq:Pscat}) and absorption (\ref{eq:Pabs}), let us first consider a modified inner product $\beta_{vw}$ between $\mathbf{j}_{{\scriptscriptstyle \pm }v}$ and $\mathbf{\bbar M}\cdot\mathbf{j}_{{\scriptscriptstyle \mp }w}$, where $\mathbf{\bbar M}$ is some matrix.  
\begin{align}
\beta_{vw}\equiv \int [\mathbf{j}_{{\scriptscriptstyle \pm }v} (\mathbf{r'})]^* \! \cdot 
[\mathbf{\bbar M}(\mathbf{r'},\mathbf{r})] \cdot \mathbf{j}_{{\scriptscriptstyle \mp}w}(\mathbf{r})\mathrm{dr'}^3\mathrm{dr}^3 \label{eq:betadef}
\end{align}
We can now utilise $\hat R$ being unitary, and substitute its operation with the corresponding scalar matrix representations $\hat R \mathbf{j}_{{\scriptscriptstyle \pm}v}(\mathbf r)= \mathrm{ D}_{\Gamma{\scriptscriptstyle \! \pm}}^{\!(\hat R)}\;\mathbf{j}_{{\scriptscriptstyle \pm}v}(\mathbf r)$, to write $\beta_{vw}$ in a new form. 
\begin{align}
\beta_{vw} &= \int [\mathbf{j}_{{\scriptscriptstyle \pm }v} (\mathbf{r'})]^* \! \cdot (\hat R^{-1} \hat R)\cdot
[\mathbf{\bbar M}(\mathbf{r'},\mathbf{r})] \cdot (\hat R^{-1} \hat R)\cdot \mathbf{j}_{{\scriptscriptstyle \mp}w}(\mathbf{r})\mathrm{dr'}^3\mathrm{dr}^3 \nonumber \\
&= \int [\mathrm{ D}_{\Gamma{\scriptscriptstyle \! \pm}}^{\!(\hat R)}  \mathbf{j}_{{\scriptscriptstyle \pm}v} (\mathbf{r'})]^*\! \cdot\!(\hat R\cdot [\mathbf{\bbar M}( \mathbf{r'},\mathbf{r})]  \cdot \hat R^{-1}\!)\!\cdot [\mathrm{ D}_{\Gamma{\scriptscriptstyle \! \mp}}^{\!(\hat R)}  \mathbf{j}_{{\scriptscriptstyle \mp}w}(\mathbf{r})]\mathrm{dr'}^3\mathrm{dr}^3 \label{eq:coordTransBeta}
\end{align}
Notably, if $\mathbf{\bbar M}$ satisfies $\mathbf{\bbar M}(\hat R \mathbf r, \hat R \mathbf r') = \hat R \cdot \mathbf{\bbar M} (\mathbf r,\mathbf r')\cdot \hat R^{-1} $, and we commute scalars $\mathrm D_\Gamma$ through $\boldsymbol{\bbar M}$, we can go one step further.  
\begin{align}
\beta_{vw} &= \int [\mathrm{ D}_{\Gamma{\scriptscriptstyle \! \pm}}^{\!(\hat R)}  \mathbf{j}_{{\scriptscriptstyle \pm}v} (\mathbf{r'})]^* \! \cdot \Big [ \mathrm{ D}_{\Gamma{\scriptscriptstyle \! \mp}}^{\!(\hat R)} \, [\mathbf{\bbar M}(\hat R \mathbf r, \hat R \mathbf r')  \cdot  \mathbf{j}_{{\scriptscriptstyle \mp}w}(\mathbf{r})] \Big ] \mathrm{dr'}^3\mathrm{dr}^3  \label{eq:coordTransBeta}
\end{align}
We can now set up an equation analogous to (\ref{eq:provenorthogonality}) by considering the sum of (\ref{eq:coordTransBeta}) over all  symmetry operations $\hat R$ in $C_n$.
\begin{align}
N_{C_n} \,\beta_{vw} = \sum \limits_{\hat R} \beta_{vw} = 0 \;,\;\; \text{using  (\ref{eq:wonderfulorthogonality}) and (\ref{eq:coordTransBeta})} \label{eq:polindepReview}
\end{align}
Here $N_{C_n}$ denotes the number of symmetry operations in $C_n$.
We now note that (\ref{eq:polindepReview}) can only be true if $\beta_{vw}=0$, because there is always the identity operation in $C_n$, and hence $N_{C_n}\geq 1$. 
Comparing (\ref{eq:betadef}) to (\ref{eq:polindepReview}) provides the following statement:
\begin{align}
 \int [\mathbf{j}_{{\scriptscriptstyle \pm}v} (\mathbf{r'})]^* \! \cdot 
[\mathbf{\bbar M}(\mathbf{r'},\mathbf{r})] \cdot \mathbf{j}_{{\scriptscriptstyle \mp}w}(\mathbf{r})\mathrm{dr'}^3\mathrm{dr}^3 = 0 \label{eq:symmImportant}
\end{align}
We are now able to put everything together.  
It follows from (\ref{eq:symmImportant}), with  (\ref{eq:CnUse2}),  (\ref{eq:Cndecomp}) and (\ref{eq:unitaryrotation}), that our modified inner products between eigenmodes are conserved under rotations  (\ref{eq:CnUse2}) by $\theta$. 
\begin{align}
\int [\mathbf{j}^{(\theta)}_{{\scriptscriptstyle \Gamma},v} (\mathbf{r'})]^* \! \cdot [\mathbf{\bbar M}(\mathbf{r'},\mathbf{r})]  \cdot \mathbf{j}^{(\theta)}_{{\scriptscriptstyle \Gamma},w} (\mathbf{r})\mathrm{dr'}^3\mathrm{dr}^3 
=
\int [\mathbf{j}_{{\scriptscriptstyle \Gamma},v} (\mathbf{r'})]^* \! \cdot [\mathbf{\bbar M}(\mathbf{r'},\mathbf{r})]  \cdot \mathbf{j}_{{\scriptscriptstyle \Gamma},w} (\mathbf{r})\mathrm{dr'}^3\mathrm{dr}^3 
\label{eq:polindepPart1}
\end{align}
By combining (\ref{eq:linearcombos}) and (\ref{eq:polindepPart1}),  we reach a similar conclusion for $\boldsymbol{\scriptstyle\mathcal{J}}_\Gamma$ under rotations \mbox{by $\theta$.}
\begin{align}
\int [\boldsymbol{\scriptstyle\mathcal{J}}^{(\theta)}_{\Gamma} (\mathbf{r'})]^* \! \cdot [\mathbf{\bbar M}(\mathbf{r'},\mathbf{r})]  \cdot \boldsymbol{\scriptstyle\mathcal{J}}^{(\theta)}_{\Gamma} (\mathbf{r})\mathrm{dr'}^3\mathrm{dr}^3 
=
 \int [\boldsymbol{\scriptstyle\mathcal J}_{\Gamma} (\mathbf{r'})]^* \! \cdot [\mathbf{\bbar M}(\mathbf{r'},\mathbf{r})]  \cdot \boldsymbol{\scriptstyle\mathcal J}_{\Gamma} (\mathbf{r})\mathrm{dr'}^3\mathrm{dr}^3
\label{eq:polindepPart2}
\end{align}
Two important choices of $\mathbf{\bbar M}$ are $\big( \mathbf {\bbar{G}_{0}}(\mathbf{r},\mathbf{r'}) - \mathbf{\bbar{\,L}}  \frac{\delta (\mathbf{r}-\mathbf{r'})}{k^2 }  \big)$  in (\ref{eq:Pscat})  and $\mathrm{Im} \{  (\boldsymbol{\bbar \epsilon}- \epsilon_0)^{-1}\}$ in (\ref{eq:Pabs}), for which (\ref{eq:polindepPart2}) shows that  $P_{scat}$ and $P_{abs}$, are invariant to the rotation angle $\theta$. 
This thereby completes a derivation of the result in [\ref{Nanoscale}] starting from the eigenmode degeneracy due to reciprocity presented in [\ref{LPOR}].
The conclusion is also slightly more general, given that any external field distribution $\mathbf{E_0} =  \boldsymbol{\scriptstyle\mathcal{E}}_{\Gamma}$ that transforms entirely according to the irreducible representation $\Gamma=E_1$, will be scattered and absorbed by exactly the same amount irrespective to any rotation in orientation of $\mathbf{E_0}$ about the principal axis of $C_n$. 
This result encompasses,  but is not necessarily restricted to, normally incident plane waves.  
Additionally, the argument we just presented also applies to $S_4$ symmetry groups owing to its analogous complex conjugate irreducible representations, noting that $S_4$ does not contain a $C_n$ ($n\!\geq\!3$) subgroup.

\chapter{Asymmetry and chirality  \label{chapterChiral}}

This will be the converse to the previous Chapter on symmetry, being a discussion on the absence of symmetry.  
I first define chirality as a geometric concept, then provide some basic arguments as to why this concept can have counterparts in optical scattering, and their relation to both chiral and achiral scattering objects.  
In the second section, I focus specifically on the differences in scattering from oppositely handed, circular-polarised plane waves, particularly the known effects of {\it circular dichroism} and {\it circular conversion dichroism}, and the specific geometric symmetries that {suppress} them. 
I then discuss {\it circular dichroism in absorption}, first discussed in [\ref{LPOR}], as a further distinct effect that originates from differing optical responses between reciprocal plane waves.

\section{Defining chirality in optics}

A physical object is said to be {\it chiral} if it cannot be arranged through any number of rotations and translations into coincidence with its mirror image, implying the given object is a different object to its mirror image. 
This is a definition of chirality, and we can use it to also define left and right {\it enantiomers} as the distinct pair of mirror images for a single chiral object.  
We now relate chirality to symmetry.
Any mirror image operation applied to a given object can be described by two consecutive operations acting on the coordinate system: a reflection plane $\hat \sigma$ as the mirror, and an arbitrary rotation $\hat C_{\boldsymbol \theta}$ to account for the orientation of the mirror. 
These two operations acting on coordinates $(x,y,z)$ can next be written as transformation matrices, $\mathbf{\bbar D}^{(\hat \sigma)}$ and $\mathbf{\bbar D}^{(\hat C_{\boldsymbol\theta})}$, in (\ref{eq:Dsigma}) and (\ref{eq:Dtheta}), which notably commute with each other: $\mathbf{\bbar D}^{(\hat C_{\boldsymbol \theta})} \mathbf{\bbar D}^{(\hat \sigma)}  = \mathbf{\bbar D}^{(\hat \sigma)} \mathbf{\bbar D}^{(\hat C_{\boldsymbol\theta})}\!\!$.
This is seen by writing $\mathbf{\bbar D}^{(\hat C_{\boldsymbol\theta})}\!\!$ as the product of three rotations about each axis, as in  (\ref{eq:Dtheta}), then directly confirming the commutivity for each of these rotation matrix with $\mathbf{\bbar D}^{(\hat \sigma)}$\!.
\begin{gather}
\mathbf{\bbar D}^{(\hat \sigma)}  =\left(\begin{array}{ccc}
-1				&				0				&		 		0					\\			
0				&				1				&				0			 		\\
0				&				0				&				1	
\end{array}\right) \label{eq:Dsigma}\\
\mathbf{\bbar D}^{(\hat C_{\boldsymbol\theta})}  =
\left(\begin{array}{ccc}
1 				&				0 				&		 		0					\\			
0				&				\cos\theta_x &				\sin\theta_x 		\\
0				&				-\sin\theta_x &				\cos\theta_x 	
\end{array}\right) \!\!
\left(\begin{array}{ccc}
\cos\theta_y	&				0 				&		 		-\sin\theta_y		\\			
0				&				1				 &				 0					\\
\sin\theta_y	&				0				 &				\cos\theta_y 	
\end{array}\right) \!\!
\left(\begin{array}{ccc}
\cos\theta_z	&				\sin\theta_z	&		 		0					\\			
-\sin\theta_z	&				\cos\theta_z &				0			 		\\
0				&				0				&				1	
\end{array}\right) \label{eq:Dtheta}
\end{gather}
Given our initial definition of chirality permits any arbitrary rotation after performing the mirror image, where we have just  seen that arbitrary rotation commutes with reflection, the orientation of the reflection plane isn't important when defining the two enantiomers of a chiral object.  
Conversely, it means an object that is symmetric about {\sl any} reflection plane will always be the same object as its mirror image, meaning it is not chiral, or rather: it is {\it achiral}.  
Similarly, point inversion, which can be written as the product of three perpendicular mirror planes and correspondingly understood as switching enantiomer handedness three times (an odd number), will also imply achirality.  
By confining ourselves to known geometric symmetry operations, we can now define chirality as being equivalent to the absence of any planes of reflection symmetry $\hat \sigma$, points of inversion symmetry $\hat{i}$, or axes of improper rotation symmetry $\hat S_n$ (rotation $\hat C_n$ and perpendicular reflection $\hat \sigma_h$).  
See the example geometries exhibiting each symmetry in Figure~\ref{fig:localChiralFields}a.

We now consider the chirality of electromagnetic fields occupying some given volume $V$, for which I will draw a distinction between {\it local chirality} ($V \rightarrow 0$) and {\it global chirality} ($V \not \rightarrow 0$).
In a spherical volume $V\rightarrow 0$, the local fields $\boldsymbol E(t)$ and $\boldsymbol B(t)$ at the sphere origin will be chiral if there is a component of $\boldsymbol E$ parallel to $\boldsymbol B$.
That is to say: fields with $\boldsymbol E \cdot \boldsymbol B \neq 0$ have no planes of reflection, axes of improper rotation, or points of inversion symmetry.
This is a consequence of $\boldsymbol E$ having opposite parity to $\boldsymbol B$ under reflection and inversion operations, which is imposed by the Maxwell curl equation~(\ref{eq:curlE}).  
The absence of symmetry is therefore illustrated in Figure~\ref{fig:localChiralFields}b by depicting $\boldsymbol B$ as the corresponding circulation of $\boldsymbol E$.
\begin{figure}[!ht]
\centerline{\includegraphics[width=\textwidth]{{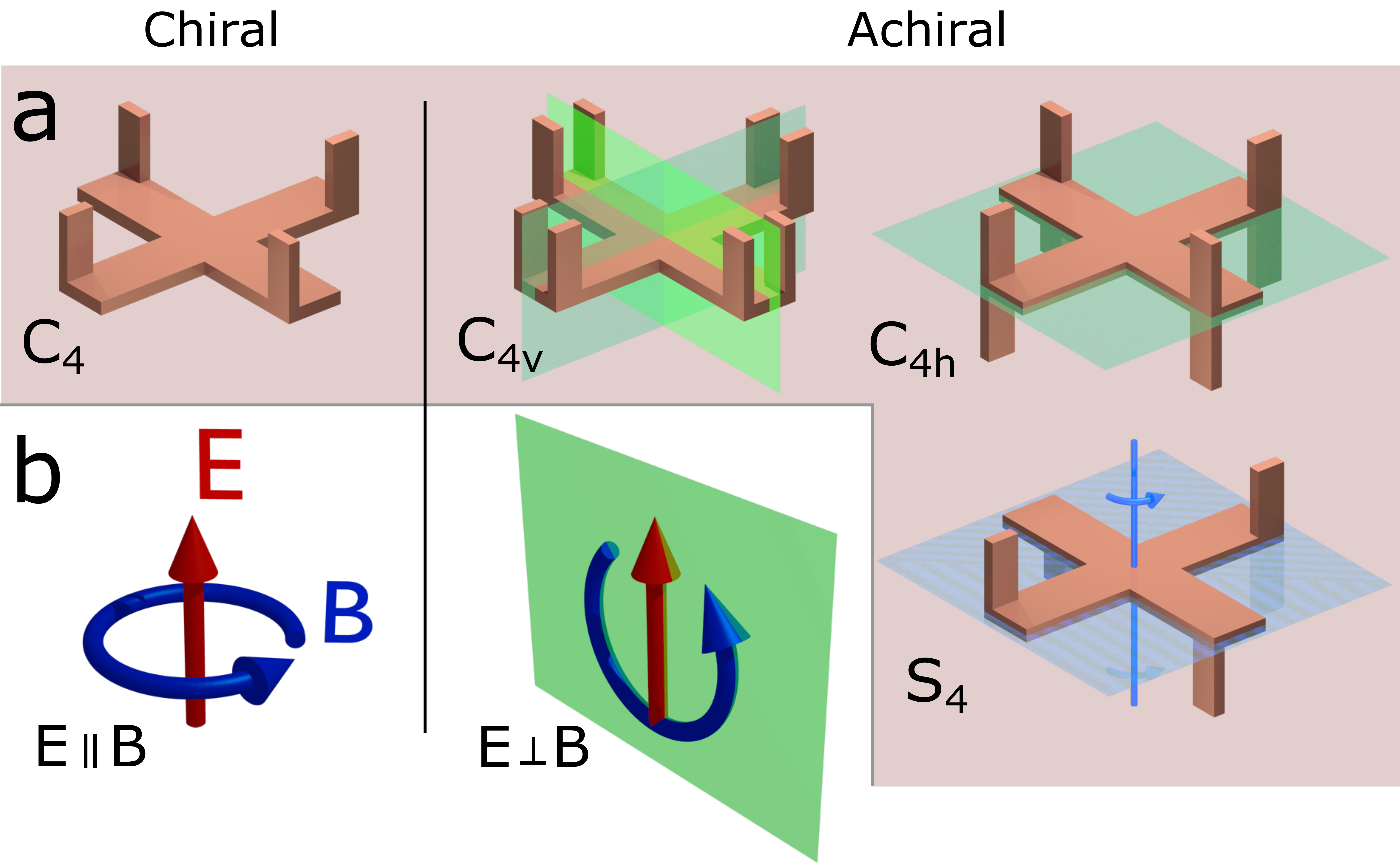}}}
\caption{(a) Examples of chiral ($C_{4}$) and achiral ($C_{4v}$,$C_{4h}$,$S_4$) geometric symmetries, and (b) chiral ($\boldsymbol E \parallel \boldsymbol B$) and achiral ($\boldsymbol E \perp \boldsymbol B$) local electromagnetic fields.  Achiral symmetry operations are marked by reflection symmetry planes ($\hat \sigma$:~green) or improper rotation axes ($\hat S_n = \hat C_n \hat \sigma_h\!\!:$~blue).  
The $\boldsymbol B$-field is depicted in this figure by the corresponding circulation of the $\boldsymbol E$-field, given $\boldsymbol B$ can be defined using a curl in (\ref{eq:curlE}).  This means the orientation of the $\boldsymbol B$ vector is normal to the plane of circulation depicted here.  }
\label{fig:localChiralFields}
\end{figure}
Chirality for local fields now obtains physical significance when considering the interaction between fields and objects that are sufficiently small to be considered as points.    
Moreover, if the local fields are {\it achiral}, the converse statement is that there is some local refection, inversion or improper rotation transformation on the coordinates of $V$ that leave these achiral local fields unchanged.  
If there is a small chiral point particle or molecule in $V$, the achiral coordinate transformation that leaves the fields unchanged would necessarily transform the small chiral point particle into its opposite enantiomer.
This means that any interaction between achiral local fields and different enantiomers of a chiral molecule is equivalent to a coordinate transformation, which we can define as being a trivial difference.  
Chiral local fields are therefore necessary to allow nontrivial differences in interaction, one key example of this being the total electromagnetic power absorbed by the molecule.  
Local field chirality is therefore a relevant concept for molecular sensing or manipulation that can distinguish between enantiomers~\cite{Barron,Hembury2008}.
However, to be more explicit, such investigations have sought to introduce a bias in the signed magnitude of local {\it helicity density} $\mathrm{Im}\{\mathbf{E}^*\!\cdot\mathbf{H}\}$, which has been show to create a net difference in both the absorption and radiation of electromagnetic fields by oppositely handed chiral molecules~\cite{Tang2010, Yoo2015}.

For nanoparticle oligomers, the relevant definition of optical chirality  also relates to the geometric concept of chirality.  However, owing to the non-negligible size of oligomers and nanostructures relative to the considered wavelength, we must attribute optical chirality in nanoparticle oligomers to field distributions in volumes $V\not \rightarrow 0$.  
This implicitly loosens the conditions for realising a chiral field distributions.  
As an example of such, the spatial field distributions $\boldsymbol E(\mathbf r,t)$ for oppositely handed, circular-polarised plane waves in a homogeneous background are mirror images, and cannot be superimposed onto each other, as seen in Figure~\ref{fig:globalChiralFields}a.
These fields therefore have a form of global chirality even though the fields are locally achiral at any single point: $\boldsymbol E(t)$ and $\boldsymbol B(t)$ are orthogonal in any plane wave.
When considering such global chirality of fields in the current model for optical scattering (\ref{eq:current equation}), the relevant volume $V\not \rightarrow 0$ is that internal to the scattering object: the induced currents $\mathbf J$, and hence their scattering response in  (\ref{eq:scatt}), will inherit the symmetry transformation properties of the applied electric field distribution $\mathbf {E_0}$ internal to $V$, as discussed following (\ref{eq:eigensymmetry}).
The significance of chirality of the internal fields is now different to that of local optical chirality.
Local optical chirality implicitly relied on the assumption that the volume $V\rightarrow 0$ is a homogeneous sphere, to disregard any variation of $V$ under mirror, inversion or improper rotation operations on the coordinate system.  
Here the volume $V\not \rightarrow 0$, which is relevant for scattering, is only invariant under the symmetry operations of corresponding the symmetry group for the given scattering object. 
In the first instance, we can consider what happens if the the scattering object is \textsl{chiral}.
The enantiomers of internal fields will exist in different enantiomers of the scattering object: any reflection, inversion, or improper rotation of the coordinate system flips the handedness of both the internal field distribution {\sl and} the scattering object.
Consequently, a left-circular polarised (LCP) plane wave in a chiral $V$ never has symmetric equivalence to that of a right-circular polarised (RCP) plane wave in the same $V$, and this permits differences between the resulting scalar quantities like total scattering (\ref{eq:Pscat}) and total absorption  (\ref{eq:Pabs}).
This thereby resembles the situation previously considered for small chiral particles and local field chirality.  
However, we can now consider the case where the scattering object is \textsl{achiral}, in which both enantiomers of chiral internal fields will exist in the same object: there is a coordinate transformation that preserves $V$ while flipping the handedness of the internal fields.  
Such enantiomers of the chiral internal fields are then equivalent to each other under a coordinate transformation, which makes scalar quantities like total scattering (\ref{eq:Pscat}) and total absorption  (\ref{eq:Pabs}) conserved for each enantiomer.  
However, while we can define equivalent enantiomers  for any  $\mathbf{E_0}$ distribution, it is not guaranteed that these enantiomers align with co-propagating LCP and RCP plane waves confined to $V$, which is the relevant comparison for effects such as circular dichroism, to be discussed in the next Section.
Moreover, if there is no sequence of symmetry operations from the symmetry of group for $V$ to transform one spiral enantiomer of a circular polarised plane wave in Figure~\ref{fig:globalChiralFields}a into its co-propagating enantiomer, the internal fields of oppositely handed plane waves will have no symmetric equivalence, in precisely the same manner as true chiral geometries.
For example, see Figure~\ref{fig:globalChiralFields}b, which illustrates the enantiomers of a circularly polarised plane wave superimposed onto the $S_4$ achiral geometry of Figure~\ref{fig:localChiralFields}a, and these are not related via symmetry operations of $S_4$ group.
This is the first tangible distinction between globally and locally chiral fields: many ``chiral'' optical effects from chiral molecules can arise from \textsl{achiral} nanoparticles, simply because the internal fields of the relevant circular-polarised plane waves are inequivalent under the symmetry transformations of the nanoparticles.  
This distinction between global and local chirality has arisen in many contexts for artificial nanostructured systems, which is what I discuss in the next section. 
\begin{figure}[!ht]
\centerline{\includegraphics[width=\textwidth]{{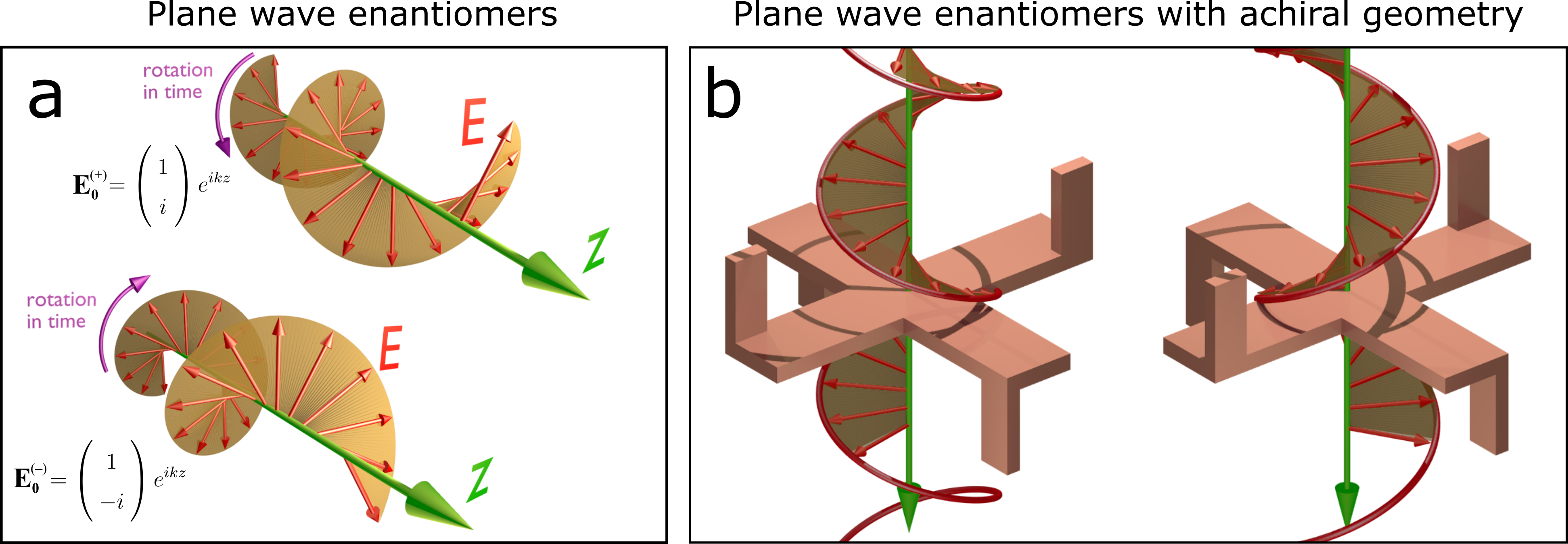}}}
\caption{Depiction of global optical chirality: (a) enantiomers formed from the electric field distributions of two oppositely handed circular-polarised plane waves (image taken from [\ref{LPOR}]), (b) enantiomers of the same plane waves when superimposed onto an achiral volume, the $S_4$ geometry in Figure~\ref{fig:localChiralFields}a.}
\label{fig:globalChiralFields}
\end{figure}

\section{Circular dichroism effects}

Here I discuss the differences experienced between LCP and RCP plane waves in: the power dissipated from the incident field (circular dichroism), and the power of the transmitted cross-polarization (circular conversion dichroism).  
This is a precursor to Section~\ref{sec:CDabs}, where I will introduce and discuss the analogous differences in power dissipated from the {\sl total} fields (circular dichroism in absorption).  \\

{\noindent \bf Circular dichroism} is one of the typical hallmarks of chiral optical response: a difference in the extinction of co-propagating LCP and RCP plane waves.
Circular dichroism is often seen as a measure of the magnitude of chiral optical response, given extinction is the total amount of light removed from the incident field (\ref{eq:Pext}) and a measure for the magnitude of total optical interaction.
One of the highest densities of optical circular dichroism is achieved with chiral nanostructures, producing 35\% absolute difference between the co-polarised transmission power of LCP and RCP, relative to the incident light power, at a thickness of $\lambda/6$~\cite{Cui2014}.
However, \textsl{achiral} nanoparticles have been observed to exhibit upwards of 80\% extinction contrast between LCP and RCP for specific propagation directions~\cite{Sersic2012}.
Circular dichroism is therefore not specific to chiral geometries.  
I now formally quantify symmetry conditions for circular dichroism in extinction.
For notation of circular-polarised plane waves, I will use $+$ and $-$ to denote LCP and RCP relative to a specified $\mathbf k$, meaning $\mathbf E_{\mathbf 0}^{(\pm)}=\frac{|\mathbf E_{\mathbf 0}|}{\sqrt{2}}\Big(\!\!{\begin{array}{c}\scriptstyle 1 \\[-0.75ex] \scriptstyle\pm i \end{array}}\!\! \Big)\,e^{i \mathbf k \cdot \mathbf r}$, where the two indices denote a right-handed orthonormal basis in the plane whose normal is directed along $+\mathbf k$.  
We begin by incorporating the result of Section~\ref{sec:tSym}, where it was derived that incident fields $\mathbf{E_0}^*$ and $\mathbf{E_0}$ both experience the same extinction, which then constrains circular dichroism \mbox{as follows:}\\[-4ex]
\begin{itemize}
\item[]\small
{Circular dichroism in extinction is forbidden if any sequence of geometric symmetry operations can arrange the distribution of $\,\mathbf E_{\mathbf 0}^{(\pm)}$  in $V$ to be proportional to the distribution in $V$ of {\sl either} $\mathrm{(i)\!:}$ $\,\mathbf E_{\mathbf 0}^{(\mp)}\!$, or $\mathrm{(ii)\!:}$ $[\mathbf E_{\mathbf 0}^{(\mp)}]^*$}.  
\end{itemize}~\\[-4ex]
Neither condition (i) or (ii) will constrain a {chiral} scattering object, as they both require an achiral symmetry operation to relate $\mathbf E_{\mathbf 0}^{(\pm)}$ to $\mathbf E_{\mathbf 0}^{(\mp)}$ in any $V$.  In other words, these conclusions are only relevant for an {achiral} scattering object.
Specifically, (i) is the standard constraint that a sequence of geometric symmetry operations will allow us to relate co-propagating LCP and RCP plane waves as a coordinate transformation, for instance: a reflection symmetry plane parallel to the propagation direction, analogous to Figure~\ref{fig:globalChiralFields}a.  
However, (ii) is an additional and important constraint from reciprocity, because it forbids circular dichroism due to inversion symmetry and symmetry under reflection perpendicular to the propagation direction.  
A geometric inversion operation $\hat i$ performs $\hat i \, \mathbf E_{\mathbf 0}^{(\pm)} = -[\mathbf E_{\mathbf 0}^{(\mp)}]^*$,  and we can apply (ii).   
A reflection operation $\hat \sigma_h$  about a symmetry plane perpendicular to the propagation direction will perform $\hat \sigma_h \, \mathbf E_{\mathbf 0}^{(\pm)} =[\mathbf E_{\mathbf 0}^{(\pm)}]^*$, and we can apply (ii).
We can now define minimum symmetries that will suppress circular dichroism:
\begin{itemize}
\item
The $C_i$=$S_2$ symmetry group forbids circular dichroism for any plane wave \mbox{using (ii), {\it i.e.}}
\begin{align}
\mathbf E_{\mathbf 0}^{(\pm)} \equiv\Bigg( 
\!\!\begin{array}{c} 
\pm 1 \\  i
\end{array}\!\!\Bigg) e^{i \mathbf k \cdot \mathbf r}
\scriptbelow{\longleftrightarrow}{\begin{array}{c}\text{\scriptsize spatial} \\ \text{\scriptsize inversion}\end{array}}
\Bigg( \!\!\begin{array}{c} 
\mp 1 \\ - i
\end{array}\!\!\Bigg) e^{- i \mathbf k \cdot \mathbf r}
\scriptbelow{\longleftrightarrow}{\begin{array}{c}\text{\scriptsize \text{reciprocal}} \\ \text{\scriptsize plane waves}\end{array}}
\Bigg( \!\!\begin{array}{c} 
\mp 1 \\  i
\end{array}\!\!\Bigg) e^{i \mathbf k \cdot \mathbf r}
\label{eq:elephant}
\end{align}
\item{The $C_{1h}$=$S_1$ symmetry group forbids circular dichroism for plane waves propagating either parallel, using (i), or perpendicular, using (ii), to the symmetry plane}
\end{itemize} 
I have called $C_i$ and $C_{1h}$ groups the \textsl{minimum} symmetries because they are the only groups that contain only an identity operation and a single achiral symmetry operation. 
The only achiral symmetry operation we appear to miss is improper rotation, but $\hat S_1$ is equal to reflection $\hat \sigma_h$ in $C_{1h}$, $\hat S_2$ is equal to inversion $\hat i$ in $C_{i}$, and $\hat S_3$ onwards exists only in larger symmetry groups given $\hat S_n\neq \hat S_n^{-1}$ for $n\!\geq\!3$, which implies there must exist at least one more symmetry operation $\hat R = \hat S_n^{-1}$ in order to form a group.
\begin{table}[!ht]
\centerline{
\begin{tabular}{c c}
\begin{tabular}{|c|c|c||c|}
\hline & &  & \\[-2ex]
& \large$\hat E$ &\large $\hat i$  & {\small examples for} $[x,y,z]$\\
\hline & &  & \\[-2ex]
\large $A_g$ & 1 & 1 &  $ x^2\!,\;y^2\!,\;z^2 $\\[1ex]
\large $A_u$ & 1 & -1 &  $x,\; y,\; z $\\[1ex]\hline
\end{tabular}
&
\begin{tabular}{|c|c|c||c|}
\hline & &  & \\[-2ex]
& \large$\hat E$ &\large $\hat \sigma_h$  & {\small examples for} $[x,y,z]$\\
\hline & &  & \\[-2ex]
\large $A$ & 1 & 1 &  $x,\; y$\\[1ex]
\large $B$ & 1 & -1 &  $z$\\[1ex]\hline
\end{tabular}
\end{tabular}}
\caption{Character tables for the (left)  $C_{i}$ symmetry group, and (right) $C_{1h}$ symmetry group.}
\label{tab:C2Ci}
\end{table}
\newpage
{\noindent  \bf Circular conversion dichroism} is the difference in the conversion efficiency of an LCP or RCP plane wave into the cross-polarised RCP or LCP plane wave.  
As previously discussed, circular dichroism in extinction is forbidden in a two-dimensional surface due to its implicit $C_{1h}$ symmetry, yet there were observations of a difference in the total transmitted power under illumination by LCP and RCP plane waves from such a surface, which was explained to be due to circular conversion dichroism~\cite{Fedotov2006, Singh2009}.
In other words, the observed difference in the total transmitted power was a difference in the power of the cross-polarised transmission, while the co-polarised transmission was conserved. 
A symmetry that will forbid circular conversion dichroism is $C_{n}$ ($n\!\geq\!3$) discrete rotational symmetry, provided $\mathbf k$ is parallel to the principal axis of $C_{n}$.
This was first derived by Fernandez-Corbaton~\cite{Fernandez-Corbaton2013}, in the context of the conservation of helicity $|\mathbf{E}\pm \sqrt{\frac{\mu_0}{\epsilon_0}}\mathbf H|^2$ in transmission, but we can also quickly re-illustrate the result here from the perspective of symmetry operations acting on plane waves.  
We begin by noting that a rotation by any angle $\phi$ about the propagation axis of a circularly polatized plane wave is equivalent to a phase shift.
\begin{align}
\Big ( \! \begin{array}{cc} \cos(\phi) & -\sin(\phi)  \\  \sin(\phi) & \cos(\phi) 
\end{array} \! \Big) \Big ( \!\!  \begin{array}{c} 1 \\ \pm i \end{array} \!\! \Big) = \Big ( \!\!  \begin{array}{c} 1 \\ \pm i \end{array} \!\! \Big) \, e^{\mp i \phi}
\end{align}
As such, an LCP or RCP plane wave $ \mathbf E_{\mathbf 0}^{(\pm)}$ with $+\mathbf{\hat k}$ parallel to the principal axis of $C_n$ will transform according to the  $E_{\pm1}$ irreducible representation of $C_n$.  
This follows because a symmetric rotation by $\frac{2\pi}{n}$ of the coordinate system is equivalent to applying a uniform phase shift of $ e^{\mp i \frac{2\pi}{n}}$ to $ \mathbf E_{\mathbf 0}^{(\pm)}$, and extending this to multiples of $\frac{2\pi}{n}$ writes out the $E_{\pm1}$ transformation properties, {\it e.g.} see Table~\ref{table:C3} for $n=3$. 
The physical locations lying on the principal axis $\mathbf r=\mathbf z$ remain static with any symmetric rotation operation in $C_n$, {\it i.e.} $\hat C_n  \mathbf z = \mathbf z$, hence the total fields $ \mathbf E(\mathbf z)$ then transform as $\mathbf E(\mathbf z) \rightarrow \hat C_n \mathbf E(\hat C_n \mathbf z)  = \mathbf{\bbar D}^{(\hat C_n)} \mathbf E(\mathbf z)$.
Given rotations of the coordinate system are equivalent to phase changes of $ \mathbf E_{\mathbf 0}^{(\pm)}$, as according to $E_{\pm1}$, it follows that $\mathbf E(\mathbf z)$ must also transform under rotations as the same uniform phase shifts, and it therefore transforms according to $E_{\pm1}$.
A cross-polarised plane wave $\mathbf E_{\mathbf 0}^{(\mp)}$ can then only exist in $\mathbf E(\mathbf z)$  if it transforms according to $E_{\pm1}$, following the orthogonality in (\ref{eq:provenorthogonality}), if it propagates in the $-\mathbf{\hat k}$ direction.
As such, there is no conversion of circular polarisation in transmission, being $\mathbf E_{\mathbf 0}^{(\mp)}$ in the $+\mathbf{\hat k}$ direction.
A similar argument also follows to conclude there is no conserved polarisation in reflection, $\mathbf E_{\mathbf 0}^{(\pm)}$ in the $-\mathbf{\hat k}$ direction.  
Notably, $C_n$ ($n\geq3$) is a chiral symmetry group, meaning it contains no achiral symmetry operations. 
As such, rotationally symmetry scattering objects that are also chiral, can exhibit circular dichroism in extinction while exhibiting no circular conversion dichroism.

\newpage
\section{Modal interference and circular dichroism in absorption}
\label{sec:CDabs}

{Circular dichroism in absorption} is the difference in the loss due to material absorption (\ref{eq:Pabs}) experienced by LCP and RCP plane waves, and it can exist separate of the two forms circular dichroism introduced in the previous section.
Moreover, the work in [\ref{LPOR}] showed circular dichroism in absorption in a system with $C_{2mh}$ ($2 m\!\geq\!3$) symmetry that forbids both: circular dichroism in extinction ($C_i \subset C_{2mh}$), and circular conversion dichroism ($C_{2m} \subset C_{2mh}$) for normal incidence plane waves.
I have referred to this effect as circular dichroism in \textsl{absorption}, in recognition that dissipative material loss is necessary to ensure that the scattering is not equal to the extinction.
Notably, however, the absence of circular dichroism in extinction, and simultaneous presence of circular dichroism in absorption, means circular dichroism must also be occurring in scattering as the difference of extinction and absorption. 
The secondary conclusion of [\ref{LPOR}] was that the observed circular dichroism in the absorption required nonorthogonal eigenmodes, and the magnitude of the dichroism could be enhanced through Fano interference, because it implies the existence of highly nonorthogonal eigenmodes as discussed in Chapter~\ref{chapterEigen}.   
In this section, I will explore further why nonorthogonal eigenmodes were necessary for circular dichroism in the absorption of LCP and RCP plane waves.

We firstly recognise that the $E_1$ irreducible representation of $C_n$ ($n\geq 3$), which we used in the previous section to describe the symmetry transformation properties for normally incident plane waves under the corresponding rotations, is separated in $C_{nh}$ into either: $E_{1u}$ and $E_{1g}$ irreducible representations if $n$ is even,  or $E'_{1}$ and $E''_{2}$ if $n$ is odd.  
These irreducible representations respectively corresponding to whether the given quantity is even or odd parity under  inversion $\hat i$ ($n$ even), or reflection $\hat \sigma_h$ ($n$ odd).  
For ease of notation, I will now just focus on $E_{1u}$, but analogous arguments will apply to $E_{1g}$, $E'_1$ and $E''_1$.
Now, as was the case with the $E_1$ irreducible representation of $C_n$ in Chapter~\ref{chapterSymmetry}, the eigenmodes $\mathbf j_{v}$  of  $E_{1u}$ occur in degenerate pairs $\mathbf j_{v+}$ and $\mathbf j_{v-}$ between the complex conjugate irreducible representations $E_{1u+}$ and $E_{1u-}$.
We can consider the $\mathbf k$ of $\mathbf{E}_{\mathbf 0}^{(\pm)} $ to be aligned with the $\pm$  notation such that $\mathbf{E}_{\mathbf 0}^{(\pm)} $ is a linear combination of eigenmodes $\mathbf j_{v\pm}$.
\begin{align}	
 \mathbf{E}_{\mathbf 0}^{(\pm)} =
\sum \limits _v\lambda_v a_{v\pm} \mathbf j_{v\pm} + \underset{\text{$E_{1g}$ component}}{\underbrace{\sum \limits _v \lambda'_v a'_{v\pm} \mathbf{j'}_{w\pm}}}  \label{eq:vv0}
\end{align}
Here the $a_v$ cannot be calculated using (\ref{eq:excitationCoefficient}), because $\mathbf j_{v\pm}$ and $\mathbf j_{v\pm}^*$ belong to  different irreducible representations $E_{1u\mp}$ and $E_{1u\mp}$, and  are therefore orthogonal using (\ref{eq:provenorthogonality}).
This means that $\int \mathbf j_{v\pm} \cdot \mathbf j_{v\pm} = 0$, which breaks the approach to calculate the excitation amplitude $a_{v\pm}$ through unconjugated projections in (\ref{eq:excitationCoefficient}).
Instead, the $a_{v\pm}$ can be calculated using: $\int \mathbf j_{v\pm} \cdot \mathbf j_{w\mp} = 0$ if $\lambda_v \neq \lambda_w$, as derived in (12) of [\ref{LPOR}].
\begin{align}
\lambda_v a_{v\pm} = \frac{\int  \mathbf{j}_{v\mp}\! \cdot {\mathbf{E}_\mathbf{0}^{(\pm)}}  }{\int  \mathbf{j}_{v\mp}\! \cdot  \mathbf{j}_{v\pm}}  \label{eq:newexcitation}
\end{align}
By relating co-propagating LCP and RCP plane waves through the inversion symmetry operation $   \mathbf{E}_{\mathbf 0}^{(\pm)} =- [\hat i \mathbf{E}_{\mathbf 0}^{(\mp)}]^*$, combined the unitarity of the inversion operation $\hat i \cdot \hat i = \hat E$, (\ref{eq:newexcitation}) is actually implying the excitation amplitude $a_{v\pm}$  is proportional to the true projection of  $\mathbf{E}_{\mathbf 0}^{(\mp)}$ onto $\mathbf{j}_{v\mp}$.
\begin{align}
\lambda_v a_{v\pm} = \frac{\int  [{\mathbf{E}_\mathbf{0}^{(\mp)}} ]^* \! \cdot \mathbf{j}_{v\mp}  }{\int  \mathbf{j}_{v\mp}\! \cdot  \mathbf{j}_{v\pm}} \label{eq:cool}
\end{align}
This is sufficient to now show that the extinction due to each eigenmode will be conserved for LCP and RCP plane waves, by substituting $\mathbf J^{(\pm)} = \sum \limits_v \lambda_v^{-1}[\lambda_v a_{v\pm} \mathbf j_{v\pm}] $ with (\ref{eq:cool}) into the expression for extinction (\ref{eq:Pext}).
\begin{align}
P_{\mathrm{ext}}  &=  \frac{1}{2}\! 
 \int  \mathrm{Re}\{ [\mathbf{E}_{\mathbf 0}^{(\pm)}]^* \!\cdot \mathbf{J}^{(\pm)}\}  
\nonumber \\
&=  \sum \limits_v \frac{1}{2} \mathrm{Re}\Big \{
\lambda_v^{-1} \frac{
\big( \int [\mathbf{E}_\mathbf{0}^{(\pm)}]^* \! \cdot \mathbf{j}_{v\pm} \big)
 \big(  {\int  [{\mathbf{E}_\mathbf{0}^{(\mp)}} ]^* \! \cdot \mathbf{j}_{v\mp}  }\big)  
}
{{\int  \mathbf{j}_{v\mp}\! \cdot  \mathbf{j}_{v\pm}}}  \Big \}\label{eq:extinction definition}
\end{align}
Note that the dependence on $+$ or $-$ has disappeared from the extinction for each independent $v$, rather than the sum of all $v$.  
As such, the degenerate eigenmodes $\mathbf j_{v+}$ and $\mathbf j_{v-}$ are associated with the same extinction under their respective excitation from $\mathbf{E}_{\mathbf 0}^{(+)} $ and $\mathbf{E}_{\mathbf 0}^{(-)} $.
Yet these two eigenmodes are importantly not chiral enantiomers of each other, $\mathbf j_{v\pm}$  is proportional to the same $\mathbf j_{v\pm}$ under inversion, and inversion is an achiral symmetry operation.   
Additionally, if the eigenmodes are nonorthogonal, there is no reciprocal relationship $\mathbf j_{v\mp} \propto \mathbf j_{v\pm}^*$ as per the discussion on circular dichroism in extinction, because this would correspond to $\mathbf j_{v\pm}$ being orthogonal to all eigenmodes by following the result $\int \mathbf j_{v\pm} \cdot \mathbf j_{w\mp} = 0$ for $\lambda_v \neq \lambda_w$, which we previously used to define (\ref{eq:newexcitation}).  
Therefore, by nature of $\mathbf j_v$ being nonorthogonal to some $\mathbf j_w$, we know that $\mathbf j_{v+}$ and $\mathbf j_{v-}$ must be nontrivially distinct current distributions, having no symmetric relationship through spatial symmetry operations, nor through time-reversal.  
Consequently, by making an oligomer of coupled nanoparticles with both strongly dissipative nanoparticles and strongly radiative nanoparticles, we can translate any differences in the distribution of $\mathbf j_{v+}$ and $\mathbf j_{v-}$ into differences of absorption and scattering.  
This is precisely what is done in [\ref{LPOR}], using nanoparticle oligomer with small nanoparticles (dissipative component) surrounding a large nanoparticle (radiative component).

This scenario was explored further in [\ref{Photonics}], where we instead considered the difference in absorption and scattering of co-polarised plane waves propagating in opposite directions, irrespective of polarization.
The previous circular dichroism in absorption and scattering between LCP and RCP plane waves of an object with $C_{nh}$ symmetry, is likened to a difference in the scattering and absorption of reciprocal plane waves for a scattering object with $C_n$ symmetry. 
For such a situation, there is a simple one-dimensional analogy of an ensemble of lossy and radiative nanoparticles: an ideal 100\% absorbing surface next to an ideal 100\% reflective surface, see Figure~\ref{fig:model}a.
\begin{figure}[!ht]
\centerline{\includegraphics[width=0.6\textwidth]{{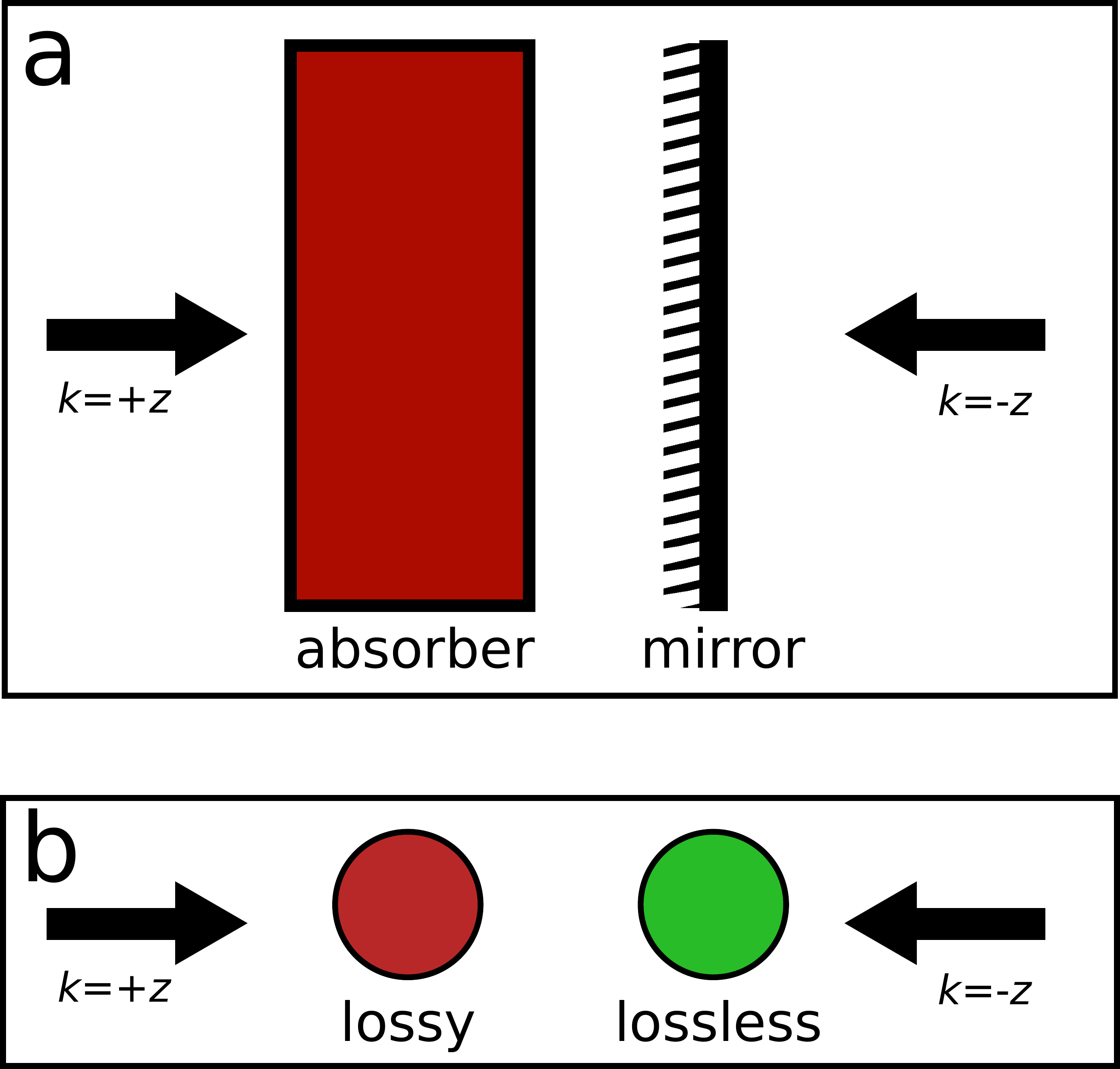}}}
\caption{An illustration showing the (a) one-dimensional analogy of a perfect mirror and perfect absorber, which can interchanges absorption and scattering loss for reciprocal plane waves (opposite propagation directions). 
 (b) Recreating this same effect with two spheres, considered in Figure~2 of [\ref{Photonics}].}
\label{fig:model}
\end{figure}
Reciprocal plane waves propagating in opposite directions are absorbed if incident on the absorbing surface first or scattered if incident on the reflective surface first; both cases have 100\% extinction, satisfying reciprocity, but the portion of absorption and scattering varies completely between reciprocal plane waves.
In [\ref{Photonics}], this to the situation was liken to the two spheres illustrated in Figure~\ref{fig:model}b, one absorbing and one radiative, which could be modelled as point electric dipoles. 
Using equally sized gold spheres, but artificially increasing the dissipative absorption of gold in one sphere, differences in scattering and absorption appeared between reciprocal plane waves.
To make the effect stronger with realistic material parameters, the geometry was then simply adjusted to be two parallel crosses ($C_{4v}$ symmetry) that could support much larger dipole moments than spheres at a comparable separations.
The arm thickness of one cross was then varied to make it a better or worse radiator, and thereby control its dissipative absorption to resemble the lossy sphere in Figure~\ref{fig:model}b.  
This geometry then predicted comparable differences in absorption between reciprocal plane waves, see Figure~3 of [\ref{Photonics}], as the circular dichroism in absorption seen in Figure~4 of [\ref{LPOR}].  
By then adding reflection planes to increase symmetry of the cross geometry to $C_{nv}$ ($n\geq3$), the absorption and scattering behaviour did not depend\footnote{$C_n$ symmetry to remove linear polarization dependence (Chapter~\ref{chapterSymmetry}), and $\sigma_v$ reflection symmetry parallel to the propagation direction to remove dependence on LCP and RCP.} on the polarization of this incident field, and thereby only depended  on the propagation direction of the plane wave.

\chapter{Conclusions}

It is apparent that the field of metamaterials has been maturing; operational principles are gradually becoming clearer, and functional devices more frequent.  
Many challenges now tend toward implementation in manufacturing and technology. 
However, a frontier the research community can significantly impact is developing understanding of the physics in metamaterial operation, and removing existing ambiguities. 
This particular pursuit will be relevant for future opportunities that now lie in how the concepts of metamaterials, and the advancements in fabrication, can be utilised to explore new physics. 
In this thesis I have focused on developing both principles and modelling approaches for the collective optical behaviour of nanoparticle oligomers.
These systems have provided an avenue to granularly quantify the formation of collective resonances from constituent elements, which has enabled delivery of the following list of key results over the course of my studies:  
\begin{itemize}
\item{
A new model for Fano resonances in optical scattering systems was presented, and is based on the interference between the nonorthogonal eigenmodes. 
This was then implemented to correctly describe Fano resonances occurring in both plasmonic and high-refractive-index dielectric nanoparticle oligomer systems.~Chapter~\ref{chapterEigen};~[\ref{PRA}].
}
\item{
The derivation of collective resonances in plasmonic and dielectric nanoparticle systems from their resonant subsystems, was shown to be translatable to the coupled dipole model while maintaining quantitatively accurate predictions.  
This was demonstrated to enable simplified calculations, interpretation and analysis of Fano resonances in symmetric arrangements of three plasmonic or dielectric nanoparticles.~Chapter~\ref{chapterEigen};~[\ref{PRB}].
}
\item{
High-refractive-index dielectric nanoparticle oligomers were shown to enable interference effects and Fano resonances between multiple collective magnetic resonances, and thereby behave analogous to two-channel point sources of magnetisation that do not occur naturally.~Chapter~\ref{chapterEigen};~[\ref{ACS}].
}
\item{
Polarisation-independent scattering and absorption losses, while still allowing nontrivial variation of the near-field, was shown to be enforced by symmetry through the combination of $n$-fold cyclic $C_n$ ($n\geq3$) discrete rotational symmetry and reciprocal eigenmode degeneracy.~Chapter~\ref{chapterSymmetry};~[\ref{Nanoscale},\,\ref{LPOR}].
}
\item{
A new form of circular dichroism due to the interaction of nonorthogonal resonances was presented, and impacts the ratio of radiative scattering loss to dissipative absorption loss.  
This was explained to be analogous to differences in scattering and absorption between reciprocal plane waves.~Chapter~\ref{chapterChiral};~[\ref{Photonics},\,\ref{LPOR}].  
}
\end{itemize}


\bibliographystyle{ieeetr}
\bibliography{../MasterBibliography/bibliography}

\begin{thebibliography}{100}

\bibitem{Huang2010}
B.~Huang, H.~Babcock, and X.~Zhuang, ``Breaking the diffraction barrier:
  Super-resolution imaging of cells,'' {\em Cell}, vol.~143, pp.~1047--1058,
  2010.

\bibitem{Sydor2015}
A.~M. Sydor, K.~J. Czymmek, E.~M. Puchner, and V.~Mennella, ``Super-resolution
  microscopy: From single molecules to supramolecular assemblies,'' {\em Trend
  Cell Biol.}, vol.~25, pp.~730--748, 2015.

\bibitem{Miller2009}
D.~A.~B. Miller, ``Device requirements for optical interconnects to silicon
  chips,'' {\em Proc. IEEE}, vol.~97, no.~7, pp.~1166--1185, 2009.

\bibitem{Agrell2016}
E.~Agrell, M.~Karlsson, A.~R. Chraplyvy, D.~J. Richardson, P.~M. Krummrich,
  P.~Winzer, K.~Roberts, J.~K. Fischer, , S.~J. Savory, B.~J. Eggleton,
  M.~Secondini, F.~R. Kschischang, A.~Lord, J.~Prat, I.~Tomkos, J.~E. Bowers,
  S.~Srinivasan, M.~Brandt-Pearce, and N.~Gisin, ``Roadmap of optical
  communications,'' {\em J. Opt.}, vol.~18, p.~063002, 2016.

\bibitem{Xia2011}
D.~Xia, Z.~Ku, S.~C. Lee, and S.~R.~J. Brueck, ``Nanostructures and functional
  materials fabricated by interferometric lithography,'' {\em Adv. Mater.},
  vol.~23, pp.~147--179, 2011.

\bibitem{Huang2011}
Z.~Huang, N.~Geyer, P.~Werner, J.~de~Boor, and U.~G\:osele, ``Metal-assisted
  chemical etching of silicon: A review,'' {\em Adv. Mater.}, vol.~23,
  pp.~285--308, 2011.

\bibitem{Liu2013_ChemRev}
Y.~Liu, J.~Goebla, and Y.~Yin, ``Templated synthesis of nanostructured
  materials,'' {\em Chem. Soc. Rev.}, vol.~42, p.~2610, 2013.

\bibitem{Vogel2015}
N.~Vogel, M.~Retsch, C.-A. Fustin, A.~del Campo, and U.~Jonas, ``Advances in
  colloidal assembly: The design of structure and hierarchy in two and three
  dimensions,'' {\em Chem. Rev.}, vol.~115, pp.~6265--6311, 2015.

\bibitem{Purcell1946}
E.~M. Purcell, ``Spontaneous emission probabilities at radio frequencies,''
  {\em Phys. Rev.}, vol.~69, p.~681, 1946.

\bibitem{Tam2007}
F.~Tam, G.~P. Goodrich, B.~R. Johnson, and N.~J. Halas, ``Plasmonic enhancement
  of molecular fluorescence,'' {\em Nano Lett.}, vol.~7, no.~2, pp.~496--501,
  2007.

\bibitem{Yoo2015}
S.~Yoo and Q.-H. Park, ``Chiral light-matter interaction in optical
  resonators,'' {\em Phys. Rev. Lett.}, vol.~114, p.~203003, 2015.

\bibitem{Fernandez-Corbaton2016}
I.~Fernandez-Corbaton, M.~Fruhnert, and C.~Rockstuhl, ``Objects of maximum
  electromagnetic chirality,'' {\em Phys. Rev. X}, vol.~6, p.~031013, 2016.

\bibitem{Stewart2008}
M.~E. Stewart, C.~R. Anderton, L.~B. Thompson, J.~Maria, S.~K. Gray, J.~A.
  Rogers, and R.~G. Nuzzo, ``Nanostructured plasmonic sensors,'' {\em Chem.
  Rev.}, vol.~108, pp.~494--521, 2008.

\bibitem{Li2015}
M.~Li, S.~K. Cushingab, and N.~Wu, ``Plasmon-enhanced optical sensors: a
  review,'' {\em Analyst}, vol.~140, pp.~386--406, 2015.

\bibitem{Zijlstra2012}
P.~Zijlstra, P.~M.~R. Paulo, and M.~Orrit, ``Optical detection of single
  non-absorbing molecules using the surface plasmon resonance of a gold
  nanorod,'' {\em Nat. Nanotech.}, vol.~7, pp.~379--382, 2012.

\bibitem{Bauer2014}
T.~Bauer, S.~Orlov, U.~Peschel, P.~Banzer, and G.~Leuchs, ``Nanointerferometric
  amplitude and phase reconstruction of tightly focused vector beams,'' {\em
  Nat. Photon.}, vol.~8, pp.~23--27, 2014.

\bibitem{Rotenberg2014}
Rotenberg and L.~Kuipers, ``Mapping nanoscale light fields,'' {\em Nat.
  Photon.}, vol.~8, pp.~919--926, 2014.

\bibitem{Park2014}
H.~Park and K.~B. Crozier, ``Multispectral imaging with vertical silicon
  nanowires,'' {\em Sci. Rep.}, vol.~3, p.~2460, 2013.

\bibitem{Bomzon2002}
Z.~Bomzon, G.~Biener, V.~Kleiner, and E.~Hasman, ``Radially and azimuthally
  polarized beams generated by space-variant dielectric subwavelength
  gratings,'' {\em Opt. Lett.}, vol.~27, no.~5, pp.~285--287, 2002.

\bibitem{Yu2011}
N.~F. Yu, P.~Genevet, M.~A. Kats, F.~Aieta, J.~P. Tetienne, F.~Capasso, and
  Z.~Gaburro, ``Light propagation with phase discontinuities: Generalized laws
  of reflection and refraction,'' {\em Science}, vol.~334, no.~6054, p.~333,
  2011.

\bibitem{Arbabi2015}
A.~Arbabi, Y.~Horie, M.~Bagheri, and A.~Faraon, ``Dielectric metasurfaces for
  complete control of phase and polarization with subwavelength spatial
  resolution and high transmission,'' {\em Nat. Mat.}, vol.~10, pp.~937--944,
  2015.

\bibitem{Kruk2016}
S.~Kruk, B.~Hopkins, I.~I. Kravchenko, A.~Miroshnichenko, D.~N. Neshev, and
  Y.~S. Kivshar, ``Broadband highly efficient dielectric metadevices for
  polarization control,'' {\em APL Photon.}, vol.~1, p.~030801, 2016.

\bibitem{Shcherbakov2014}
M.~R. Shcherbakov, D.~N. Neshev, B.~Hopkins, A.~S. Shorokhov, I.~Staude, E.~V.
  Melik-Gaykazyan, M.~Decker, A.~A. Ezhov, A.~E. Miroshnichenko, I.~Brener,
  A.~A. Fedyanin, and Y.~S. Kivshar, ``Enhanced third-harmonic generation in
  silicon nanoparticles driven by magnetic response,'' {\em Nano Lett.},
  vol.~14, no.~11, pp.~6488--6492, 2014.

\bibitem{Celebrano2015}
M.~Celebrano, X.~Wu, M.~Baselli, S.~Großmann, P.~Biagioni, A.~Locatelli, C.~D.
  Angelis, G.~Cerullo, R.~Osellame, B.~Hecht, L.~Du\`o, F.~Ciccacci, and
  M.~Finazzi, ``Mode matching in multiresonant plasmonic nanoantennas for
  enhanced second harmonic generation,'' {\em Nat. Nanotech.}, vol.~10,
  pp.~412--417, 2015.

\bibitem{Shcherbakov2015}
M.~R. Shcherbakov, P.~P. Vabishchevich, A.~S. Shorokhov, K.~E. Chong, D.-Y.
  Choi, I.~Staude, A.~E. Miroshnichenko, D.~N. Neshev, A.~A. Fedyanin, and
  Y.~S. Kivshar, ``Ultrafast all-optical switching with magnetic resonances in
  nonlinear dielectric nanostructures,'' {\em Nano Lett.}, vol.~15, no.~10,
  p.~6985–6990, 2015.

\bibitem{Yang2015}
Y.~Yang, W.~Wang, A.~Boulesbaa, I.~I. Kravchenko, D.~P. Briggs, A.~Puretzky,
  D.~Geohegan, and J.~Valentine, ``Nonlinear Fano-resonant dielectric
  metasurfaces,'' {\em Nano Lett.}, vol.~15, no.~11, pp.~7388--7393, 2015.

\bibitem{Pfeiffer2013}
C.~Pfeiffer and A.~Grbic, ``Metamaterial huygens' surfaces: Tailoring wave
  fronts with reflectionless sheets,'' {\em Phys. Rev. Lett.}, vol.~110,
  p.~197401, May 2013.

\bibitem{StrattonChu1939}
J.~A. Stratton and L.~J. Chu, ``Diffraction theory of electromagnetic waves,''
  {\em Phys. Rev.}, vol.~56, pp.~99--107, 1939.

\bibitem{Pendry1999}
J.~B. Pendry, A.~J. Holden, D.~J. Robbins, and W.~J. Stewart, ``Magnetism from
  conductors and enhanced nonlinear phenomena,'' {\em IEEE Trans. Microw.
  Theory Tech.}, vol.~47, no.~11, pp.~2075--2084, 1999.

\bibitem{Pors2013}
A.~Pors, M.~G. Nielsen, and S.~I. Bozhevolnyi, ``Broadband plasmonic half-wave
  plates in reflection,'' {\em Opt. Lett.}, vol.~38, pp.~513--515, 2013.

\bibitem{Zhao2013}
Y.~Zhao and A.~Al\`u, ``Tailoring the dispersion of plasmonic nanorods to
  realize broadband optical meta-waveplates,'' {\em Nano Lett.}, vol.~13,
  pp.~1086--1091, 2013.

\bibitem{Ni2013}
X.~Ni, S.~Ishii, A.~V. Kildishev, and V.~M. Shalaev, ``Ultra-thin, planar,
  babinet-inverted plasmonic metalenses,'' {\em Light Sci. Appl.}, vol.~2,
  p.~e72, 2013.

\bibitem{Evlyukhin2010}
A.~B. Evlyukhin, C.~Reinhardt, A.~Seidel, B.~S. Luk'yanchuk, and B.~N.
  Chichkov, ``Optical response features of si-nanoparticle arrays,'' {\em Phys.
  Rev. B}, vol.~82, p.~045404, 2010.

\bibitem{Nieto-Vesperinas2010}
M.~Nieto-Vesperinas, J.~J. Saenz, R.~Gomez-Medina, and L.~Chantada, ``Optical
  forces on small magnetodielectric particles,'' {\em Opt Express}, vol.~18,
  no.~11, p.~11428, 2010.

\bibitem{Garcia-Etxarri2011}
A.~Garc\'ia-Etxarri, R.~G\'omez-Medina, L.~S. Froufe-P\'erez, C.~L\'opez,
  L.~Chantada, F.~Scheffold, J.~Aizpurua, M.~Nieto-Vesperinas, and J.~J.
  S\'aenz, ``Strong magnetic response of submicron silicon particles in the
  infrared,'' {\em Optics Express}, vol.~19, pp.~4815--4826, 2011.

\bibitem{Ginn2012}
J.~C. Ginn, I.~Brener, D.~W. Peters, J.~R. Wendt, J.~O. Stevens, P.~F. Hines,
  L.~I. Basilio, L.~K. Warne, J.~F. Ihlefeld, P.~G. Clem, and M.~B. Sinclair,
  ``Realizing optical magnetism from dielectric metamaterials,'' {\em Phys.
  Rev. Lett.}, vol.~108, p.~097402, 2012.

\bibitem{Kuznetsov2012}
A.~I. Kuznetsov, A.~E. Miroshnichenko, Y.~H. Fu, J.~B. Zhang, and
  B.~Luk'yanchuk, ``Magnetic light,'' {\em Sci. Rep.}, vol.~2, p.~492, 2012.

\bibitem{Evlyukhin2012}
A.~B. Evlyukhin, S.~M. Novikov, U.~Zywietz, R.~L. Eriksen, C.~Reinhardt, S.~I.
  Bozhevolnyi, and B.~N. Chichkov, ``Demonstration of magnetic dipole
  resonances of dielectric nanospheres in the visible region,'' {\em Nano
  Lett.}, vol.~12, pp.~3749--3755, 2012.

\bibitem{Decker2015}
M.~Decker, I.~Staude, M.~Falkner, J.~Dominguez, D.~N. Neshev, I.~Brener,
  T.~Pertsch, and Y.~S. Kivshar, ``High-efficiency dielectric huygens’
  surfaces,'' {\em Adv. Opt. Mater.}, vol.~3, pp.~813--820, 2015.

\bibitem{Khorasaninejad2016}
M.~Khorasaninejad, W.~T. Chen, R.~C. Devlin, J.~Oh, A.~Y. Zhu, and F.~Capasso,
  ``Metalenses at visible wavelengths: Diffraction-limited focusing and
  subwavelength resolution imaging,'' {\em Science}, vol.~352, pp.~1190--1194,
  2016.

\bibitem{Chong2016}
K.~E. Chong, L.~Wang, I.~Staude, A.~R. James, J.~Dominguez, S.~Liu, G.~S.
  Subramania, M.~Decker, D.~N. Neshev, I.~Brener, and Y.~S. Kivshar†,
  ``Efficient polarization-insensitive complex wavefront control using
  huygens’ metasurfaces based on dielectric resonant meta-atoms,'' {\em ACS
  Photon.}, vol.~3, no.~4, pp.~514--519, 2016.

\bibitem{Chong2015}
K.~E. Chong, I.~Staude, A.~James, J.~Dominguez, S.~Liu, S.~Campione, G.~S.
  Subramania, T.~S. Luk, M.~Decker, I.~B. Dragomir N. Neshev~and, and Y.~S.
  Kivshar, ``Polarization-independent silicon metadevices for efficient optical
  wavefront control,'' {\em Nano Lett.}, vol.~15, no.~8, pp.~5369--5374, 2015.

\bibitem{Pendry2006}
J.~B. Pendry, D.~Schurig, and D.~R. Smith, ``Controlling electromagnetic
  fields,'' {\em Science}, vol.~312, pp.~1780--1782, 2006.

\bibitem{Joannopoulos2008book}
J.~D. Joannopoulos, S.~G. Johnson, J.~N. Winn, and R.~D. Meade, {\em Photonic
  Crystals: Molding the Flow of Light}.
\newblock Princeton University Press, 2008.

\bibitem{Hentschel2010}
M.~Hentschel, M.~Saliba, R.~Vogelgesang, H.~Giessen, A.~P. Alivisatos, and
  N.~Liu, ``Transition from isolated to collective modes in plasmonic
  oligomers,'' {\em Nano Lett.}, vol.~10, pp.~2721--2726, 2010.

\bibitem{FeynmanI}
R.~P. Feynman, R.~B. Leighton, and M.~L. Sands, ``The Feynman lectures on
  physics: Volume I.'' \url{www.feynmanlectures.caltech.edu}.

\bibitem{Gitman2016}
D.~M. Gitman, A.~E. Shabad, and A.~A. Shishmarev, ``A note on “measuring
  propagation speed of coulomb fields” by r. de sangro, g. finocchiaro, p.
  patteri, m. piccolo, g. pizzella,'' {\em Eur. Phys. J. C}, vol.~76, p.~261,
  2016.

\bibitem{Jackson}
J.~Jackson, {\em Classical Electrodynamics (3rd ed.)}.
\newblock New York: Wiley, 1998.

\bibitem{Kholmetskii2014}
A.~Kholmetskii, O.~Missevitch, and T.~Yarman, ``Electric/magnetic dipole in an
  electromagnetic field: force, torque and energy,'' {\em Eur. Phys. J. Plus},
  vol.~129, p.~215, 2014.

\bibitem{Figueroa-O'Farrill1998}
J.~Figueroa-O'Farrill, ``Electromagnetic duality for children.''
  \url{http://www.maths.ed.ac.uk/~jmf/Teaching/EDC.html}, October 1998.

\bibitem{Devaney1974}
A.~J. Devaney and E.~Wolf, ``Multipole expansions and plane wave
  representations of the electromagnetic field,'' {\em J. Math. Phys.},
  vol.~15, p.~234, 1974.

\bibitem{Grahn2012}
P.~Grahn, A.~Shevchenko, and M.~Kaivola, ``Electromagnetic multipole theory for
  optical nanomaterials,'' {\em New J. Phys.}, vol.~14, p.~093033, 2012.

\bibitem{GrahnThesis}
P.~Grahn, {\em Multipole excitations in optical meta-atoms}.
\newblock PhD thesis, Aalto University, March 2012.

\bibitem{Bohren1983}
C.~F. Bohren and D.~R. Huffman, {\em Absorption and scattering of light by
  small particles}.
\newblock New York: Wiley, 1983.

\bibitem{Mie1908}
G.~Mie, ``Beitrage zur optik truber medien,'' {\em Ann. Phys.}, vol.~25,
  pp.~377--445, 1908.

\bibitem{Sheinfux2014}
H.~H. Sheinfux, I.~Kaminer, Y.~Plotnik, G.~Bartal, and M.~Segev,
  ``Subwavelength multilayer dielectrics: Ultrasensitive transmission and
  breakdown of effective-medium theory,'' {\em Phys. Rev. Lett.}, vol.~113,
  p.~243901, 2014.

\bibitem{Evlyukhin2011}
A.~B. Evlyukhin, C.~Reinhardt, and B.~N. Chichkov, ``Multipole light scattering
  by nonspherical nanoparticles in the discrete dipole approximation,'' {\em
  Phys. Rev. B}, vol.~84, p.~235429, 2011.

\bibitem{Yaghjian1980}
A.~D. Yaghjian, ``Electric dyadic green‘s functions in the source region,''
  {\em Proc. IEEE}, vol.~68, no.~2, pp.~248--263, 1980.

\bibitem{Markel1995}
V.~A. Markel, ``Antisymmetrical optical states,'' {\em J. Opt. Soc. Am. B},
  vol.~12, no.~10, pp.~1783--1791, 1995.

\bibitem{Draine1988}
B.~T. Draine, ``The discrete-dipole approximation and its application to
  interstellar graphite grains,'' {\em ‎Astrophys. J.}, vol.~333,
  pp.~848--872, 1988.

\bibitem{LandauLifshitzVol5}
L.~D. Landau, E.~M. Lifshitz, and L.~P. Pitaevski{\u\i}, {\em Statistical
  Physics, Part 1}, vol.~5 of {\em Course of theoretical physics}.
\newblock Pergamon Press Ltd., third~ed., 1980.

\bibitem{Draine1994}
B.~T. Draine and P.~J. Flatau, ``Discrete-dipole approximation for scattering
  calculations,'' {\em J. Opt. Soc. Am. A}, vol.~11, pp.~1491--1499, Apr 1994.

\bibitem{Jin_FEM}
J.~Jin, {\em The Finite Element Method in Electromagnetics}.
\newblock John Wiley and Sons Inc., second~ed., 2002.

\bibitem{Agrawal2017}
A.~Agrawal, T.~Benson, R.~M. {De La Rue}, and G.~Wurtz, eds., {\em Recent
  Trends in Computational Photonics}, vol.~204.
\newblock Springer International Publishing, 2017.

\bibitem{Mulholland1994}
G.~W. Mulholland, C.~F. Bohren, and K.~A. Fuller, ``Light scattering by
  agglomerates: Coupled electric and magnetic dipole method,'' {\em Langmuir},
  vol.~10, no.~8, p.~2533, 1994.

\bibitem{Merchiers2007}
O.~Merchiers, F.~Moreno, F.~Gonzalez, and J.~M. Saiz, ``Light scattering by an
  ensemble of interacting dipolar particles with both electric and magnetic
  polarizabilities,'' {\em Phys. Rev. A.}, vol.~76, no.~4, p.~043834, 2007.

\bibitem{Raab2005}
R.~E. Raab and O.~L. {de Lange}, {\em Multipole Theory in Electromagnetism}.
\newblock Oxford University Press, 2005.

\bibitem{Chen2011}
J.~Chen, J.~Ng, Z.~Lin, and C.~T. Chan, ``Optical pulling force,'' {\em Nat.
  Photon.}, vol.~5, pp.~531--534, 2011.

\bibitem{Miroshnichenko2015}
A.~E. Miroshnichenko, A.~B. Evlyukhin, Y.~F. Yu, R.~M. Bakker, A.~Chipouline,
  A.~I. Kuznetsov, B.~Luk'yanchuk, B.~N. Chichkov, and Y.~S. Kivshar,
  ``Nonradiating anapole modes in dielectric nanoparticles,'' {\em Nat.
  Commun.}, vol.~6, p.~8069, 2015.

\bibitem{Stratton}
J.~A. Stratton, {\em Electromagnetic Theory}.
\newblock McGraw-Hill Book Company, Inc., 1941.

\bibitem{Onsager1930}
L.~Onsager, ``Reciprocal relations in irreversible processes. II.,'' {\em Phys.
  Rev.}, vol.~37, pp.~405--426, 1931.

\bibitem{Onsager1931}
L.~Onsager, ``Reciprocal relations in irreversible processes. II.,'' {\em Phys.
  Rev.}, vol.~38, pp.~2265--2279, 1931.

\bibitem{Callen1951}
H.~B. Callen and T.~A. Welton, ``Irreversibility and generalized noise,'' {\em
  Phys. Rev.}, vol.~83, p.~34, 1951.

\bibitem{Bai2000}
Z.~Bai, J.~Demmel, J.~Dongarra, A.~Ruhe, and H.~van~der Vorst, eds., {\em
  Templates for the Solution of Algebraic Eigenvalue Problems}.
\newblock Society for Industrial and Applied Mathematics, 2000.

\bibitem{Gantmacher1959}
F.~Gantmacher, {\em The Theory of Matrices}.
\newblock Chelsea Publishing Company, 1959.

\bibitem{Craven1969}
B.~Craven, ``Complex symmetric matrices,'' {\em J. Austral. Math. Soc.},
  vol.~10, pp.~341--354, 1969.

\bibitem{Prodan2003}
E.~Prodan, C.~Radloff, N.~J. Halas, and P.~Nordlander, ``A hybridization model
  for the plasmon response of complex nanostructures,'' {\em Science},
  vol.~302, pp.~419--422, 2003.

\bibitem{Turner2010}
M.~D. Turner, M.~M. Hossain, and M.~Gu, ``The effects of retardation on plasmon
  hybridization within metallic nanostructures,'' {\em New J. Phys.}, vol.~12,
  p.~083062, 2010.

\bibitem{Prodan2004}
E.~Prodan and P.~Nordlander, ``Plasmon hybridization in spherical
  nanoparticles,'' {\em J. of Chem. Phys.}, vol.~120, pp.~5444--5454, 2004.

\bibitem{Nordlander2004}
P.~Nordlander, C.~Oubre, E.~Prodan, K.~Li, and M.~I. Stockman, ``Plasmon
  hybridization in nanoparticle dimers,'' {\em Nano Lett.}, vol.~4,
  pp.~899--903, 2004.

\bibitem{Albella2015}
P.~Albella, T.~Shibanuma, and S.~A. Maier, ``Switchable directional scattering
  of electromagnetic radiation with subwavelength asymmetric silicon dimers,''
  {\em Sci. Rep.}, vol.~5, p.~18322, 2015.

\bibitem{Zywietz2015}
U.~Zywietz, M.~K. Schmidt, A.~B. Evlyukhin, C.~Reinhardt, J.~Aizpurua, and
  B.~N. Chichkov, ``Electromagnetic resonances of silicon nanoparticle dimers
  in the visible,'' {\em ACS Photon.}, vol.~2, pp.~913--920, 2015.

\bibitem{Yan2015}
J.~Yan, P.~Liu, Z.~Lin, H.~Wang, H.~Chen, C.~Wang, and G.~Yang, ``Directional
  Fano resonances in a silicon nanoparticle dimer,'' {\em ACS Nano}, vol.~9,
  pp.~2968--2980, 2015.

\bibitem{Shibanuma2016}
T.~Shibanuma, P.~Albella, and S.~A. Maier, ``Unidirectional light scattering
  with high efficiency at optical frequencies based on low-loss dielectric
  nanoantennas,'' {\em Nanoscale}, vol.~8, pp.~14184--14192, 2016.

\bibitem{Giannini2011}
V.~Giannini, Y.~Francescato, H.~Amrania, C.~C. Phillips, and S.~A. Maier,
  ``Fano resonances in nanoscale plasmonic systems: A parameter-free modeling
  approach,'' {\em Nano Lett.}, vol.~11, pp.~2835--2840, 2011.

\bibitem{Francescato2012}
Y.~Francescato, V.~Giannini, and S.~A. Maier, ``Plasmonic systems unveiled by
  Fano resonances,'' {\em ACS Nano}, vol.~6, no.~2, pp.~1830--1838, 2012.

\bibitem{Lovera2013}
A.~Lovera, B.~Gallinet, P.~Nordlander, and O.~J. Martin, ``Mechanisms of Fano
  resonances in coupled plasmonic systems,'' {\em ACS Nano}, vol.~7 (5),
  pp.~4527--4536, 2013.

\bibitem{Fan2010_NL}
J.~A. Fan, K.~Bao, C.~Wu, J.~Bao, R.~Bardhan, N.~J. Halas, V.~N. Manoharan,
  G.~Shvets, P.~Nordlander, and F.~Capasso, ``Fano-like interference in
  self-assembled plasmonic quadrumer clusters,'' {\em Nano Lett.}, vol.~10,
  pp.~4680--4685, 2010.

\bibitem{Fano1961}
U.~Fano, ``Effects of configuration interaction on intensities and phase
  shifts,'' {\em Phys. Rev.}, vol.~124, pp.~1866--1878, 1961.

\bibitem{Miroshnichenko2010}
A.~E. Miroshnichenko, S.~Flach, and Y.~S. Kivshar, ``Fano resonances in
  nanoscale structures,'' {\em Rev. Mod. Phys.}, vol.~82, pp.~2257--2298, 2010.

\bibitem{Joe2006}
Y.~S. Joe, A.~M. Satanin, and C.~S. Kim, ``Classical analogy of Fano
  resonances,'' {\em Phys. Scr.}, vol.~74, pp.~259--266, 2006.

\bibitem{Gallinet2011}
B.~Gallinet and O.~J.~F. Martin, ``Ab initio theory of Fano resonances in
  plasmonic nanostructures and metamaterials,'' {\em Phys. Rev. B}, vol.~83,
  p.~235427, 2011.

\bibitem{Gallinet2012}
B.~Gallinet, {\em Fano Resonances in Plasmonic Nanostructures: Fundamentals,
  Numerical Modeling and Applications}.
\newblock PhD thesis, \'Ecole Polytechnique F\'ed\'erale de Lausanne, June
  2012.

\bibitem{Rahmani2013_lpor}
M.~Rahmani, B.~Luk'yanchuk, and M.~Hong, ``Fano resonance in novel plasmonic
  nanostructures,'' {\em Laser Photon. Rev.}, vol.~7, no.~3, pp.~329--349,
  2013.

\bibitem{Miroshnichenko2012}
A.~E. Miroshnichenko and Y.~S. Kivshar, ``Fano resonances in all-dielectric
  oligomers,'' {\em Nano Lett.}, vol.~12, pp.~6459--6463, 2012.

\bibitem{Forestiere2013}
C.~Forestiere, L.~D. Negro, and G.~Miano, ``Theory of coupled plasmon modes and
  Fano-like resonances in subwavelength metal structures,'' {\em Phys. Rev. B},
  vol.~88, p.~155411, 2013.

\bibitem{Frimmer2012}
M.~Frimmer, T.~Coenen, and A.~F. Koenderink, ``Signature of a Fano resonance in
  a plasmonic metamolecule’s local density of optical states,'' {\em Phys.
  Rev. Lett.}, vol.~108, p.~077404, 2012.

\bibitem{Heiss2001}
W.~D. Heiss, ``Exceptional points of non-hermitian operators,'' {\em J. Phys.
  A: Math Gen.}, vol.~37, p.~2455, 2001.

\bibitem{Dembowski2001}
C.~Dembowski, H.-D. Graf, H.~L. Harney, A.~Heine, W.~D. Heiss, H.~Rehfeld, and
  A.~Richter, ``The physics of exceptional points,'' {\em Phys. Rev. Lett.},
  vol.~86, pp.~787--790, 2001.

\bibitem{Heiss2012}
W.~D. Heiss, ``The physics of exceptional points,'' {\em J. Phys. A: Math.
  Theor.}, vol.~45, p.~444016, 2001.

\bibitem{Bronson2009}
R.~Bronson and G.~B. Costa, eds., {\em Matrix Methods}.
\newblock Boston: Academic Press, third~ed., 2009.

\bibitem{Shorokhov2016}
A.~S. Shorokhov, E.~V. Melik-Gaykazyan, D.~A. Smirnova, B.~Hopkins, K.~E.
  Chong, D.-Y. Choi, M.~R. Shcherbakov, A.~E. Miroshnichenko, D.~N. Neshev,
  A.~A. Fedyanin, and Y.~S. Kivshar, ``Multifold enhancement of third-harmonic
  generation in dielectric nanoparticles driven by magnetic Fano resonances,''
  {\em Nano Lett.}, vol.~16, pp.~4857--4861, 2016.

\bibitem{Harris1989}
D.~C. Harris and M.~D. Bertolucci, {\em Symmetry and spectroscopy: an
  introduction to vibrational and electronic spectroscopy}.
\newblock Dover Publications, Inc., 1989.

\bibitem{Cotton1989}
F.~A. Cotton, {\em Chemical Applications of Group Theory}.
\newblock John Wiley and Sons, third~ed., 1989.

\bibitem{Dresselhaus2008}
M.~Dresselhaus, G.~Dresselhaus, and A.~Jorio, {\em Group Theory: Application to
  the Physics of Condensed Matter}.
\newblock Springer, 2008.

\bibitem{Potton2004}
R.~J. Potton, ``Reciprocity in optics,'' {\em Rep. Prog. Phys.}, vol.~67,
  pp.~717--754, 2004.

\bibitem{LandauLifshitzVol8}
L.~D. Landau, E.~M. Lifshitz, and L.~P. Pitaevski{\u\i}, {\em Electrodynamics
  of Continuous Media}, vol.~8 of {\em Course of theoretical physics}.
\newblock Pergamon Press Ltd., second~ed., 1984.

\bibitem{FeynmanII}
R.~P. Feynman, R.~B. Leighton, and M.~L. Sands, ``The Feynman lectures on
  physics: Volume II.'' \url{www.feynmanlectures.caltech.edu}.

\bibitem{Barron}
L.~D. Barron, {\em Molecular Light Scattering and Optical Activity}.
\newblock Cambridge University Press, 2004.

\bibitem{Hembury2008}
G.~A. Hembury, V.~V. Borovkov, and Y.~Inoue, ``Chirality-sensing supramolecular
  systems,'' {\em Chem. Rev.}, vol.~108, no.~1, pp.~1--73, 2008.

\bibitem{Tang2010}
Y.~Tang and A.~E. Cohen, ``Optical chirality and its interaction with matter,''
  {\em Physical review letters}, vol.~104, no.~16, p.~163901, 2010.

\bibitem{Cui2014}
Y.~Cui, L.~Kang, S.~Lan, S.~Rodrigues, and W.~Cai, ``Giant chiral optical
  response from a twisted-arc metamaterial,'' {\em Nano Lett.}, vol.~14, no.~2,
  pp.~1021--1025, 2014.

\bibitem{Sersic2012}
I.~Sersic, M.~A. {van de Haar}, F.~B. Arango, and A.~F. Koenderink, ``Ubiquity
  of optical activity in planar metamaterial scatterers,'' {\em Phys. Rev.
  Lett.}, vol.~108, p.~223903, 2012.

\bibitem{Fedotov2006}
V.~A. Fedotov, P.~L. Mladyonov, S.~L. Prosvirnin, A.~V. Rogacheva, Y.~Chen, and
  N.~I. Zheludev, ``Asymmetric propagation of electromagnetic waves through a
  planar chiral structure,'' {\em Phys. Rev. Lett.}, vol.~97, p.~167401, 2006.

\bibitem{Singh2009}
R.~Singh, E.~Plum, C.~Menzel, C.~Rockstuhl, A.~K. Azad, R.~A. Cheville,
  F.~Lederer, W.~Zhang, and N.~I. Zheludev, ``Terahertz metamaterial with
  asymmetric transmission,'' {\em Phys. Rev. B}, vol.~80, p.~153104, 2009.

\bibitem{Fernandez-Corbaton2013}
I.~Fernandez-Corbaton, ``Forward and backward helicity scattering coefficients
  for systems with discrete rotational symmetry,'' {\em Opt. Express}, vol.~21,
  no.~24, p.~29885, 2013.

\end{thebibliography}

\appendix 
\chapter{Thesis publications \label{pubs}} 

\section{Optically isotropic responses induced by discrete rotational symmetry of nanoparticle clusters \label{Nanoscale}}

\subsection*{Summary}
The most robust and general approach to eliminate dependence of scattering on the incident polarization of light will come from the overall symmetry of the scattering geometry, given this is not affected by the operating wavelength or the material properties. 
There had previously been no comprehensive investigations on this topic except for some example case studies on specific symmetries, particularly given the intrinsic absorption spectra was neglected almost unanimously.
In this work, we performed a rigorous investigation on symmetry induced polarisation-independent scattering and absorption.  
Discrete $n$-fold ($n\geq3$) rotational symmetry is shown to provide polarisation-independent optical responses, not only in far-field properties, such as extinction, but also for intrinsic absorption, which is usually considered to result from a near-field that is itself highly polarisation-dependent 
This paper clarifies both physically and mathematically the uncertainty around the topic of symmetry induced polarisation-independence.

\subsection*{Notes and errata}
\begin{itemize}
\item{While presented derivations use the dipole model, this model can be generalised to continuous current distributions by simply performing a substitution  of $\mathbf{J}(\mathbf r) \mathrm{dr}^3 = - i \omega \mathbf{p}_{\scriptscriptstyle (\mathbf r)}$, as was done in Chapter~\ref{chapterModels}.  
Furthermore, given we consider both electric and magnetic dipoles simultaneously, the derivations of this work actually account for magnetisation current density defined as $\mathbf{J_{\!m}}\!(\mathbf r) \mathrm{dr}^3 = - i \omega \mathbf{m_r}$.
The conclusions and derivations of this work will therefore apply generally to any object consisting both permittivity and permeability, which is not made evident in the work itself.  }
\item{The curl of the dyadic Green's function is a factor of ${ik}$ larger than it should be, however all equations are correct once making the substitution: $ \boldsymbol \nabla \times \hat{G}^{0}(\mathbf r_i, \mathbf r_j) \rightarrow \frac{1}{ik}\boldsymbol \nabla \times  \hat{G}^{0}(\mathbf r_i, \mathbf r_j) $.}
\end{itemize}

\newpage

\includepdf[pages=-, offset=8mm 0, noautoscale = true, scale = 0.9]{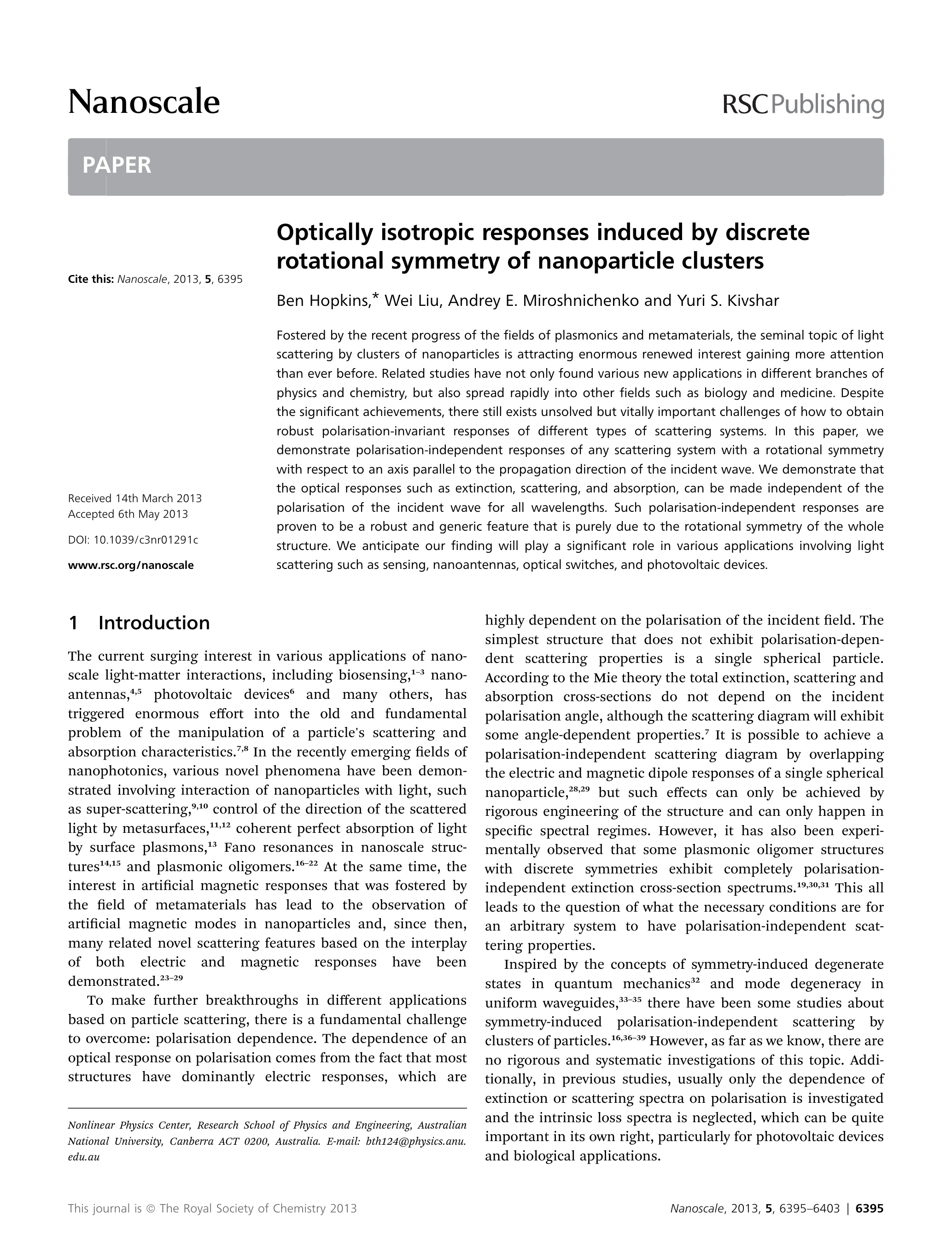}

\section{Revisiting the physics of Fano resonances for nanoparticle oligomers \label{PRA}}

\subsection*{Summary}
The interference phenomena known as Fano resonances generated significant interest in metal nanoparticle oligomers, where they offered characteristically thin spectral lineshapes and significant field enhancement. 
Yet Fano resonances were predicted to occur also in all-dielectric oligomers~\cite{Miroshnichenko2012}, which suggested a mechanism other than plasmonic hybridisation of modes was able to lead to Fano resonances. 
In this work, we revisit the optical responses from general nanoparticle oligomers and acknowledge the non-Hermitian interactions, which allows us to rigorously show that Fano resonances can be described purely from the overlap interference between nonorthogonal collective eigenmodes. 
This unifies the understanding of Fano resonances in both plasmonic and dielectric nanoparticle oligomers, and demonstrates that low-loss dielectric oligomers can provide comparable outcomes to what had been an exclusively plasmonic pursuit.

\subsection*{Notes and errata}
\begin{itemize}
\item
{The expression for the extinction cross section in equation (23) is missing a factor of $\mu_0$, the correct expression is (\ref{eq:extinctionDip}).
}
\item
{The discussion of FIG.~1 suggests a maximum of four, doubly degenerate, eigenmodes can be excited by a plane wave normally incident on an rotationally symmetric ring of three or more electric and magnetic dipole pairs.  
This is true provided one does not define eigenmodes as simultaneously containing both electric and magnetic dipoles, whereupon coupling between electric and magnetic dipoles allows these dipoles to be oriented parallel to the principal rotation axis.
It subsequently follows that more than four eigenmodes, when defined according to the discussion in Chapter~\ref{chapterEigen}, can be excited by a normal-incidence plane wave.  
However, the related derivation of the maximum two, doubly degenerate, eigenmodes being excitable for each rotationally symmetric ring of three or more electric \textsl{xor} magnetic dipoles, following equations~(2)-(10), remains unchanged. 
}
\item{The curl of the dyadic Green's function is a factor of ${ik}$ larger than it should be, however all equations are correct once making the substitution: $ \boldsymbol \nabla \times \hat{G}^{0}(\mathbf r_i, \mathbf r_j) \rightarrow \frac{1}{ik}\boldsymbol \nabla \times  \hat{G}^{0}(\mathbf r_i, \mathbf r_j) $.}
\end{itemize}
\newpage
\includepdf[pages=-, offset=8mm 0, noautoscale = true, scale = 0.9]{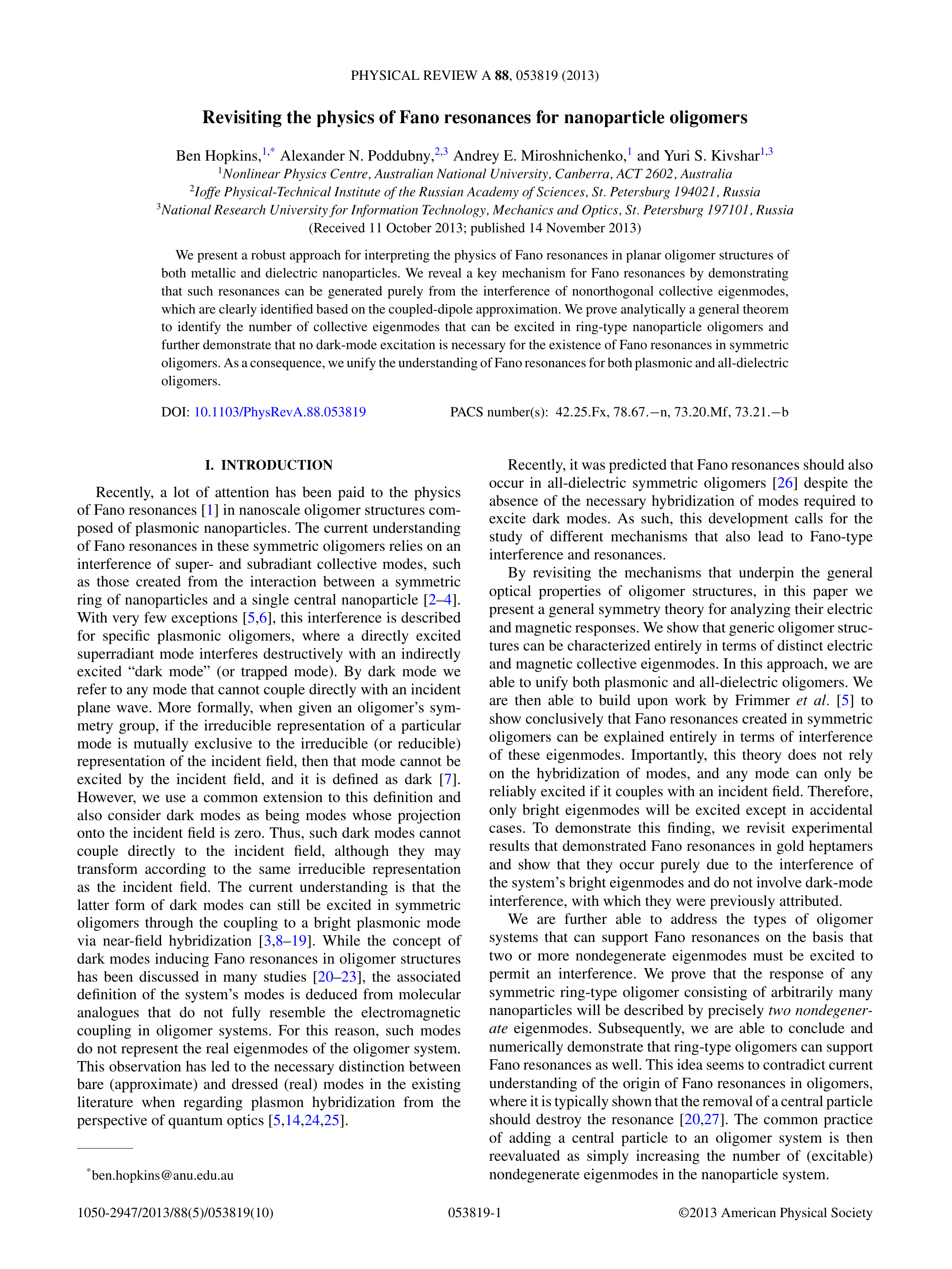}

\section[Interplay of magnetic responses in all-dielectric oligomers to realize magnetic Fano]{Interplay of magnetic responses in all-dielectric oligomers to realize magnetic Fano resonances \label{ACS}}

\subsection*{Summary}
Optically induced magnetic response was one of the key motivations in the development of metamaterials, while there was also a more focused interest in optically magnetic responses that exhibit plasmonic-like interference features, particularly Fano resonances.
Here it is demonstrated that oligomers made of high-index dielectric nanoparticles can support multiple dimensions of magnetic dipolar response; the magnetic dipole moment of an individual dielectric nanoparticle is shown to interact with the collective magnetic response of a nanoparticle oligomer.  
Using these two channels of magnetic dipolar response, a sharp and magnetic-magnetic Fano resonance was demonstrated in a symmetric high-index dielectric nanoparticle quadrumer. 
Such interference features had previously only been produced by the interaction between the electric and magnetic responses of asymmetric structures or nanoparticle clusters, demonstrating that high-index dielectric nanoparticle oligomers are a fully functional magnetic counterpart for plasmonic nanoparticle systems.

\subsection*{Notes and errata}
\begin{itemize}
\item
{
In equations (4) and (5), and impacting Figure 4, there are no additional scaling factors added between $\mathbf p$ and $\mathbf m$, and between $\mathbf E$ and $\mathbf H$, which were necessary to define eigenmodes with single units, as used in (\ref{eq:DmodelEig}).  
The substitution $\epsilon_0 = \mu_0 = c_0 = 1$ here will provide the same eigenmodes as defined in Chapter~\ref{chapterEigen}. 
}
\item{The curl of the dyadic Green's function is a factor of ${ik}$ larger than it should be, however all equations are correct once making the substitution: $\boldsymbol \nabla \times \hat G_0 \rightarrow \frac{1}{ik}\boldsymbol \nabla \times \hat G_0  $.}
\end{itemize}
\newpage
\includepdf[pages=-, offset=8mm 0, noautoscale = true, scale = 0.9]{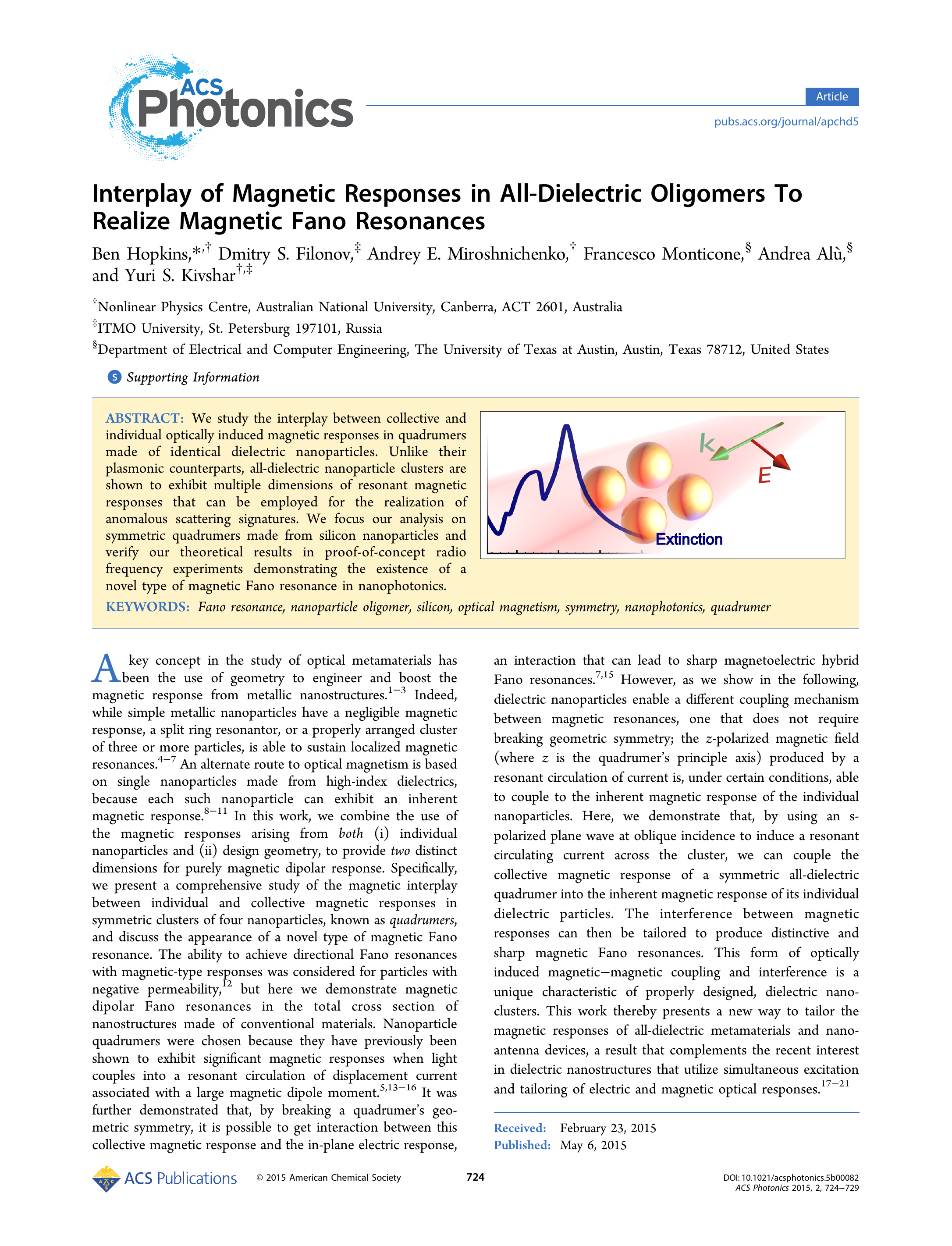}

\includepdf[pages=-, offset=8mm 0, noautoscale = true, scale = 0.9]{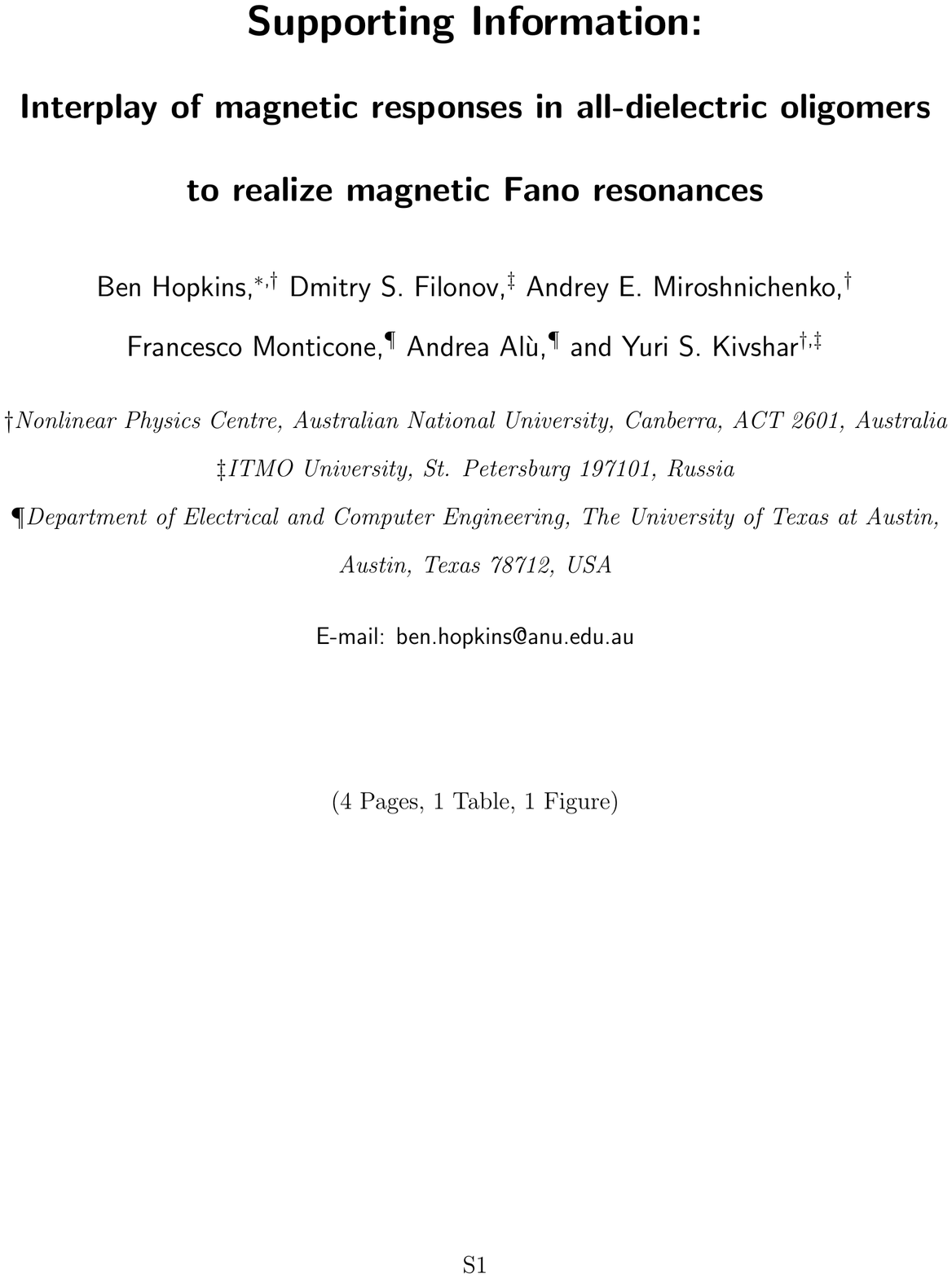}

\section{Fano resonance enhanced nonreciprocal absorption and scattering of light \label{Photonics}}

\subsection*{Summary}
It is well recognized that the total losses from light incident on a optical scattering system does not change if you invert the propagation direction; a consequence of reciprocity as discussed in Section~\ref{sec:tSym}.  
An optical device that distinguishes between opposite propagation directions therefore typically requires one breaks reciprocity by typically using strong magnetic fields with magnetic materials, or an optical systes with properties that depend on a DC current.  
However, we show here that asymmetric plasmonic nanostructures can exhibit abrupt differences in absorption and scattering properties for light that propagates in opposite directions, \textsl{without breaking reciprocity}.  
Such behaviour is derived to occur in the presence of nonorthogonal eigenmodes, and we show it can thereby be driven and enhanced through Fano resonances.   
This thereby highlights a new avenue by which nanoscale optical devices are able to measure and distinguish between oppositely propagating light without requiring external magnetic fields or magnetic materials, to thereby avoid a number of fabrication and operation challenges.  

\subsection*{Notes and errata}
\begin{itemize}
\item
{In the eigenmode equation (4), $\mathbf J$ should be replaced with $\boldsymbol v_i$.  The correct expression is (\ref{eq:eigenmode equation}).
}
\item{
The use of the word `nonreciprocal' is incorrect given the considered system is entirely reciprocal, the work instead refers to a difference of optical response due to reciprocal plane waves.  
}
\end{itemize}
\newpage
\includepdf[pages=-, offset=8mm 0, noautoscale = true, scale = 0.9]{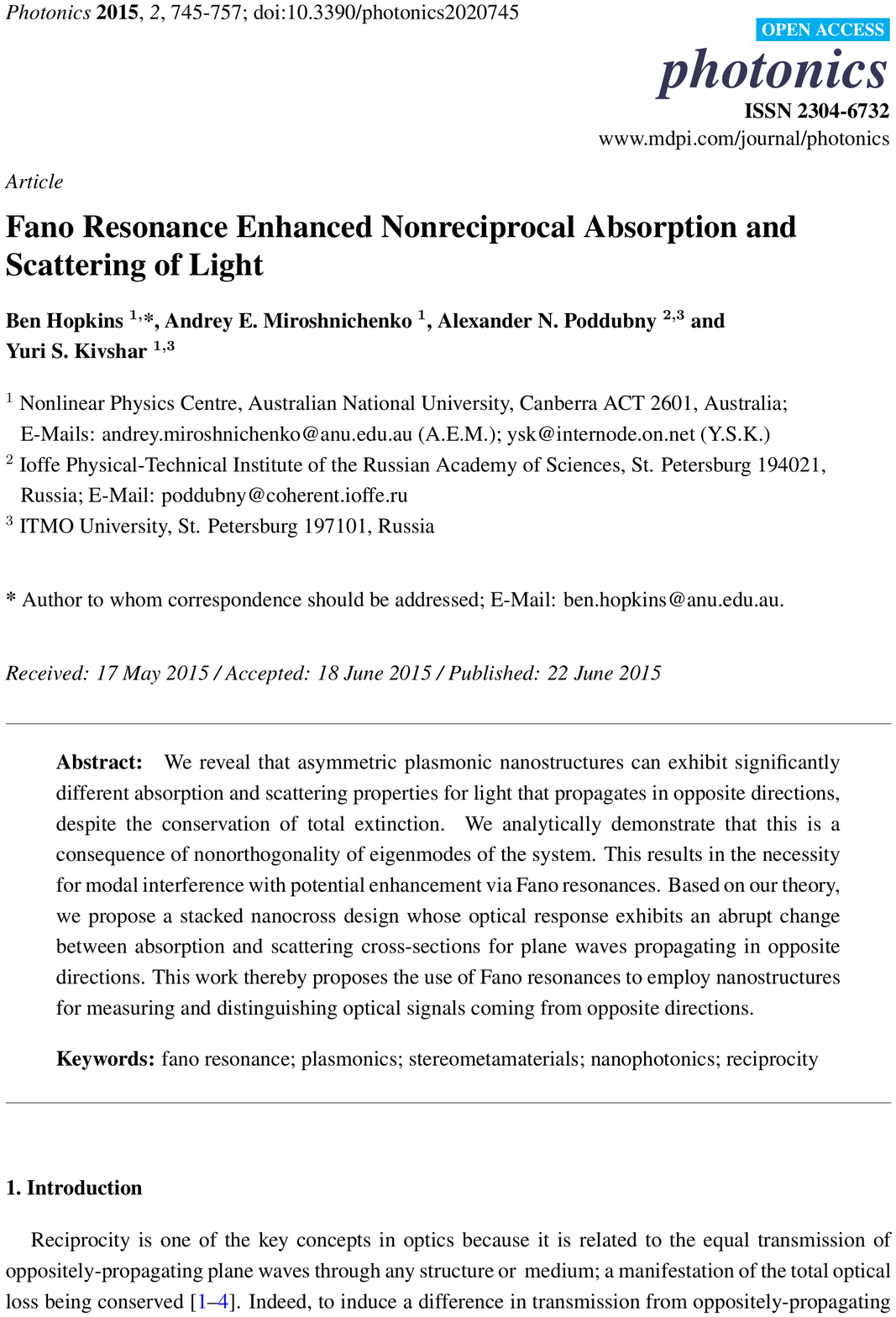}

\section[Hybridisation and the origin of Fano resonances in symmetric trimers]{Hybridisation and the origin of Fano resonances in symmetric nanoparticle trimers \label{PRB}}

\subsection*{Summary}
The analysis of optical scattering properties for nanostructures is commonly based on the concept of mode hybridisation, particularly when considering plasmonic materials.  
Yet, in the vast majority of studies, hybridisation of plasmons is performed in only a qualitative manner, largely due to complexity of the quantitative calculation. 
 Here we derive and present a simplified approach to the hybridisation procedure using the dipole model to investigate the optical response of a symmetric trimer. 
We are able to analytically follow the path by which hybridisation creates polarization-independent Fano resonances in symmetric nanoparticle trimers, a feature which has not yet been predicted in analyses that use plasmonic hybridisation theory.  
We are further able to show that the use of high-index dielectric particles provides an additional space of bianisotropic responses, which makes them more susceptible to Fano resonances compared to their plasmonic counterparts.  
The derived propensity of high-index dielectric trimers for Fano resonances is then demonstrated experimentally with full agreement to our theory.
We are subsequently able to provide a validation of our hybridisation procedure and demonstrate that simple dielectric nanoparticle systems can lead to pronounced Fano resonances.

\subsection*{Notes and errata}
\begin{itemize}
\item
{
In section III, the scaling factors necessary to define eigenmodes with single units, are not included between $\mathbf p$ and $\mathbf m$, nor between $\mathbf E$ and $\mathbf H$.  
Consequently, the eigenmodes of  section III depend on the choice of $\epsilon_0$ and $\mu_0$, unlike the eigenmodes as discussed in Chapter~\ref{chapterEigen}.
One can perform the substitution  $\epsilon_0 = \mu_0 = c_0 = 1$ in the expressions of section III to obtain the same eigenmodes as defined in Chapter~\ref{chapterEigen}. 
}
\item{
The expression for the doubly degenerate eigenmodes of a dielectric trimer is given as a function of the respective eigenvalue.  
It should be stated that the eigenvalue can be found relatively straightforwardly from the respective \textsl{characteristic equation}.  
An example of this is provided in \mbox{\S10.3.3} of~\cite{Agrawal2017}.
}
\item{The curl of the dyadic Green's function is defined to be a factor of ${ik}$ larger than it should be in (B2), however all equations are correct once performing the substitution $\boldsymbol \nabla \times \hat G_0 \rightarrow \frac{1}{ik}\boldsymbol \nabla \times \hat G_0  $.}
\end{itemize}
\newpage
\includepdf[pages=-, offset=8mm 0, noautoscale = true, scale = 0.9]{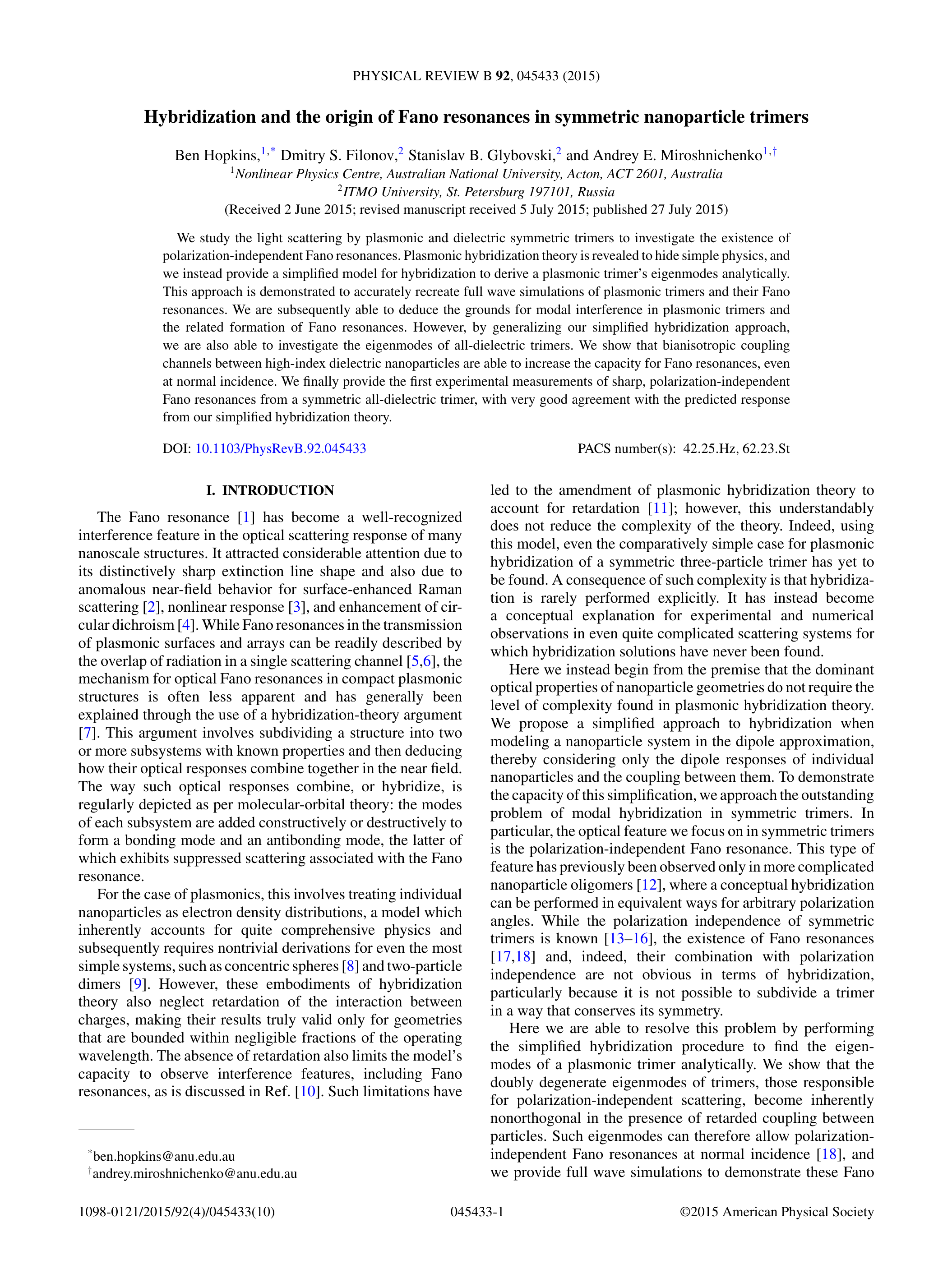}

\section{Circular dichroism induced by Fano resonances in planar chiral oligomers \label{LPOR}}
\subsection*{Summary}
The widespread desire for strong optical interactions with chiral molecules and proteins, present in many forms of biologicals and pharmaceuticals, has made nanoscale chiral materials and particles some of most imminently applicable devices in the broader field of optical metamaterials.  
However, it is relatively well-accepted that surfaces which are chiral in only two-dimensions cannot exhibit circular dichroism at normal incidence, a common metric used to gauge the total chiral response of a scattering object or medium.   
In this work, a new form of circular dichroism that \textsl{can exist} in planar systems  is presented, and is able to change the dominant loss mechanism of a plasmonic nanostructure between far-field radiation and near-field material absorption, both of which are key (and easily observable) forms of chiral response.  
To introduce this as a new circular dichroism effect, a ground-up description of its origin is provided, and it is shown that this form of circular dichroism is inherently linked to modal interference, and it can therefore be amplified through Fano resonances.  
This allows us to intuitively identify and demonstrate planar chiral oligomers as an obvious nanoparticle geometry that will exhibit this circular dichroism effect.

\subsection*{Notes and Errata}
\begin{itemize}
\item
{The equation (14) for the dipole polarisability of a current voxel is missing a factor of $\epsilon_0$; the correct expression is in the text between (\ref{eq:Pabs_dip}) and (\ref{eq:Pabs}).
The cascaded effects of this mistake are that the expressions for absorption in equation (16) is also missing a factor of $\epsilon_0$; the correct expression is (\ref{eq:Pabs}) after relating power $P$ to cross-section $\sigma$ as: $\frac{1}{2}\sqrt{\frac{\epsilon_0}{\mu_0}}\, |\mathbf{E_0}|^2 $.  
}
\end{itemize}
\newpage

\includepdf[pages=2-, offset=8mm 0, noautoscale = true, scale = 0.9]{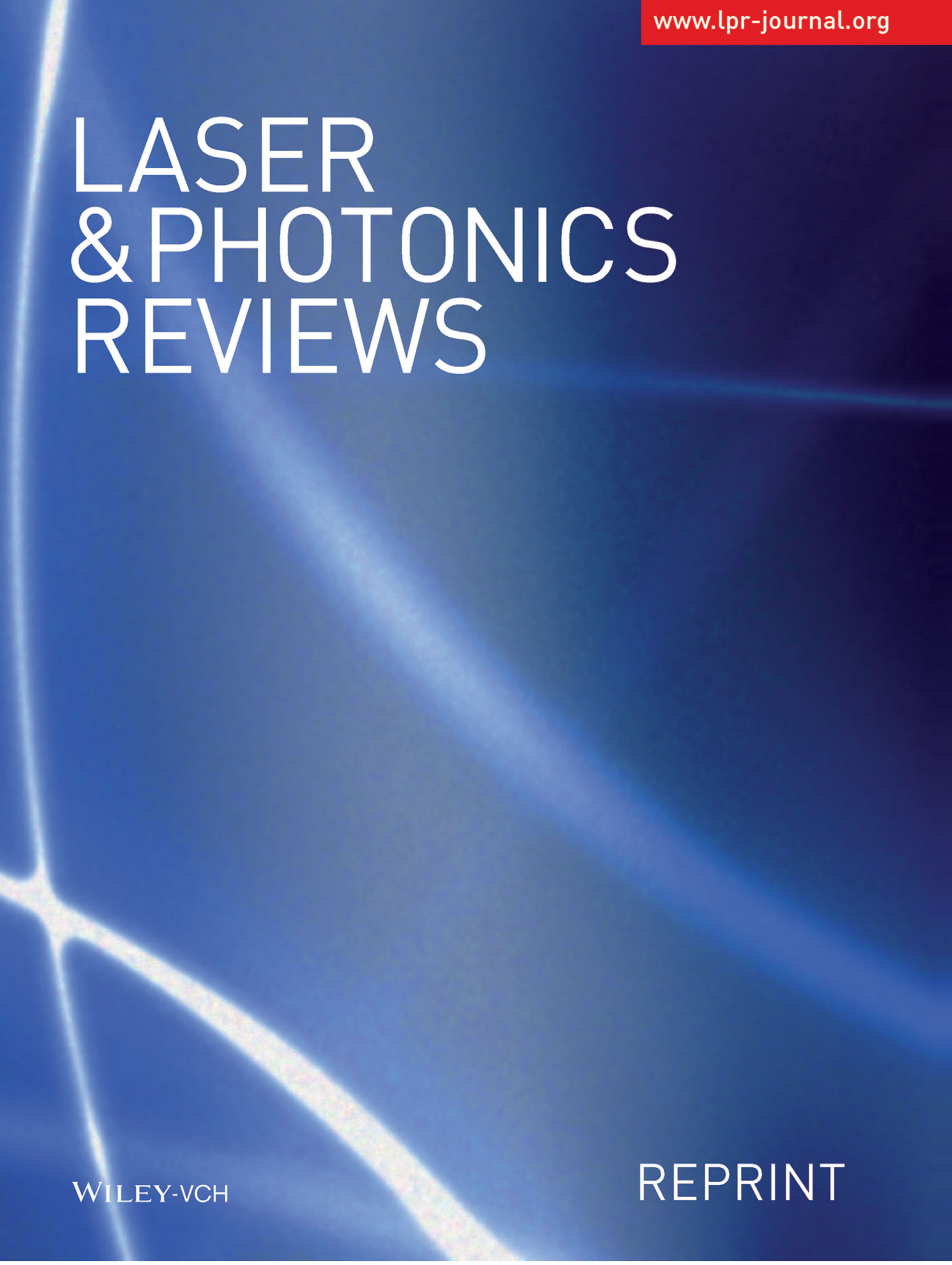}

\chapter{List of all publications} 

\begin{itemize}
\item{B. Hopkins, W. Liu, A.E. Miroshnichenko and Y.S. Kivshar, Optically isotropic responses induced by discrete rotational symmetry of nanoparticle clusters, {\it Nanoscale}~{\bf 5}, 6359-6403 (2013)}
\item{M. Rahmani, E. Yoxall, B. Hopkins, Y. Sonnefraud, Y.S. Kivshar, M. Hong, C. Phillips, S.A. Maier and A.E. Miroshnichenko*, Plasmonic Nanoclusters with Rotational Symmetry: Polarisation-Invariant Far-Field Response vs Changing Near-Field Distribution, {\it ACS Nano}~{\bf 7}:12, 11138-11146 (2013)}
\item{B. Hopkins, A.N. Poddubny, A.E. Miroshnichenko and Y.S. Kivshar, Revisiting the physics of Fano resonances for nanoparticle oligomers, {\it Phys. Rev. A}~{\bf 88}:5, 053819 (2013) }
\item{D.S. Filonov, A.P. Slobozhanyuk, A.E. Krasnok, P.A. Belov, E.A. Nenasheva, B. Hopkins, A.E. Miroshnichenko and Y.S. Kivshar, Near-field mapping of Fano resonances in all-dielectric oligomers, {\it Appl. Phys. Lett.}~{\bf 104}, 021104 (2014)}
\item{K.E. Chong, B. Hopkins, I. Staude, A.E. Miroshnichenko, J. Dominguez, M. Decker, D.N. Neshev, I. Brener and Y.S. Kivshar, Observation of Fano Resonances in All-Dielectric Nanoparticle Oligomers, {\it Small}~{\bf 10}:10, 1985-1990 (2014)}
\item{M.R. Shcherbakov, D.N. Neshev, B. Hopkins, A.S. Shorokhov, I. Staude, E.V. Melik-Gaykazyan, M. Decker, A.A. Ezhov, A.E. Miroshnichenko, I. Brener, A.A. Fedyanin and Y.S. Kivshar, Enhanced Third-Harmonic Generation in Silicon Nanoparticles Driven by Magnetic Response, {\it Nano Lett.}~{\bf 14}:11, 6488-6492 (2014)}
\item{M.R. Shcherbakov, A.S. Shorokhov, D.N. Neshev, B. Hopkins, I. Staude, E.V. Melik-Gaykazyan, A.A. Ezhov, A.E. Miroshnichenko, I. Brener, A.A. Fedyanin and Y.S. Kivshar, Nonlinear Interference and Tailorable Third-Harmonic Generation from Dielectric Oligomers, {\it ACS Photon.}~{\bf 2}, 578−582 (2015)}
\item{B. Hopkins, D.S. Filonov, A.E. Miroshnichenko, F. Monticone, A. Alù and Y.S. Kivshar, Interplay of Magnetic Responses in All-Dielectric Oligomers To Realize Magnetic Fano Resonances, {\it ACS Photon.}~{\bf 2}, 724−729 (2015).}
\item{B. Hopkins, A.E. Miroshnichenko, A.N. Poddubny and Y.S. Kivshar, Fano Resonance Enhanced Nonreciprocal Absorption and Scattering of Light, {\it Photonics}~{\bf 2}, 745-757 (2015)}
\item{B. Hopkins, D.S. Filonov, S.B. Glybovski and A.E. Miroshnichenko, Hybridization and the origin of Fano resonances in symmetric nanoparticle trimers, {\it Phys. Rev. B}~{\bf 92}, 045433 (2015)}
\item{B. Hopkins, A.N. Poddubny, A.E. Miroshnichenko and Y.S. Kivshar, Circular dichroism induced by Fano resonances in planar chiral oligomers, {\it Laser Photon. Rev.}~{\bf 10}:1, 137-146 (2016)}
\item{G. Geraci, B. Hopkins, A.E. Miroshnichenko, B. Erkihun, D.N. Neshev, Y.S. Kivshar, S.A. Maier and M. Rahmani, Polarisation-independent enhanced scattering by tailoring asymmetric plasmonic systems, {\it Nanoscale}~{\bf 8}, 6021-6027 (2016)}
\item{S. Kruk, B. Hopkins, I. Kravchenko, A.E. Miroshnichenko, D.N. Neshev and Y.S. Kivshar, Broadband highly efficient dielectric metadevices for polarisation control, {\it APL Photon.}~{\bf 1}, 030801 (2016)}
\item{A.S. Shorokhov, E.V. Melik-Gaykazyan, D.A. Smirnova, B. Hopkins, K.E. Chong, D.-Y. Choi, M.R. Shcherbakov, A.E. Miroshnichenko, D.N. Neshev, A.A. Fedyanin and Y.S. Kivshar, Multifold enhancement of third-harmonic generation in dielectric nanoparticles driven by magnetic Fano resonances, {\it Nano Lett.}~{\bf 16}:8, 4857-4861 (2016)}
\end{itemize}

\end{document}